\documentclass[10pt,journal,twocolumn]{IEEEtran}
\usepackage[noadjust]{cite}
\usepackage{multirow} 
\usepackage{algpseudocode}
\usepackage{algorithm}
\usepackage{rotating}
\usepackage{kantlipsum} 
\usepackage{commath}
\allowdisplaybreaks
\usepackage{mathtools}  
\usepackage{verbatim}
\usepackage{xr-hyper} 
\usepackage{enumitem}

\usepackage{epstopdf}
\usepackage{wrapfig}
\usepackage{latexsym}
\usepackage{amssymb}
\usepackage{amsthm}
\usepackage{amsfonts}
\usepackage{amsmath} 
\usepackage{graphicx}
\usepackage{latexsym}
\usepackage{booktabs}
\usepackage{support-caption} 
\usepackage{subcaption} 
\usepackage{xspace}
\usepackage[normalem]{ulem} 

\usepackage[T1]{fontenc}

\usepackage{breqn}

\algnewcommand\algorithmicinput{\textbf{INPUT:}}
\algnewcommand\INPUT{\item[\algorithmicinput]}
\algnewcommand\algorithmicoutput{\textbf{OUTPUT:}}
\algnewcommand\OUTPUT{\item[\algorithmicoutput]}

\newcommand{\dq}[1]{``#1''}

\usepackage[dvipsnames]{xcolor}

\newcommand{\commentBy}[3]{\textcolor{#1}{\textbf{#2:} #3}}

\newif\ifcommentson
\commentsontrue

\newcommand{\plv}[1]{\ifcommentson \commentBy{RoyalPurple}{PLV}{#1} \fi}
\newcommand{\ste}[1]{\ifcommentson \commentBy{blue}{SS}{#1} \fi}

\newcommand{\numCoveredCategories}{8\xspace}

\newcommand{\MonitoringPapers}{Eight\xspace}

\newcommand{\tePapers}{twenty two\xspace}

\newcommand{\FailurePapers}{Nine\xspace}

\newcommand{\ccaPapers}{sixteen\xspace}

\newcommand{\PathencPapers}{Eight\xspace}

\newcommand{\networkProgPapers}{eight\xspace}

\newcommand{\perfevalPapers}{four\xspace}

\newcommand{\numTotalPapers}{88\xspace}

\newcommand{\numRFCStandardization}{14\xspace}

\newcommand{\numDrafttandardization}{50\xspace}

\newif\ifextended
\newif\ifshortver

\shortvertrue

\newcommand{\extended}[1]{\ifextended \ifshortver \textcolor{purple}{#1} \else \textcolor{black}{#1} \fi  \fi}
\newcommand{\shortver}[1]{\ifshortver \ifextended \textcolor{blue}{#1} \else \textcolor{black}{#1} \fi \fi}

\newif\ifrevision

\revisiontrue

\newcommand{\hide}[1]{}

\newcommand{\optional}[1]{\ignorespaces}
\begin{document}

\bstctlcite{IEEEexample:BSTcontrol}

\title{Segment Routing: a Comprehensive Survey of Research Activities, Standardization Efforts and Implementation Results}

\author{Pier Luigi Ventre, Stefano Salsano, Marco Polverini, Antonio Cianfrani,\\
Ahmed Abdelsalam, Clarence Filsfils, Pablo Camarillo, Francois Clad
\IEEEcompsocitemizethanks{\protect
\IEEEcompsocthanksitem P.L. Ventre is with with the Department of Electronic Engineering at the University of Rome Tor Vergata - Rome, Italy, E-mail: pier.luigi.ventre@uniroma2.it.
\IEEEcompsocthanksitem S. Salsano is with the Department of Electronic Engineering at the University of Rome Tor Vergata and with the Consorzio Nazionale Interuniversitario per le Telecomunicazioni (CNIT) - Rome, Italy, E-mail: stefano.salsano@uniroma2.it.
\IEEEcompsocthanksitem M. Polverini and A. Cianfrani are with Department of Information Engineering, Electronics and Telecommunications at the University of Rome Sapienza - Rome, Italy, E-mail: \{marco.polverini, antonio.cianfrani\}@uniroma1.it.
\IEEEcompsocthanksitem A. Abdelsalam, C. Filsfils, P. Camarillo and F. Clad are with Cisco Systems, E-mail: \{ahabdels, cfilsfil, pcamaril, fclad\}@cisco.com
}
\vspace{2ex}
\textbf{\\Revision R2 - June 2020}
\vspace{-3ex}
}

\markboth{\shortver{Submitted to IEEE Communications Surveys \& Tutorials}}
{P.L. Ventre \MakeLowercase{\textit{et al.}}: Segment Routing: A comprehensive survey of research activities, standardization efforts and implementation results}

\IEEEtitleabstractindextext{
\begin{abstract} 
Fixed and mobile telecom operators, enterprise network operators and cloud providers strive to face the challenging demands coming from the evolution of IP networks (e.g. huge bandwidth requirements, integration of billions of devices and millions of services in the cloud). Proposed in the early 2010s, Segment Routing (SR) architecture helps face these challenging demands, and it is currently being adopted and deployed. SR architecture is based on the concept of source routing and has interesting scalability properties, as it dramatically reduces the amount of state information to be configured in the core nodes to support complex services. SR architecture was first implemented with the MPLS dataplane and then, quite recently, with the IPv6 dataplane (SRv6). IPv6 SR architecture (SRv6) has been extended from the simple steering of packets across nodes to a general \textit{network programming} approach, making it very suitable for use cases such as Service Function Chaining and Network Function Virtualization. 
In this paper we present a tutorial and a comprehensive survey on SR technology, analyzing standardization efforts, patents, research activities and implementation results. We start with an introduction on the motivations for Segment Routing and an overview of its evolution and standardization. Then, we provide a tutorial on Segment Routing technology, with a focus on the novel SRv6 solution. We discuss the standardization efforts and the patents providing details on the most important documents and mentioning other ongoing activities. We then thoroughly analyze research activities according to a taxonomy. We have identified \numCoveredCategories main categories during our analysis of the current state of play: Monitoring, Traffic Engineering, Failure Recovery, Centrally Controlled Architectures, Path Encoding, Network Programming, Performance Evaluation and Miscellaneous. We report the current status of SR deployments in production networks and of SR implementations (including several open source projects). Finally, we report our experience from this survey work and we identify a set of future research directions related to Segment Routing.
\end{abstract}

}
\maketitle
\IEEEdisplaynontitleabstractindextext
\IEEEpeerreviewmaketitle

\section{Introduction}
\label{sec:intro}

\IEEEPARstart{S}egment Routing (SR) is based on the loose Source Routing concept. A node can include an ordered list of instructions in the packet headers. These instructions steer the forwarding and the processing of the packet along its path in the network.

The single instructions are called \textit{segments}, a sequence of instructions can be referred to as a \textit{segment list} or as an \textit{SR Policy}. Each segment can enforce a topological requirement (e.g. pass through a node) or a service requirement (e.g. execute an operation on the packet). The term \textit{segment} refers to the fact that a network path towards a destination can be split into segments by adding intermediate way-points. The segment list can be included by the original source of the packet or by an intermediate node. When the segment list is inserted by an intermediate node, it can be removed by another node along the path of the packet, supporting the concept of \textit{tunneling} through an \textit{SR domain} from an \textit{ingress} node to an \textit{egress} node
. 

The implementation of the Segment Routing Architecture requires a data plane which is able to carry the segment lists in the packet headers and to properly process them. Control plane operations complement the data plane functionality, allowing the allocation of segments (i.e. associating a segment identifier to a specific instruction in a node) and the distribution of the segment identifiers within an SR domain.

As for the data plane, two instances of SR Architecture have been designed and implemented, SR over MPLS (SR-MPLS) and SR over IPv6 (SRv6). With SR-MPLS, no change to the MPLS forwarding plane is required \cite{id-segment-routing-mpls}. SRv6 is based on a new type of IPv6 routing header called SR Header (SRH) \cite{rfc8754}.

Temporally, SR-MPLS has been the first instance of SR Architecture, while the recent interest and developments are focusing on SRv6. In particular, the IPv6 data plane for SR is being extended to support the so-called SRv6 Network Programming Model \cite{id-srv6-network-prog}. According to this model, Segment Routing functions can be combined to achieve an end-to-end (or edge-to-edge) networking objective that can be arbitrarily complex. This is appealing for implementing complex services such as Service Function Chaining. SRv6 can be used as an \textit{overlay} tunneling mechanism directly exposed and used by servers (similar to VXLAN tunneling) and as an \textit{underlay} transport mechanism in network backbones (supporting Traffic Engineering and Resilience services). In light of this, SRv6 can simplify network architectures avoiding the use of different protocol layers. 

As for the SR Control Plane operations, they can be based on a distributed, centralized or hybrid architecture. In the distributed approach, the routing protocols are used to signal the allocation of segments, and the nodes take independent decisions to bind packets to the segment lists. In the centralized approach, an SR controller allocates the segments, takes the decision on which packets need to be associated to which SR Policy and configures the nodes accordingly. Very often, an hybrid approach which consists of the combination of the previous strategies is used (see for example \cite{ventre2018sdn}). 

The goal of this paper is to provide a comprehensive survey on Segment Routing technology, including all the achieved results and the ongoing work. Hereafter (section \ref{sec:sr-evolution}), we start with the historical context behind the development of Segment Routing. In section \ref{sec:arch}, we provide an introduction to the main concepts of Segment Routing architecture. We consider both the SR-MPLS and the SRv6 data planes, but we focus more deeply on SRv6 which is currently attracting a lot of interest. In section~\ref{sec:standard}, we provide a classification and a discussion of the standardization efforts together with an analysis of the most relevant patents. We provide a comprehensive review of research activities in section~\ref{sec:research}, covering \numTotalPapers scientific papers. The most relevant implementation results and the status of ongoing deployments are reported in section~\ref{sec:tools}. We report in section~\ref{sec:lesson} lessons learned and our experience. We highlight future research directions and open issues in section \ref{sec:future}.  

\subsection{Segment Routing roots and evolution}
\label{sec:sr-evolution}

The Source Routing approach consists of the inclusion of the route of the packet as a list of hops in the packet header. This has two variants. \textit{Strict} Source Routing requires the specification of the full sequence of hops from the source to the destination. \textit{Loose} source routing consists of specifying a list of nodes that represent \textit{way points} to be crossed (in their order) before reaching the destination.

These two variants of Source Routing have been considered among the possible solutions for packet routing and forwarding since the early phases of the design of packet switching technologies. In particular, they have been considered in the original definition of IPv4 protocol in the late 1970s. RFC 791 \cite{rfc791}, which defined IPv4 in 1981, included the \textit{Strict Source and Record Route} (SSRR) and the \textit{Loose Source and Record Route} (LSRR) options in the IPv4 packet header. These options have been rarely used in IPv4 networks, also due to security issues. Packets carrying the SSRR or LSRR options are typically filtered (dropped) by IPv4 routers on the Internet.

Segment Routing follows the \textit{loose} variant of Source Routing, using the same approach of IPv4 Loose Source Routing, but it is specifically based on MPLS or IPv6 data planes. 

The research and standardization activities on Segment Routing originated in the late 2000s. Their main goal was to overcome some scalability issues and limitations \cite{rfc5439}, which had been identified in the traffic engineered Multi Protocol Label Switching (MPLS-TE) solutions used for IP backbones. In particular it was observed that MPLS-TE requires explicit state to be maintained at all hops along an MPLS path; this may lead to scalability problems in the control-plane and in the data-plane. Moreover, the per-connection traffic steering model of MPLS-TE does not easily exploit the load balancing offered by Equal Cost MultiPath (ECMP) routing. On the other hand, Segment Routing can steer traffic flows along traffic engineered paths with no per-flow state in the nodes along the path and exploit ECMP routing within each segment. 

In the early 2010s, the IETF started the \dq{Source Packet Routing in Networking} Working Group (SPRING WG) to deal with Segment Routing. The activity of the SPRING WG has included the identification of Use Cases and Requirements for Segment Routing (for example, \cite{rfc7855}, \cite{rfc8355} and \cite{rfc8354} have become IETF RFCs). In July 2018, the SPRING WG has issued the \dq{Segment Routing Architecture} document (RFC 8402 \cite{rfc8402}) along with another informational RFC on monitoring use cases \cite{rfc8403}. More recently (December 2019) other 3 RFCs have been issued, while many other documents are still under discussion by the WG, as it will be analyzed later in this paper.

Looking at the scientific bibliography, the seminal paper on the Segment Routing Architecture is \cite{filsfils2015segment}. Published in 2014, it provides an overview of the motivations for SR, describes a set of important use cases and illustrates the architecture. The basic concepts proposed in \cite{filsfils2015segment} have been elaborated and refined in the RFC 8402 \cite{rfc8402}. 

Currently, Segment Routing is receiving a lot of interest from operators for its applications in different types of networks (transport backbones, access networks, datacenters and 5G networks).

The MPLS based data plane (SR-MPLS) relies on the well established MPLS technology. SR-MPLS can be seen as an improvement and a simplification of the traditional MPLS control plane, so it is beneficial to operators with an already deployed MPLS infrastructure. The IPv6 based data plane (SRv6) is gaining traction as it offers the possibility to combine overlay and underlay networking services and features only using the IPv6 technology. The SRv6 \textit{network programming model} offers unprecedented flexibility in designing and operating network services, so SRv6 is an attractive choice for operators that are deploying new networks or planning the evolution of their networking architectures.

We conclude this short historical review by noting that a very large number of patents (about 900) have been registered related to Segment Routing, as it is possible to verify with a cursory on-line search. We  provide an overview and analysis on the most relevant patents in section \ref{sec:standard}, they further demonstrate the high interest of vendors and service providers in SR technology.

\section{The Segment Routing (SR) architecture}
\label{sec:arch}

This section includes a short tutorial on the main aspects of SR architecture. Our goal is to provide a common ground and a conceptual framework reference for the survey. RFC 8402 \cite{rfc8402} represents the most important source of information for SR architecture. The work in \cite{sr-ietf-journal} (published in 2017) provides a short and effective introduction to Segment Routing architecture, with a focus on the MPLS data plane. The survey paper \cite{abdullah2018segment} has a section about SR architecture, which attempts to give more details related to both data plane (for SR-MPLS) and control plane aspects.  

\begin{figure}
    \centering
    \includegraphics[width=0.45\textwidth]{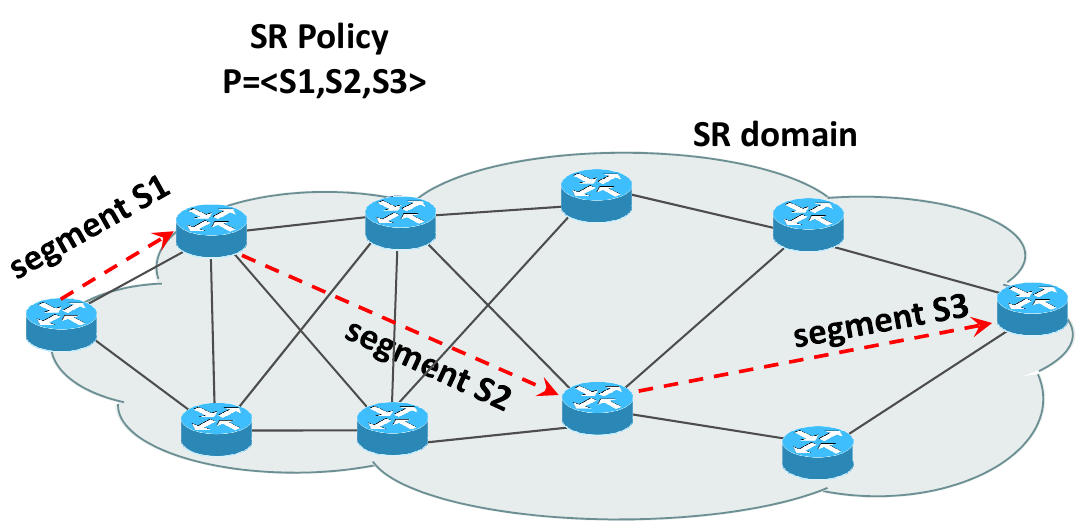}
    \caption{SR policy and segments}
    \label{fig:sr_policy_and_segments}
    \vspace{-3ex}
\end{figure}

Following the RFC 8402, let us start by discussing the general concepts of SR, which are independent from the specific data plane (MPLS or IPv6). A simple example of an SR path composed of three \textit{segments} (S1,S2,S3) is shown in Fig.~\ref{fig:sr_policy_and_segments}. We can refer to the list of segments as an \textit{SR Policy}: the Segment Routing policy P=<S1,S2,S3> consists of steering the packets through node S1, then through node S2 and then to the destination S3. The ordered list of segments (segment list) is inserted in the packet headers by the source node of the SR Policy. The Segment Routing domain (\textit{SR domain}) is the set of nodes participating in the source-based routing model.

\begin{figure}
    \centering
    \includegraphics[width=0.45\textwidth]{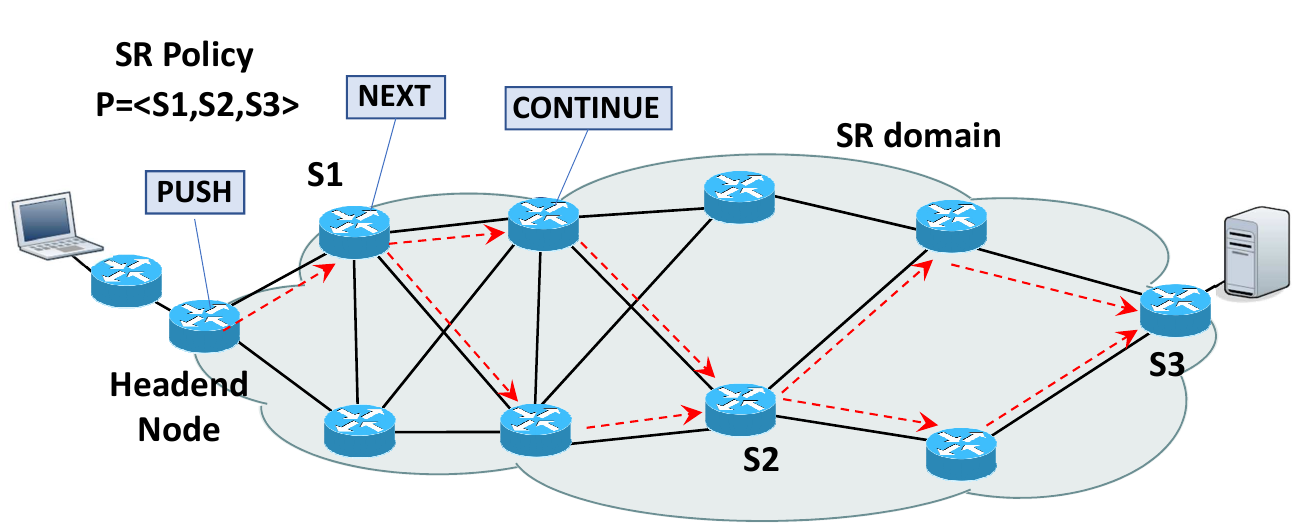}
    \caption{Segment Routing operations}
    \label{fig:sr_operations}
    \vspace{-3ex}
\end{figure}

A segment is described by a Segment Identifier (\textit{Segment ID} or \textit{SID}). For the MPLS data plane, a SID is an MPLS label, while for the IPv6 data plane a SID is an IPv6 address. As shown in Fig.~\ref{fig:sr_operations}, the segment list is added to the packet headers by a \textit{headend} node that \textit{steers} the packets of a flow onto the SR policy. The headend node can be the originator of the packet (e.g. an host) or an intermediate node that performs a classification of the traffic and associates the SR policies to the packets (as in Fig.~\ref{fig:sr_operations}). In other words, the hosts \textit{can} be part of an SR domain, but this is not required and depends on the overall scenario in which SR is applied. It is expected that all nodes in an SR domain are managed by the same administrative entity. For example, a Service Provider backbone can constitute an SR domain and the headend node will be the ingress edge router of the backbone (in this case, the hosts are not part of the SR domain). 

Three basic operations on SIDs and segment lists have been defined for a generic SR data plane: PUSH, NEXT and CONTINUE. In the examples of Fig.~\ref{fig:sr_policy_and_segments} and Fig.~\ref{fig:sr_operations}, We assume for simplicity that S1 and S2 represent topological instructions and S3 is the destination node of the SR policy P, so that the policy P instructs the packet to cross two nodes identified by the SIDs S1 and S2 (in this order) and then to reach the node identified by the SID S3. 

The PUSH operation consists of the insertion of a segment on top of the segment list, i.e. as the new first segment of the SR policy. In order to build the SR policy P described above, the headend node executes the PUSH operations in this order: PUSH(S3), PUSH(S2), PUSH(S1). In an SR packet, the segment that specifies the instruction to be executed is called the \textit{active} segment. In the considered example with the SR policy P, the headend node will send the packet with active segment S1.

The NEXT operation is executed by a node that has processed the active segment and considers the next segment of the SR policy to be executed. In our example, the node identified with SID S1 receives the packet and performs the NEXT operation. The next segment is S2, which becomes the active segment so that the packet is forwarded toward S2. The NEXT operation also covers the case of the last node of an SR policy, in which the NEXT operation usually results in processing the packet according to regular IP forwarding. 

Finally, the CONTINUE operation is performed by nodes that are in the path between two segments. For example, the intermediate nodes in the path between S1 and S2 perform the CONTINUE operation. The path between S1 and S2 is not prescribed by the SR policy and will be chosen considering the regular IP routing toward node S2 in the SR domain. If there are multiple equal cost paths between nodes S1 and S2 (as in Fig.~\ref{fig:sr_operations}) and the ECMP (Equal Cost MultiPath) mechanism is supported by the IP routing in the SR domain, it can be conveniently exploited by Segment Routing. 

The segments can be classified into \textit{Global Segments} and \textit{Local Segments}. Global Segments correspond to instructions that are globally valid in an SR domain. Local segments correspond to instructions that are valid within a single node.

The typical example of a global segment is an instruction to forward packets toward a given destination IP network or a destination IP node. Considering that an IGP (Interior Gateway Protocol) routing protocol (e.g. OSPF or ISIS) is used in the SR domain, these instructions are called \textit{IGP-prefix segment} and \textit{IGP-node segment} (or simply prefix segment and node segment). All nodes in the SR domain can execute the prefix segment or node segment instructions by considering the path toward the destination network or destination node in their routing table.

The most important example of local segment is the instruction to forward a packet to a node identified as adjacent by the IGP routing protocol. This corresponds to sending the packet on a specific outgoing interface and can be executed only by a specific node. This instruction is called \textit{IGP-Adjacency Segment}. Thanks to the use of IGP-Adjacency segments, it is possible to prove that any path across an SR domain can be expressed by an SR Policy (which can include a combination of global and local segments) \cite{pmsr}. Local segments can also be used to represent service instructions to be executed in a given node. The mapping of global and local segments into Segment Identifiers (SIDs) and the distribution of the SIDs in an SR domain are different for SR-MPLS and SRv6 and will be discussed in the next subsections.

The \textit{IGP-Anycast Segment} is an IGP-Prefix segment that corresponds to an anycast prefix, i.e. a prefix advertised by a set of routers that can be used for High Availability or Load Balancing purposes. 


The \textit{Binding Segment} is used to associate an SR policy (i.e. a Segment List) to a SID (called Binding SID or \textit{BSID}) in a given node. This means that the node that processes the Binding SID replaces this segment with a Segment List: a packet received with the BSID as active segment will be steered according to the associated SR policy. In this way, the packet classification can be executed by a node X that adds the Binding SID in the SR Header. The node X does not need to know the details of the SR policy to be applied (i.e. the Segment List). Thanks to the BSID, the packet will be forwarded to a node Y, which is able to apply the SR Policy.

As an example, the node X can classify traffic for a given destination network N that requires \dq{low latency} and traffic for the same destination network N that requires \dq{low loss}. Node Y is an ingress node of a backbone that provides connectivity toward the network N. Two SR policies (\dq{low latency} and \dq{low loss}) are used to forward traffic toward the network N across the backbone. The respective lists of segments can change over time, based on Traffic Engineering considerations. Upon these changes, the node Y is re-configured to apply the current SR Policy to the packets identified by the Binding SID. Node X does not need to be reconfigured, as the Binding SIDs remain constant over time. This approach improves the scalability, resilience and service independence of the solutions based on Segment Routing.

Table~\ref{table-sr-mappings} summarizes the mapping of the SR concepts into the two data planes (MPLS and IPv6) and will be discussed in the next two subsections.

It is interesting to note how some requirements that led to the definition of the SR solutions are currently fulfilled by \dq{Over the top} (OTT) providers to deliver services with a lower degree of flexibility, using tunneling technologies such as GRE (Generic Routing Encapsulation) \cite{rfc2784} and VXLAN (Virtual eXtensible LAN) \cite{rfc7348}. These technologies allow to encapsulate traffic and forward packets toward remote nodes according to an overlay logical topology. Unfortunately, they come with a penalty, for example a protocol like VXLAN needs an L3 underlay to transport traffic and loses full L3 forwarding capabilities such as ECMP forwarding \cite{line}. They are not really forms of source routing and do not allow the user to define way-points to which traffic can be directed. To further elaborate, in \cite{line} it is reported an interesting use case where multi-tenancy in a datacenter fabric has been implemented using SRv6 as overlay/underlay instead of the commonly used technologies such as VXLAN with a drastic simplification of the architecture.

As of Service Function Chaining (SFC) scenarios, the Network Service Header (NSH) \cite{rfc8300} is a solution that works on top of the tunneling technologies. Therefore, NSH can be used in conjunction with Segment Routing, when SR is used only as a tunneling mechanism (enhanced with Traffic Engineering features). On the other hand, NSH can be seen as an alternative to Segment Routing for implementing the Service Function Chaining functionality. In this respect, \cite{metaswitch} and \cite{mayer2019efficient} elaborate on Service Function Chaining scenarios where SRv6 would allow for a full replacement of the NSH layer, leading to a simplification of the infrastructure and a reduction in the burden on the devices.

\begin{table}
\caption{\\Mapping SR concepts into SR-MPLS and SRv6}
\label{table-sr-mappings}
\begin{tabular}{|l|l|l|}
\hline
\textbf{Generic SR}                                          & \textbf{SR-MPLS} & \textbf{SRv6}                                                                                                                               \\ \hline
SR Policy                                                    & Label Stack      & \begin{tabular}[c]{@{}l@{}}Segment List (of IPv6\\ addresses) in the SR Header\end{tabular}                                                 \\ \hline
Active Segment                                               & Topmost Label    & \begin{tabular}[c]{@{}l@{}}IPv6 address indicated\\ in the IPv6 Destination Address\end{tabular}                                                  \\ \hline
\begin{tabular}[c]{@{}l@{}}PUSH\\ Operation\end{tabular}     & Label Push       & \begin{tabular}[c]{@{}l@{}}Adding an IPv6 in the Segment\\ List in the SR Header\end{tabular}                                               \\ \hline
\begin{tabular}[c]{@{}l@{}}NEXT\\ Operation\end{tabular}     & Label POP        & \begin{tabular}[c]{@{}l@{}}Decrementing the Segment Left\\ field, copying the active segment\\ in the IPv6 Destination Address\end{tabular} \\ \hline
\begin{tabular}[c]{@{}l@{}}CONTINUE\\ Operation\end{tabular} & Label Swap       & \begin{tabular}[c]{@{}l@{}}Forwarding according to IPv6\\ Destination Address\end{tabular}                                                  \\ \hline
\end{tabular}
\end{table}

\subsection{MPLS data plane (SR-MPLS)}
\label{sec:mpls-data plane}

The MPLS data plane (SR-MPLS) is specified in \cite{id-segment-routing-mpls}. For SR-MPLS, Segment Routing does not require any change to the MPLS forwarding plane. An SR Policy is instantiated through the MPLS Label Stack: the Segment IDs (SIDs) of a Segment List are inserted as MPLS Labels. 
The classical forwarding functions available for MPLS networks allow implementing the SR operations. The PUSH operation corresponds to the Label Push function, i.e. pushing an MPLS label on a packet. The NEXT operation corresponds to the Label Pop function, i.e. removing the topmost label. The CONTINUE operation corresponds to the Label Swap function, i.e. associating an incoming label with an outgoing interface and outgoing label and forwarding the packet on the outgoing interface. The encapsulation of an IP packet into an SR-MPLS packet is performed at the edge of an SR-MPLS domain, reusing the MPLS Forwarding Equivalent Class (FEC) concept. A Forwarding Equivalent Class (FEC) can be associated with an SR Policy.

The mapping of Segments to MPLS Labels (SIDs) is a critical process in the SR-MPLS data plane. In general cases, different routers in the SR domain could have different available ranges of labels to be used for Segment Routing. Therefore, each router can advertise its own available label space to be used for Global Segments called \textit{SRGB - Segment Routing Global Block} (in general, this label space can even be composed of a set of non-contiguous blocks). For this reason, in the SR domain Global Segments are identified by an index, which has to be re-mapped into a label, taking into account the node that will process the label. This methodology has been also covered by the patent \cite{US9559954B2} and defines the following.

\begin{figure}
    \centering
    \includegraphics[width=0.48\textwidth]{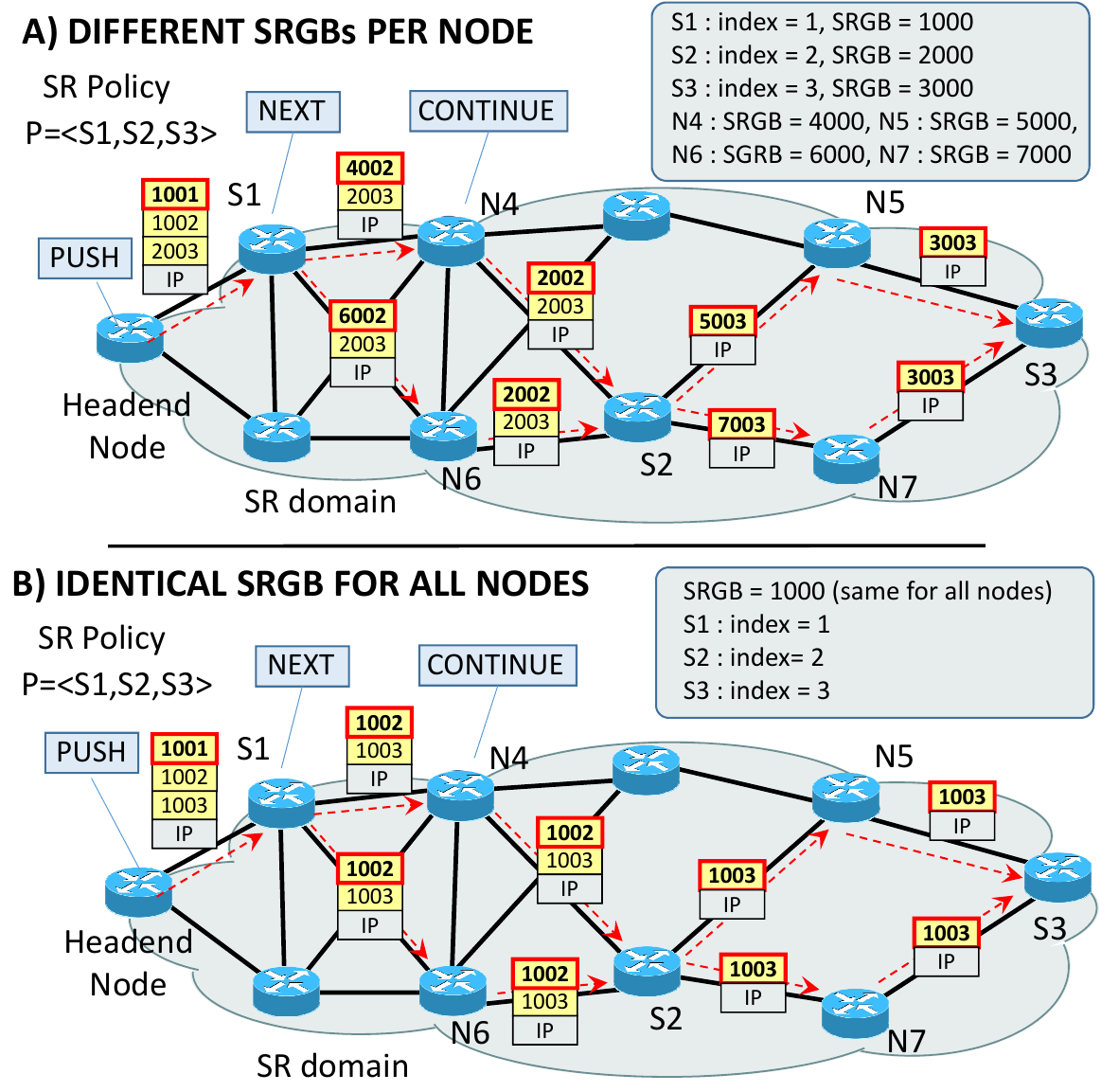}
    \caption{SR-MPLS data plane: mapping segments to labels using the SRGB}
    \label{fig:mpls-data plane}
    \vspace{-3ex}
\end{figure}

Assuming that the SRGB of a node is a label range starting from 10000, for a Global Segment with index X, the node needs to receive the label 10000+X. As an example, in Fig.~\ref{fig:mpls-data plane}~A we consider how to implement the SR policy described in Fig.~\ref{fig:sr_operations} using the SR-MPLS data plane. We assume that different nodes are using different SRGBs. The SRGBs of the nodes and the segment index associated to the segments S1, S2 and S3 are shown in the gray rectangle. The headend node needs to consider in advance which is the SRGB of the nodes that will perform the NEXT operation to the segments, because the label for the next segments needs to be crafted accordingly. In particular, the initial label for segment S2 set by the headend node will be 1002, i.e. the SRGB of node S1 (1000) plus the index for segment S2 (2). Node S1 will have to modify the label to 4002 if the packet is forwarded to node N4 (whose SRGB is 4000) or to label 6002 if the packet is forwarded to node N6 (whose SRGB is 6000). Both nodes N4 and N6 will remap (swap) the label to 2002 when forwarding the packets to S2. The initial label for node S3 set by the headend node is 2003, i.e. the SRGB of node S2 (2000) plus the index for segment S3 (3). This label will reach node S2 unmodified, then it will be properly processed by node S2 that will remap (swap) it considering the SRGB of the next hop in the path toward node S3.

This remapping process complicates operations and troubleshooting. There are also services (e.g. involving anycast segments) that cannot be realized if different SRGBs are used by different nodes. For this reason, \cite{rfc8402} strongly recommends that an identical range of labels (SRGB) is used in all routers, so that a Global Segment will always be mapped to the same SID (MPLS label) in all nodes. In Fig~\ref{fig:mpls-data plane}~B we present the mapping of the same SR policy  under the suggested operating mode in which an identical SRGB is used in all nodes. We observe that the MPLS labels do not need to be remapped, so that the same label consistently identifies the same global segment throughout the SR domain.

\subsection{IPv6 data plane (SRv6)}
\label{sec:ipv6 data plane}

\begin{figure}
    \centering
    \includegraphics[width=0.8\columnwidth]{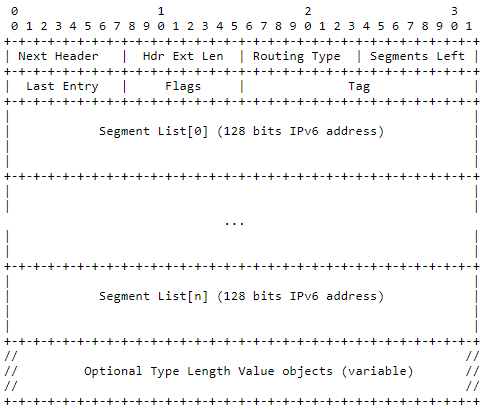}
    \caption{Segment Routing Header}
    \label{fig:sr-header}
\end{figure}

For the IPv6 data plane (SRv6), a new type of IPv6 Routing Extension Header, called Segment Routing Header (SRH) has been defined in \cite{rfc8754}. The format of the SRH is shown in Fig.~\ref{fig:sr-header}. The SRH contains the Segment List (SR Policy) as an ordered list of IPv6 addresses: each address in the list is a SID. A dedicated field, referred to as \textit{Segments Left}, is used to maintain the pointer to the active SID of the Segment List. 

\begin{figure}
    \centering
    \includegraphics[width=0.48\textwidth]{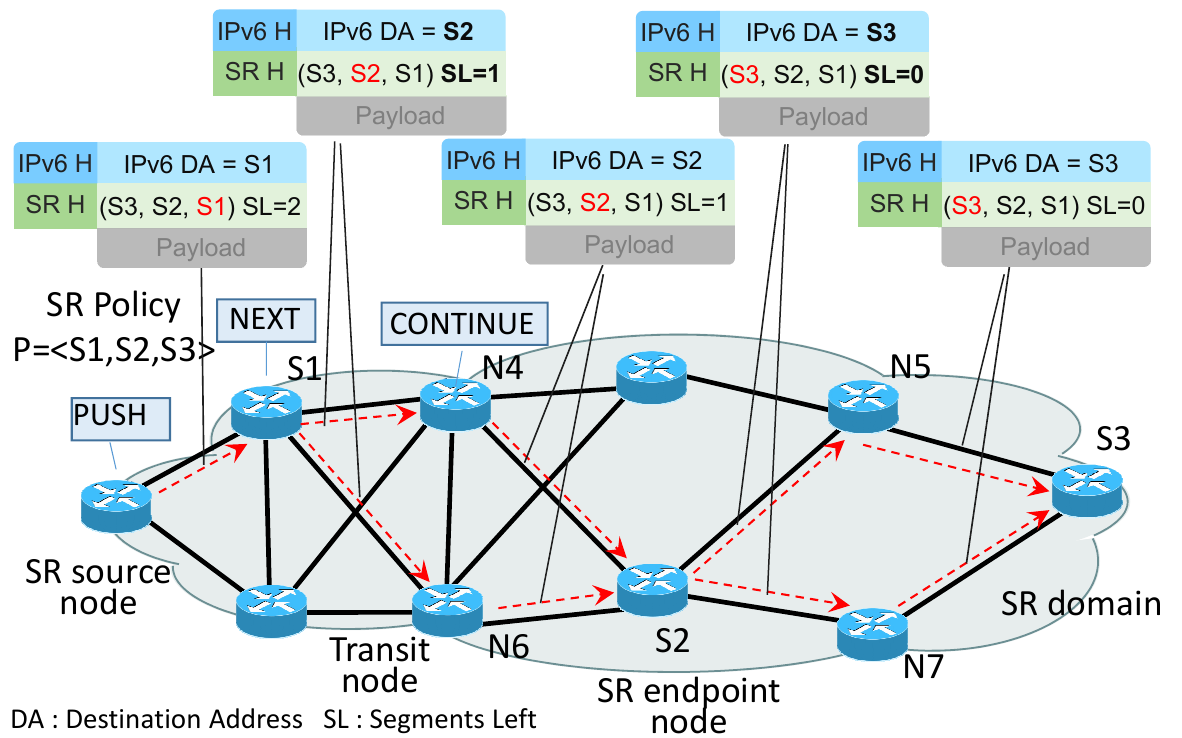}
    \caption{SRv6 data plane operations}
    \label{fig:srv6-data plane}
    \vspace{-3ex}
\end{figure}

In order to explain the SRv6 data plane, we consider three categories of nodes: Source SR nodes, Transit nodes and SR Segment Endpoint nodes. A Source SR node corresponds to the \textit{headend} node discussed above. It can be a host originating an IPv6 packet, or an SR domain ingress router encapsulating a received packet. In Fig.~\ref{fig:srv6-data plane} we consider the latter case, the Source SR node is an edge router that encapsulates a packet (which can be IPv6, IPv4 or even a Layer 3 frame) into an outer IPv6 packet and inserts the SR Header (SRH) as a Routing Extension Header in the outer IPv6 header. The encapsulated packet is indicated as Payload in Fig.~\ref{fig:srv6-data plane}. The Segment List in the SRH is composed of S1, S2 and S3 which are stored in reverse order (the fist SID is S3, the last segment in the SR policy). The Segment Left field is set to 2, so that the active segment is S1, represented in red in the figure. The Source SR node sets the first SID of the SR Policy (S1) as IPv6 Destination Address of the packet. These operations correspond to a sequence of the PUSH operations described above. 

The SR Segment Endpoint node receives packets whose IPv6 destination address is locally configured as a segment. The SR Segment Endpoint node inspects the SR header: it detects the new active segment, i.e. the next segment in the Segment List, modifies the IPv6 destination address of the outer IPv6 header and forwards the packet on the basis of the IPv6 forwarding table. These operations correspond to the NEXT operation described above. In Fig.~\ref{fig:srv6-data plane}, we can see that S1 is the first SR Endpoint node, it decrements the Segment Left fields to 1, making S2 the active segment, and sets S2 as IPv6 destination address.

A Transit node forwards the packet containing the SR header as a normal IPv6 packet, i.e. on the basis of the (outer) IPv6 destination address, because the IPv6 destination address does not locally match with a segment. These operations correspond to the CONTINUE operation. In Fig.~\ref{fig:srv6-data plane}, nodes N4, N5, N6 and N7 are Transit nodes, which perform a regular forwarding of the packet toward the IPv6 destination address. Note that in SRv6 the Transit nodes do not need to be SRv6 aware, as every IPv6 router can act as an SRv6 Transit node. 

In the given example, the PUSH operation is performed by encapsulating a packet (IPv6, IPv4 or Layer 2 frame) into an outer IPv6 packet with a Segment Routing Header. Another possibility is to perform the \textit{insertion} of an SRH as a new header between the IPv6 header and the Next Header (e.g. the Trasport Layer Header, TCP or UDP), without encapsulating the packet in a new IPv6 packet. This option only applies to IPv6 packets and it is especially suited in case the source host is acting as Source SR node (Headend node). 

In addition to the basic operations (PUSH/ NEXT/ CONTINUE), the \textit{SRv6 Network Programming} model \cite{id-srv6-network-prog} describes a set of functions that can be associated to segments and executed in a given SRv6 node. Examples of such functions are: different types of packet encapsulation (e.g. IPv6 in IPv6, IPv4 in IPv6, Ethernet in IPv6), corresponding decapsulation, lookup operation on a specific routing table (e.g. to support VPNs). The list of functions described in \cite{id-srv6-network-prog} (discussed in section~\ref{sec:sr_net_prog}) is not meant to be exhaustive, as any function can be associated to a segment identifier in a node. Obviously, the definition of a standardized set of segment routing functions facilitates the deployment of SR domains with interoperable equipment from multiple vendors.

According to \cite{id-srv6-network-prog}, we can revisit the notion of Segment IDentifier (SID) taking into account that IPv6 addresses are used as SIDs in SRv6. A 128 bit SID can be logically split into three fields and interpreted as LOCATOR:FUNCTION:ARGS (in short LOC:FUNCT:ARG) where LOC includes the L most significant bits, FUNCT the following F bits and ARG the remaining A bits, where 128=L+F+A.

The LOC corresponds to an IPv6 prefix (for example with a length of 48, 56 or 64 bits) that can be distributed by the routing protocols and provides the reachability of a node that hosts a number of functions. The length L of the locator is not fixed and can be chosen by each operator for its own SR domain (also independently for different nodes). All the different functions residing in a node can share the same locator and have a different FUNCT code, so that their SIDs will be different. From the routing point of view, the solution is very scalable, as a single prefix is distributed for a node, which implements a potentially large number of functions, with limited impact on the routing tables of the nodes in the SR domain. The ARG bits can be used to provide information (arguments) to a function. They are optional: if A=0, the SID can be simply decomposed in two fields as LOC:FUNCT, and 128=L+F.

A SID, that splites into LOC:FUNCT:ARG, is a global segment if the LOC prefix is routable in the SRv6 domain, which is the typical case. It is also possible to define local segments in SRv6, i.e. non routable SIDs that can be executed by a node and need to be preceded by a global SID used to forward the packet to the node. In any case, the use of such local SIDs can be avoided as the FUNCT part of a global SID in the form LOC:FUNCT:ARG can represent the needed local function.

\subsection{SRv6 Network Programming Model}
\label{sec:sr_net_prog}

The SRv6 Network Programming model is defined in \cite{id-srv6-network-prog}. It consists of combining functions that can reside in different nodes to achieve \textit{a networking objective that goes beyond mere packet routing}. The functions described in \cite{id-srv6-network-prog} can support valuable services and features such as layer 3 and layer 2 VPNs, traffic engineering and fast rerouting. The Network Programming model offers the possibility to implement virtually any service by combining the basic functions in a \textit{network program} that is embedded in the packet header.

As shown in Fig.~\ref{fig:sr-header}, the SRH can include an optional section that carries Type Length Value (TLV) objects. These TLV objects can be defined to transport information that needs to be elaborated by one or more segments of an SR policy (Segment List). For example, the so-called HMAC TLV can be added and used to verify that an SRH header has been created by an authorized node and that the segment list is not modified in transit. Another potential use of TLV objects is for exchanging Operation and Maintenance (OAM) information among the nodes of the SR domain.

The draft \cite{id-srv6-network-prog} defines two different sets of SRv6 behaviors, known as \textit{SR policy headend} and \textit{endpoint} behaviors. With reference to Fig.~\ref{fig:srv6-data plane}, \textit{SR policy headend} behaviors are executed in the SR source nodes, while \textit{endpoint behaviors} in SR endpoint nodes (e.g. S1, S2, S3). \textit{SR policy headend} behaviors steer received packets into the SRv6 policy, matching the packet attributes. Each SRv6 policy has a list of SIDs to be attached to the matched packets. Note that in an earlier version of \cite{id-srv6-network-prog}, the \textit{SR policy headend behaviors} were referred to as \textit{transit} behaviors, which was misleading because the same attribute (\textit{transit}) was applied to the SR source nodes and to the transit nodes not fulfilling any operation. On the other hand, an SRv6 \textit{endpoint} behavior, also known as behavior associated with a SID, represents a function to be executed on packets at a specific location in the network. Such function can be a simple routing instruction, but also any advanced network function (e.g., firewall, NAT).

Table~\ref{tab:sr-behaviors} reports a non-exhaustive list of SRv6 behaviors, listing the documents that provide their description.
The \textit{H.Encaps} behavior encapsulates an IPv6 packet, which becomes the inner packet of an IPv6-in-IPv6 packet. The outer IPv6 header carries the SRH header, which includes the SIDs list. The \textit{H.Encaps.L2} behavior is the same as the \textit{H.Encaps} behavior, with the difference that \textit{H.Encaps.L2} encapsulates the full received layer-2 frame rather than the IP packet (Ethernet over IPv6 encapsulation). The \textit{H.Insert} behavior inserts an SRH in the original IPv6 packet immediately after the IPv6 header and before the transport level header. The original IPv6 header is modified; specifically, the IPv6 destination address is replaced with the IPv6 address of the first segment in the segment list, while the original IPv6 destination address is carried in the SRH header as the last SID of the SIDs list.

The \textit{End} behavior represents the most basic SRv6 function among the endpoint behaviors. It replaces the IPv6 destination address of the packet with the next SID in the SIDs list. It then forwards the packet by performing a lookup of the updated IPv6 destination address in the routing table of the node. We will refer to the lookup in the routing table as \textit{FIB lookup}, where FIB stands for Forwarding Information Base. The \textit{End.T} behavior is a variant of the \textit{End} behavior, 
in which the FIB lookup is performed in a specific IPv6 table associated with the SID rather than in the main routing table. The \textit{End.X} behavior is another variant of the \textit{End} behavior, in which the packet is directly forwarded to a specified  layer-3 adjacency bound to the SID, without performing a FIB lookup of the IPv6 destination address.

The \textit{End.DT6} behavior pops out the SRv6 encapsulation and performs a FIB lookup of the IPv6 destination address of the exposed inner packet in a specific IPv6 table associated with the SID. It is possible to associate the default IPv6 routing table with the SID, in this case the inner IPv6 packets will be decapsulated and then forwarded on the basis of its IPv6 destination address according to the default routing of the node. The \textit{End.DX6} behavior removes the SRv6 encapsulation from the packet and forwards the resulting IPv6 packet to a specific layer-3 adjacency associated to the SID. \textit{End.DT4} and \textit{End.DX4} are respectively the IPv4 variant of \textit{End.DT6} and \textit{End.DX6}, i.e. they are used when the encapsulated packet is an IPv4 packet. The \textit{End.DX2} behavior is used for packets encapsulated at Layer 2 (e.g. with H.Encaps.L2). It pops out the SRv6 encapsulation and forwards the resulting L2 frame via an output interface associated to the SID. 

\begin{table}[htbp]
\caption{(Non-exhaustive) list of SRv6 behaviors}
\label{tab:sr-behaviors}
\centering
\begin{tabular}{|l|c|}
\hline
Behavior          &  Defined in \\
\hline
H.Encaps          & srv6-network-programming \cite{id-srv6-network-prog} \\
\hline
H.Insert          & srv6-network-programming \cite{id-srv6-network-prog}  \\
\hline
H.Encaps.L2       & srv6-network-programming \cite{id-srv6-network-prog} \\
\hline
End               & srv6-network-programming \cite{id-srv6-network-prog} \\
\hline
End.T             & srv6-network-programming \cite{id-srv6-network-prog} \\
\hline
End.X             & srv6-network-programming \cite{id-srv6-network-prog} \\
\hline
End.DT4           & srv6-network-programming \cite{id-srv6-network-prog} \\
\hline
End.DT6           & srv6-network-programming \cite{id-srv6-network-prog} \\
\hline
End.DX4           & srv6-network-programming \cite{id-srv6-network-prog} \\
\hline
End.DX6           & srv6-network-programming \cite{id-srv6-network-prog} \\
\hline 
End.DX2           & srv6-network-programming \cite{id-srv6-network-prog} \\
\hline
\hline
End.AS            & service-programming (SFC) \cite{id-sr-service-programming} \\
\hline
End.AD            & service-programming (SFC) \cite{id-sr-service-programming} \\
\hline
End.AM            & service-programming (SFC) \cite{id-sr-service-programming} \\
\hline
\hline
T.M.Tmap          & mobile-uplane \cite{id-srv6-mobile-uplane} \\
\hline
End.M.GTP4.E      & mobile-uplane \cite{id-srv6-mobile-uplane} \\
\hline
End.M.GTP6.D      & mobile-uplane \cite{id-srv6-mobile-uplane} \\
\hline
End.M.GTP6.E      & mobile-uplane \cite{id-srv6-mobile-uplane} \\
\hline
\end{tabular}
\end{table}



\subsection{Control plane for SR and relation with SDN}
\label{sec:sr_control_plane}

Control Plane operations are needed to complement the data plane functionality and provide a complete solution for Segment Routing. The Control Plane can be based on a fully distributed approach, in which the routers have the capability to take independent decisions to setup and enforce SR Policies. It can rely on a centralized SR controller that takes a decision and instructs the routers according to SDN principles, or on a combination of the two approaches (hybrid solution). 

For the SR-MPLS data plane, the definition of a fully distributed approach has been worked out within the IETF, with the definition of extensions to the IGP routing protocols (OSPF, ISIS, see \cite{ietf-ospf-segment-routing-extensions} \cite{ietf-ospf-ospfv3-segment-routing-extensions} \cite{ietf-isis-segment-routing-extensions}). 
These extensions to the routing protocols are used by each router to advertise the different types of IGP-segments (prefix, node, adjacency, anycast) and to distribute some SR configuration data. All other routers in the SR domain will receive this information by means of the IGP routing protocol. This information is needed to map the segments included in an SR policy into SIDs represented as MPLS labels in the SR-MPLS data plane. 
As we have discussed in subsection \ref{sec:mpls-data plane}, in the general case each router could allocate different ranges of labels to be used for Global Segments. The range of labels used for the global segments by a router, called \textit{SRGB - Segment Routing Global Block} is among the SR configuration information advertised using the routing protocol. We recall that it is strongly recommended to use an identical range of labels (SRGB) in all routers. 

For the IPv6 data plane, the process of advertising the IGP-prefix, IGP-node and IGP-anycast segments is simplified thanks to the use of IPv6 addresses as SIDs. In particular, there is no need to extend the IGP routing protocols to distribute these segment types, as they are natively distributed by the routing protocols. Also the definition of a Segment Routing Global Block as in the SR-MPLS is not needed, and the operations related to Global Segments can rely on IPv6 addresses that are globally routable in the SR domain. This means that the Control Plane for SRv6 can use the regular IGP routing protocols (OSPF, ISIS) to support the basic operations, while extensions are still needed (\cite{id-isis-srv6-extensions} \cite{li-ospf-ospfv3-srv6-extensions}) to distribute IGP-Adjacency segments and other SR configuration information. 

The definition of the control plane for Segment Routing has started from the SR-MPLS data plane and then the SRv6 data plane has inherited most of the functionality, which has been adapted to the new data plane. We observe that an original design goal of the control plane for Segment Routing has been to support the fully distributed approach, in which routers are capable of taking autonomous decisions. This allows the same functionality of a traditional MPLS network to be offered, which does not need a centralized SDN controller for its operations. On the other hand, we observe now a trend to focus on an hybrid approach in which distributed routing protocols coexist with an SR controller. This hybrid approach is aligned with the vision of Software Defined Networking, which aims to remove complexity from the devices and to centralize control plane function in SDN controllers.

In light of this, the Segment Routing architecture can be deployed by seeking the right balance between distributed and centralized control. The distributed control is used by the routers to exchange reachability information and evaluate the shortest paths in a traditional way; with no need to interact with the centralized controller. We observe that this is the best approach to provide connectivity in Wide Area Networks in which the control connections between the nodes and the SDN controllers are affected by non-negligible latency and failure probability. Segment Routing can be still used for Fast Reroute by pre-configuring SR policies that provide alternate paths in case of link or node failures, and these are automatically activated by the node when the failure happens.

The pre-calculation of such SR policies can be performed in a distributed mode or can be centralized in a controller; these concepts have also been explored in \cite{US9647944B2}, which patents a method for the orchestration of SR based WANs. Basic topology information and additional information for Traffic Engineering need be conveyed to the controller, as well as service related information that is advertised by nodes using distributed routing protocols. The SDN controller can receive this information in different ways. For example, it can participate to the IGP routing protocol or it can interact with routers in a proprietary way to extract their IGP databases. Otherwise, it can receive information by routers using extensions to BGP-LS (BGP-Link State) \cite{US9660897B1}. Whatever mechanism has been used to retrieve the needed information from the nodes, the SDN controller is in charge of taking decisions about the SR policies that implement advanced features or services such as Traffic Engineering, VPNs or Service Function Chaining. This approach allows the clear decoupling of the data plane operations from the service logic operating in the control plane.

The mechanisms and protocols for the SDN controller to enforce the SR policies by configuring  the nodes are left open in the definition of SR architecture. As mentioned in \cite{rfc8402}, some options are Network Configuration Protocol (NETCONF), Path Computation Element Communication Protocol (PCEP) \cite{ietf-pce-segment-routing}, and BGP. The OpenFlow protocol can be used as a mechanism to configure the SR policies only for SR-MPLS, while the processing of the SRv6 header is not supported by the latest standard version of the protocol. An Open Source implementation of a SouthBound API for SRv6 based on gRPC is reported in \cite{ventre2018sdn}. The main characteristic of the Segment Routing solution compared to other SDN solutions is that only the edge nodes need to be configured to enforce a given SR policy, while the internal nodes do not need to keep state per SR policy. This feature gives a substantial improvement in terms of scalability. 

\subsection{Segment Routing motivations and use cases}

As anticipated in the introduction section, the RFC 5439 \cite{rfc5439} has identified some scalability issues of traditional MPLS networks with Traffic Engineering support. These issues gave rise to an interest in defining a more scalable solution such as Segment Routing back in the late `00s. Several use cases and requirements for Segment Routing have been collected in a number of documents.

In \cite{rfc7855}, the main use cases identified are: MPLS tunneling (i.e. to support VPN services), Fast ReRoute (FRR) and Traffic Engineering (further classified in a number of more specific use cases). A set of Resiliency use cases is described in \cite{rfc8355}. In \cite{rfc8354}, the Segment Routing use cases for IPv6 networks are considered with a set of exemplary deployment environments for SRv6: Small Office, Access Network, Data Center, Content Delivery Networks and Core Networks.

\section{Standardization activities and patents}
\label{sec:standard}
In this section we propose a classification and description of the standardization activity related to Segment Routing and the analysis of the most relevant patents. We have classified  \numRFCStandardization Request For Comment (RFC) and \numDrafttandardization Internet Drafts. Our taxonomy is based on 7 categories and the result of the classification is shown in Table \ref{tab:standardization}. 


\begin{table}
\caption{\\Classification of the Standardization Activities documents}
\label{tab:standardization}
\begin{tabular}{|l|c|l|l|}
\hline
\multicolumn{3}{|l|}{\textbf{Architecture}}                                                                & \cite{rfc8402,id-segment-routing-policy,id-sr-policy-considerations,id-sr-policy-yang,id-segment-routing-mpls,id-srv6-network-prog,id-srv6-net-pgm-illustration,id-sr-service-programming} \\ \hline
\multicolumn{3}{|l|}{\textbf{Use-case and Requirements}}                                                   & \cite{rfc7855,rfc8355,rfc8354,id-segment-routing-msdc,id-segment-routing-central-epe,rfc8604,id-sr-for-sdwan,id-srv6-mobile-uplane,id-srv6-mobile-pocs,id-network-slicing-building-blocks,id-sr-traffic-counters,id-spring-poi-sr} \\ \hline
\multicolumn{3}{|l|}{\textbf{Fast Reroute (FRR)}}                                                          & \cite{id-segment-routing-ti-lfa,id-segment-routing-uloop,id-microloop-avoidance} \\ \hline
\multicolumn{3}{|l|}{\textbf{OAM}}                                                                         & \cite{rfc8403,rfc8287,id-srv6-oam,id-sr-traffic-accounting,id-bfd-sr-policy,id-srv6-oam} \\ \hline
\multicolumn{3}{|l|}{\textbf{Performance Measurement}}                                                     & \cite{id-sr-mpls-pm,id-udp-pm,id-spring-twamp-srpm,rfc6374,rfc7876} \\ \hline
\multirow{8}{*}{\parbox{1.3cm}{\textbf{Protocol Extensions}}} & \multirow{2}{*}{\textbf{Data Plane}}    & \textit{SR-MPLS} & \cite{ietf-spring-segment-routing-ldp-interop,ietf-spring-mpls-anycast-segments,filsfils-spring-sr-recursing-info,rfc8426,ietf-mpls-sr-over-ip} \\ \cline{3-4} 
                                              &                                         & \textit{SRv6}    & \cite{rfc8754,voyer-6man-extension-header-insertion,raza-spring-srv6-yang} \\ \cline{2-4} 
                                              & \multirow{6}{*}{\textbf{Control Plane}} & \textit{BGP}     & \cite{ietf-idr-bgp-prefix-sid,ietf-idr-segment-routing-te-policy,dawra-bess-srv6-services} \\ \cline{3-4} 
                                              &                                         & \textit{BGP-LS}  & \cite{ietf-idr-bgp-ls-segment-routing-ext,ietf-idr-te-lsp-distribution,ietf-idr-bgpls-segment-routing-epe,ietf-idr-bgp-ls-segment-routing-msd,rfc8571,ietf-idr-bgpls-srv6-ext,ketant-idr-bgp-ls-bgp-only-fabric,dawra-idr-bgp-ls-sr-service-segments} \\ \cline{3-4} 
                                              &                                         & \textit{IS-IS}   & \cite{ietf-isis-segment-routing-extensions,id-isis-srv6-extensions,ietf-lsr-flex-algo,id-lsr-flex-algo-yang,rfc8491,ietf-isis-l2bundles,rfc7810} \\ \cline{3-4} 
                                              &                                         & \textit{OSPF}    & 
                                              \begin{tabular}[c]{@{}l@{}}\cite{ietf-ospf-segment-routing-extensions,ietf-ospf-ospfv3-segment-routing-extensions,rfc8476,li-ospf-ospfv3-srv6-extensions}\\ \cite{ietf-lsr-flex-algo,id-lsr-flex-algo-yang,rfc7471}\end{tabular} \\ \cline{3-4} 
                                              &                                         & \textit{PCEP}    & \cite{ietf-pce-segment-routing,sivabalan-pce-binding-label-sid} \\ \cline{3-4} 
                                              &                                         & \textit{LISP}    & \cite{rodrigueznatal-lisp-srv6} \\ \hline
\end{tabular}
\end{table}

Hereafter we discuss the categories of the classification and then in the next subsection\extended{s} we report \shortver{an overview of the key standardization activities.}\extended{an overview of the key standardization activities and a detailed listing of all the documents.}
The first category is \textit{Architecture}, where all the documents regarding the description of the general architecture of a Segment Routing network are considered. The RFC 8402 \cite{rfc8402} falls into this category and describes the main features of SR, such as the source routing paradigm idea, the concept of SID and the definition of the two supported data planes.
In the category \textit{Use-case and Requirements} the documents describing use case scenarios for SR, e.g., use of SR in WANs, data center networks, mobility and network slicing, are inserted. Specifically, in this category there are three RFCs: i) RFC 7855 \cite{rfc7855} introducing the Source Packet Routing in Networking (SPRING), ii) RFC 8355 \cite{rfc8355} related to network resiliency using SR, and iii) RFC 8354 \cite{rfc8354} that describes how to steer IPv6 or MPLS packets over the SPRING architecture.

The third category is \textit{FRR} one, i.e, Fast Reroute realized through SR. The main standardization activity in this category is related to fast recovery after a link failure, and is referred to as Topology Independent Loop Free Alternate (TI-LFA), described in \cite{id-segment-routing-ti-lfa}. No RFC has been published in this category.
\textit{Operations, Administration, and Maintenance} (OAM) is the fourth defined category, where we include all the standardization activities related to tools used for maintenance of the network. As example, RFC 8287 \cite{rfc8287} focuses on the implementation of the ping and traceroute tools in SR-MPLS, while \cite{id-srv6-oam} does the same for SRv6.
In the \textit{Performance Measurement} category we consider all the documents describing measurement procedures related to performance parameters, such as delay and packet loss, in an SR network. We include in this category also the two specifications RFC 6374 \cite{rfc6374} and RFC 7876 \cite{rfc7876} that explain how to measure delay and packet loss in MPLS. Despite these two documents have not been produced during the standardization activities of SR, we decided to include them in Table \ref{tab:standardization} since they are massively used in the drafts for performance monitoring regarding SR.
Finally, the \textit{Protocol Extensions} category covers two different set of documents related to extensions of legacy protocols: i) data plane protocols extensions, and ii) control plane protocols extensions.
As for the data plane, we include the draft describing SR-MPLS \cite{id-segment-routing-mpls} and the RFC 8754 describing SRv6 \cite{rfc8754}.
As for the control plane, we the consider the documents on modifications to routing protocol (eg. BGP and OSPF) for the dynamic distribution of the SIDs in the SR network, or control protocol for the communication between a central controller (in case of centralized control plane) and the devices at the data plane (eg. PCEP).

As regards the patents, we have reported 18 documents. Our analysis has covered different type of documents including use case driven patents. Section \ref{sec:patents} elaborates on the surveyed documents. Among these, there are 4 patents covering the founding principles of Segment Routing: \hide{US9929946B2}\cite{US9929946B2}, \hide{US10063475B2}\cite{US10063475B2}, \hide{US9049233B2}\cite{US9049233B2} and \hide{US9049233B2}\cite{US9450829B2}. Another group of patents are related to the extensions of the routing protocols that are needed to transport SR related configurations \hide{US9660897B1}\cite{US9660897B1}, \hide{US10454821B2}\cite{US10454821B2} and \hide{US9565160B2}\cite{US9565160B2}. Then, there is a large group of documents covering specific mechanisms and use cases: \hide{US9369371B2}\cite{US9369371B2}, \hide{US9749187B2}\cite{US9749187B2}, \hide{US10182000B2}\cite{US10182000B2}, \hide{US9401858B2}\cite{US9401858B2}, \hide{US9647944B2}\cite{US9647944B2}, \hide{US9559954B2}\cite{US9559954B2}, \hide{US9485150B2}\cite{US9485150B2} and \hide{US9979601B2}\cite{US9979601B2}. Finally, during our research we have found a number of recent documents like \hide{US10601724B1}\cite{US10601724B1}, \hide{US10285155B1}\cite{US10285155B1} and \hide{US10635480B2}\cite{US10635480B2} that show the centrality of Segment Routing also in the deployment of future networks.

\subsection{Key standardization efforts}
\label{sec:key_standard}

In this subsection, we provide an overview of the most important standardization efforts, by considering 9 documents among the almost 70 listed in Table \ref{tab:standardization}. \cite{rfc8402} and \cite{id-segment-routing-policy} define key tenets of the SR architecture and discuss the benefits brought by SR in terms of scalability, privacy and security. \cite{id-sr-service-programming}, \cite{id-sr-for-sdwan} and  \cite{id-srv6-mobile-uplane} elaborate more on the support of key use cases like NFV/SFC, SD-WAN and next generation of mobile networks. \cite{id-srv6-network-prog} extends basic SR concepts and \cite{id-segment-routing-ti-lfa} provides Fast Re-route (FRR) mechanisms against single failures. Finally,  
\cite{id-segment-routing-uloop} and \cite{ietf-lsr-flex-algo} analyze the improvements of the routing stability and extensions to the routing protocols.

\cite{rfc8402} describes the Segment Routing architecture and its overall design. It defines the concept of a segment as a network instruction and presents the basic types of segments: prefix-SID, adjacency-SID, peering-SID and binding-SID. It also explains how such segments can be attached to data packets, leveraging the MPLS or IPv6 data planes, in order to steer traffic flows on any path in the network without requiring any per-flow state in the fabric.

\cite{id-segment-routing-policy} details the concept of an SR policy. It explains how Candidate Paths are defined as explicit SID-lists or as dynamically computed paths based on some optimization criteria, and how the active Candidate Path is selected. Moreover, it presents various ways of steering traffic into an SR Policy, automatically by coloring BGP service routes, remotely using a Binding-SID, or statically with route policies. The concepts described in this draft equally apply to the MPLS and SRv6 data planes.

The SR architecture is extended from the simple steering of packets across nodes to a general network programming approach in \cite{id-srv6-network-prog}. Using this framework, it is possible to encode arbitrary instructions and not only locations in a SID-list. Each SID is associated with a function to be executed at a specific location in the network. A set of basic functions are defined in~\cite{id-srv6-network-prog}, but other functions can be defined by network operators to fit their particular needs. Moreover, SID arguments allow functions to be provided additional context or their behavior to be tweaked on a per-flow basis.

\cite{id-sr-service-programming} defines the service SIDs and describes how to implement service programming (i.e. Service Function Chaining) in SR-MPLS and SRv6 enabled networks. The key tenet is to associate a SID to each network function (either physical or virtual). These service SIDs may be combined together in a SID-list and finely programmed by leveraging the network programming concept. They can also be combined with other types of SIDs to provide traffic engineering or VPN services. Service segments can be associated to legacy appliances (\textit{SR-unaware} VNFs, i.e. VNFs with no SRv6 capabilities), thanks to the SR proxy mechanisms which perform the SR processing and hide the SR information from the VNF.
The three \textit{endpoint} behaviors that has been defined in~\cite{id-sr-service-programming} for supporting Service Function Chaining are: \textit{End.AD}, \textit{End.AS} and \textit{End.AM}. The first two implement respectively a \textit{static} and a \textit{dynamic} SRv6 proxy for SR-unaware Virtual Network Function (VNF). They support IPv6 SR packets in \textit{H.Encaps} mode. The SRv6 proxy intercepts SR packets before being handed to the SR-unaware VNF, hence it can remove the SR encapsulation from packets. For packets coming back from the SR-unaware VNF, the SR proxy can restore the SRv6 encapsulation updating the SRH properly. The difference between the \textit{static} and the \textit{dynamic} proxies is that the SR information that needs to be pushed back in the packets is statically configured in the first case and it is \textit{learned} from the incoming packets in the \textit{dynamic} case. Instead, \textit{End.AM} implements the \textit{masquerading} proxy that supports SR packets travelling in \textit{H.Insert} mode.

\cite{id-sr-for-sdwan} explains how the SR technology enables underlay Service Level Agreements (SLA) for a VPN in a scalable and secure way, while ensuring service opacity. SR based VPNs are analyzed considering the case of a single provider and of multiple providers. Moreover, the draft addresses the control plane aspects of such solution, which are managed by an over the top SD-WAN controller. Finally, the benefits brought by the SR technology to VPN services are analyzed in term of scalability, privacy, billing management and security.

\cite{id-srv6-mobile-uplane} describes the applicability of SRv6 to the user plane of mobile networks. Three modes are addressed: traditional mode, enhanced mode and enhanced mode with interworking. In the traditional mode, the mobile user plane is unchanged except for the use of SRv6 as transport instead of GTP-U \cite{gtpu}. Enhanced mode uses only SRv6 and its programming framework. Finally, the enhanced mode with interworking uses SRv6 but provides also interworking functionality with legacy components still using GTP. The document describes a mechanism for end-to-end network slicing and defines the SRv6 behaviors for the SRv6 mobile user plane.
Among these behaviors, the most important ones define the functions for the coexistence of GTP-U \cite{gtpu} and SRv6. In particular, \textit{T.M.Tmap} translates a GTP-U over IPv4 packet to a SRv6 packet. Its counterpart is \textit{End.M.GTP4.E}, which maps an SRv6 packet to a GTP-U over IPv4 packet. Finally, \textit{End.M.GTP6.D} and \textit{End.M.GTP6.E} define respectively the translation of GTP-U over IPv6 packet to a SRv6 packet and SRv6 packet to a GTP-U over IPv6 packet.

Topology Independent Loop-free Alternate (TI-LFA) \cite{id-segment-routing-ti-lfa} provides Fast Re-Route (FRR) mechanisms protecting against link, node or local Shared Risk Link Groups (SRLGs) failures in SR enabled networks. For each destination in the network, a backup path is pre-computed and installed in the forwarding table, so that it is ready to be activated as soon as a failure is detected. For each destination, the backup path matches the post-convergence path, which is followed by the traffic after the network convergence. The draft analyzes also the benefits of using Segment Routing with respect to traditional FRR solutions.

\cite{id-segment-routing-uloop} describes a mechanism leveraging SR to prevent transient routing inconsistencies during the convergence period that follows a network topology modification. Instead of directly converging to a new next-hop after a topology modification, a node can temporarily steer the impacted traffic through a set of loop-free SR Policies, thus preventing it from being affected by routing inconsistencies. After the network has fully converged, the temporary SR Policies are removed with no impact on the traffic.

\cite{ietf-lsr-flex-algo} defines a set of extensions to the IGP routing protocols that enable Prefix-SIDs to be associated with operator-defined shortest path algorithms, called SR Flexible Algorithms (Flex-Algo). These algorithms are defined as an optimization metric (IGP, TE or delay) and a set of constraints (e.g., resources to be excluded from the path). Each node participating in a Flex-Algo computes the shortest paths to the Prefix-SIDs of that Flex-Algo and installs them in it forwarding table. SR Flexible Algorithms allow traffic to be steered along traffic-engineered paths such as low-latency or dual-plane disjoint path with a single Prefix-SID.

\extended{\subsection{Architecture}
\begin{itemize}
    \item Segment Routing Architecture \cite{rfc8402} RFC 8402 
    \item SR Policy Architecture \cite{id-segment-routing-policy} WG DOCUMENT
    \item SR Policy Architecture - Companion document \cite{id-sr-policy-considerations} DRAFT
    \item YANG Data Model for Segment Routing Policy \cite{id-sr-policy-yang} DRAFT
    \item Segment Routing with MPLS data plane \cite{id-segment-routing-mpls} WG DOCUMENT
    \item SRv6 Network Programming \cite{id-srv6-network-prog} and \cite{id-srv6-net-pgm-illustration} DRAFT
    \item Segment Routing for Service Programming \cite{id-sr-service-programming} DRAFT
    \item IGP Flexible Algorithm \cite{ietf-lsr-flex-algo} and \cite{id-lsr-flex-algo-yang}
\end{itemize}}

\extended{\subsection{Use-Cases and Requirements}
\begin{itemize}
    \item Source Packet Routing in Networking (SPRING) Problem Statement and Requirements \cite{rfc7855} RFC 7855
    \item Resiliency Use Cases in Source Packet Routing in Networking (SPRING) Networks \cite{rfc8355} RFC 8355 \ste{it should be updated on sr.net!} 
    \item Use Cases for IPv6 Source Packet Routing in Networking (SPRING) \cite{rfc8354} RFC 8354
    \item BGP-Prefix Segment in large-scale data centers \cite{id-segment-routing-msdc} WG DOCUMENT
    \item Segment Routing Centralized BGP Peer Engineering \cite{id-segment-routing-central-epe} WG DOCUMENT (Submitted to IESG for Publication)
    \item Interconnecting Millions Of Endpoints With Segment Routing \cite{rfc8604} RFC8604
    \item SR for SDWAN - VPN with Underlay SLA \cite{id-sr-for-sdwan} DRAFT
    \item Segment Routing IPv6 for Mobile User Plane \cite{id-srv6-mobile-uplane} DRAFT \ste{the name could be updated on sr.net}
    \item SRv6 for Mobile User-Plane 3GPP STUDY ITEM \ste{There is not a 3GPPP document with this name, there is this document ``Study on User-plane Protocol in 5GC'' which is still empty (Specification \#: 29.892)}
    \item Segment Routing IPv6 for mobile user-plane PoCs \cite{id-srv6-mobile-pocs} DRAFT \ste{the name could be updated on sr.net}
    \item Building blocks for Slicing in Segment Routing Network \cite{id-network-slicing-building-blocks} DRAFT
    \item Segment Routing Traffic Accounting Counters \cite{id-sr-traffic-counters} DRAFT
    \item Packet-Optical Integration in Segment Routing \cite{id-spring-poi-sr} DRAFT
\end{itemize}}

\extended{\subsection{Fast Reroute (FRR)}
\begin{itemize}
    \item Topology Independent Fast Reroute using Segment Routing \cite{id-segment-routing-ti-lfa} DRAFT
    \item Loop avoidance using Segment Routing \cite{id-segment-routing-uloop} DRAFT
    \item Micro-loop avoidance using SPRING \cite{id-microloop-avoidance} DRAFT EXPIRED \ste{do you want to mark expired drafts on sr.net?}
\end{itemize}}

\extended{\subsection{OAM}
\begin{itemize}
    \item A Scalable and Topology-Aware MPLS data plane Monitoring System \cite{rfc8403} RFC 8403
    \item Label Switched Path (LSP) Ping/Traceroute for Segment Routing (SR) IGP-Prefix and IGP-Adjacency Segment Identifiers (SIDs) with MPLS Data Planes \cite{rfc8287} \ste{update the name on sr.net?} RFC 8287
    \item Operations, Administration, and Maintenance (OAM) in Segment Routing Networks with IPv6 Data plane (SRv6) \cite{id-srv6-oam} \ste{update the name on sr.net?} DRAFT
    \item Traffic Accounting in Segment Routing Networks \cite{id-sr-traffic-accounting} DRAFT \ste{update the name on sr.net?}
    \item BFD for SR Policies \cite{id-bfd-sr-policy} DRAFT
\end{itemize}}

\extended{\subsection{Performance Measurement}
\begin{itemize}
    \item Packet Loss and Delay Measurement for MPLS Networks \cite{rfc6374} RFC 6374 \ste{is it a Segment Routing specification?} \plv{no}
    \item UDP Return Path for Packet Loss and Delay Measurement for MPLS Networks \cite{rfc7876} RFC 7876 \plv{no}
    \item Performance Measurement in Segment Routing Networks with MPLS Data Plane \cite{id-sr-mpls-pm} DRAFT
    \item UDP Path for In-band Performance Measurement for Segment Routing Networks \cite{id-udp-pm} DRAFT
\end{itemize}}

\extended{\subsection{Protocol Extensions}
\begin{itemize}
    \item IGP Flexible Algorithm WG DOCUMENT \cite{ietf-lsr-flex-algo} and \cite{id-lsr-flex-algo-yang}
\end{itemize}
\subsubsection{SR-MPLS}
\begin{itemize}
    \item Segment Routing interworking with LDP WG DOCUMENT \cite{ietf-spring-segment-routing-ldp-interop}
    \item Anycast Segments in MPLS based Segment Routing DRAFT \cite{ietf-spring-mpls-anycast-segments} 
    \item Segment Routing Recursive Information DRAFT \cite{filsfils-spring-sr-recursing-info}
    \item Recommendations for RSVP-TE and Segment Routing LSP co-existance \cite{rfc8426} RFC 8426
    \item SR-MPLS over IP DRAFT \cite{ietf-mpls-sr-over-ip}
\end{itemize}
\subsubsection{SRv6}
\begin{itemize}
    \item IPv6 Segment Routing Header (SRH) WG DOCUMENT \cite{ietf-6man-segment-routing-header}
    \item Insertion of IPv6 Segment Routing Headers in a Controlled Domain DRAFT \cite{voyer-6man-extension-header-insertion}
    \item YANG Data Model for SRv6 DRAFT \cite{raza-spring-srv6-yang}
\end{itemize}    
\subsubsection{BGP}
\begin{itemize}
    \item Segment Routing Prefix SID extensions for BGP WG DOCUMENT \cite{ietf-idr-bgp-prefix-sid}
    \item Advertising Segment Routing Traffic Engineering Policies in BGP WG DOCUMENT \cite{ietf-idr-segment-routing-te-policy}
    \item BGP Signaling of IPv6-Segment-Routing-based VPN Networks DRAFT \cite{dawra-bess-srv6-services}
\end{itemize}    
\subsubsection{BGP-LS}
\begin{itemize}
    \item BGP Link-State extensions for Segment Routing WG DOCUMENT \cite{ietf-idr-bgp-ls-segment-routing-ext}
    \item SR Policy Distribution via BGP-LS WG DOCUMENT \cite{ietf-idr-te-lsp-distribution}
    \item Segment Routing BGP Egress Peer Engineering BGP-LS Extensions WG DOCUMENT \cite{ietf-idr-bgpls-segment-routing-epe}
    \item Signaling Maximum SID Depth using Border Gateway Protocol Link-State WG DOCUMENT \cite{ietf-idr-bgp-ls-segment-routing-msd}
    \item BGP - Link State (BGP-LS) Advertisement of IGP Traffic Engineering Performance Metric Extensions RFC 8571 \cite{rfc8571}
    \item BGP Link State extensions for IPv6 Segment Routing (SRv6) DRAFT \cite{ietf-idr-bgpls-srv6-ext}
    \item BGP Link-State Extensions for BGP-only Fabric DRAFT \cite{ketant-idr-bgp-ls-bgp-only-fabric}
    \item BGP-LS Advertisement of Segment Routing Service Segments \cite{dawra-idr-bgp-ls-sr-service-segments}
\end{itemize}    
\subsubsection{IS-IS}
\begin{itemize}
    \item IS-IS Extensions for Segment Routing WG DOCUMENT \cite{ietf-isis-segment-routing-extensions}
    \item IS-IS Extensions to Support Segment Routing over IPv6 data plane DRAFT \cite{id-isis-srv6-extensions}
    \item Signaling MSD (Maximum SID Depth) using IS-IS RFC 8491 \cite{rfc8491}
    \item Advertising L2 Bundle Member Link Attributes in IS-IS WG DOCUMENT \cite{ietf-isis-l2bundles}
    \item IS-IS Traffic Engineering (TE) Metric Extensions RFC 7810 \cite{rfc7810}
\end{itemize}    
\subsubsection{OSPF}
\begin{itemize}
    \item OSPF Extensions for Segment Routing WG DOCUMENT \cite{ietf-ospf-segment-routing-extensions}
    \item OSPFv3 Extensions for Segment Routing WG DOCUMENT \cite{ietf-ospf-ospfv3-segment-routing-extensions}
    \item Signaling MSD (Maximum SID Depth) using OSPF RFC 8476 \cite{rfc8476}
    \item OSPFv3 Extensions for SRv6 DRAFT \cite{li-ospf-ospfv3-srv6-extensions}
    \item OSPF Traffic Engineering (TE) Metric Extensions RFC 7471 \cite{rfc7471}
\end{itemize}    
\subsubsection{PCEP}
\begin{itemize}
    \item PCEP Extensions for Segment Routing WG DOCUMENT \cite{ietf-pce-segment-routing}
    \item Carrying Binding Label/Segment-ID in PCE-based Networks DRAFT \cite{sivabalan-pce-binding-label-sid}
\end{itemize}    
\subsubsection{LISP}
\begin{itemize}
    \item LISP Control Plane for SRv6 Endpoint Mobility DRAFT \cite{rodrigueznatal-lisp-srv6}
\end{itemize}} 

\subsection{Relevant patents}
\label{sec:patents}

In this section we reviewed 18 among almost 1000 patents that can be found in patent search engines. It is interesting to note that SR is the central technology of the patent only in around 200 of the 1000 patents. For most of the patents there is a corresponding standardization activity (e.g. an internet draft or an RFC) promoted by the same vendor that produced the patent.

We will start by analyzing \cite{US9450829B2}, \cite{US9929946B2}, \cite{US10063475B2} and \cite{US9049233B2} which cover key tenets of the SR architecture.

Despite the fact that the name refers to MPLS and SR, \cite{US9049233B2} patents a primitive architecture for Segment Routing. It has likely been one of the first patents related to Segment Routing and basically describes a mechanism to map egress ports of the routers to labels, to advertise these data with a link state protocol and to forward data packets in the network with this information attached.

\cite{US9929946B2} is related to RFC8402 \cite{rfc8402}, and it includes the definition of a SID list and explains how a node can build such a stack of labels; it shows the forwarding behavior of the SR nodes and makes a clear distinction between the operations of an IP router, Label/MPLS router and SR router. Moreover, the patent covers the different types of SIDs, and for each them a definition is provided. Finally, different use cases for SR are illustrated.

\cite{US10063475B2} is another foundational document, which shares a lot of concepts with \cite{rfc8754}. It patents the usage of an IPv6 extension header for native implementation of Segment Routing in IP networks. The document also describes the SRv6 Extension Header along with the operations performed by the nodes to update the extension header and IPv6 header upon the reception of a packet. Instead, \cite{US9450829B2} describes the applicability of Segment Routing in a scenario where a computer network has been partitioned in multiple sub-networks.

The second group of documents mainly describe extensions to the routing protocols. For example, \cite{US9660897B1} describes extensions for BGP-LS to signal SIDs within an IGP domain. \cite{US10454821B2} describes a method for the creation and maintenance of SR policies through BGP protocol. The method relies on receiving BGP update messages with additional metadata referring to an SR policy. \cite{US9565160B2} covers several scenarios regarding the advertisement of adjacency segment identifiers. The method applies both to point-to-point links and LAN segments.

The vast majority of patents elaborate on the support of key use cases such as TE, SD-WAN, network monitoring, NFV/SDN, SFC and so on. \cite{US9369371B2} describes methods and an architecture for the monitoring of network paths which leverage SR technology. According to the patent, nodes can assemble a SID list with the aim of testing/monitoring a specific network path. It is possible find a lot of similarities with the works of the section \ref{sec:mon}.

\cite{US9749187B2} describes how SR can be used in a Label Distribution Protocol (LDP) domain and how MPLS nodes can coexist with nodes that are SR enabled. The patent also includes the definition of hybrid nodes that are capable of operating both as LDP enabled nodes and SR aware routers.

Three patents are related to fast reroute mechanisms and loop avoidance. In \cite{US10182000B2}, a method to detect and avoid loops into SR traffic engineered paths is described. It is based on the idea of defining for each tunnel a second path, which is exploited when the traffic is received from a node internal to the network. Instead, \cite{US9401858B2} defines various methods for loop avoidance upon network failures. Finally, \cite{US9485150B2} describes FRR mechanisms for Segment Routing based networks.

\cite{US9647944B2} defines a method for the orchestration of connectivity services in an SR-enabled WAN. The controller is considered outside the network. It is able to monitor the current state of the WAN and leverages a solver to perform traffic engineering. Decisions taken by the solver are enforced using SR technology. \cite{US9559954B2} describes a method for the assignment of the SIDs, which is based on assigning a global index value to each node and transmitting a base value to calculate the final segment identifiers. \cite{US9979601B2} describes an algorithm to “translate” an explicit path into a set of SIDs (like the works in the section \ref{sec:path}). The main idea of the proposed algorithm is to compute, starting from the source node, the shortest path towards the farthest node without ambiguity, i.e. with no multiple equal cost paths: a SID of the resulting Segment List is associated to each farthest node.

An analysis of the most recent patents highlights that SR represents a key technology also for the deployment of future networks in many different scenarios: i) in \cite{US10601724B1}, the implementation of a network slicing mechanism exploiting an SR-flexible algorithm is proposed; ii) \cite{US10285155B1} proposes a mechanism to report mobile node location information in 5G networks using SR; iii) in \cite{US10635480B2}, the flexibility of path encoding provided by SR is exploited to define a Virtual Machine migration procedure with zero loss \cite{desmouceauxzero}.

\section{Research activities}
\label{sec:research}

\extended{\cite{huang2018real}, \cite{gang2018throughput}, \cite{zhong2018improving}, \cite{basuki2019assuring}, \cite{spinelli2019chaining}, \cite{long2019scalable}, \cite{van2019rp1}, \cite{basuki2018sub}, \cite{jadin2019cg4sr}, \cite{wang2019segment}, \cite{li2019traffic}, \cite{liu2019load}, \cite{ch2018sdn}, \cite{yang2019qoe}, \cite{pereira2018segment}, \cite{roomisemi}, \cite{dominicini2019keysfc}, \cite{xie2019mitigating}, \cite{lacan2020xor}, \cite{li2020pasr}, \cite{kitsuwan2020single}, \cite{schuller2019failure} \cite{anbiah2019sr} \cite{odegbile2019scalable} \cite{aubry2020models}, \cite{li2020fast}, \cite{loreti2020implementation}, \cite{dong2019path}, \cite{zhang2020multipath}, \cite{tian2020traffic}, \cite{pereira2020traffic}, \cite{li2020traffic}, \cite{polverini2020theoretical}}

In this section, we describe research activities related to SR, and we provide two different classifications to characterize research papers on the basis of their main scope. We also show how to extract useful information regarding ongoing SR research activity from analyzing the relationship between the two classifications proposed.

The first classification proposed is based on the identification of seven main \emph{Research Categories}, as reported in Table~\ref{tab:res-taxonomy}.
The first one is the \emph{Monitoring} category, collecting all the works that describe and implement tools related to network monitoring activities.
Some examples include the measurements of the end to end delay over a given input route or the assessment of the volume of the traffic flows.
The second category is \emph{Traffic Engineering}, where we include all works proposing advanced routing strategies to optimize network performances.
The third category is \emph{Failure Recovery}, covering solutions to provide fast network recovery in the case of a node/link failure. Due to the time scale constraints, the works in this category are based on local mechanisms, i.e. not involving the central controller.
The fourth category defined is \emph{Centrally Controlled Architectures}, including all papers focusing on the implementation of an SR network with a centralized control plane realized on top of an underlay network (IP, SDN, MPLS). Here we point out that, despite some works classified as \emph{Traffic Engineering} being based on a centrally controlled architectures, they are not included in the \emph{Centrally Controlled Architectures} category. This is due to the fact that their main scope remains to optimize a TE goal (such the reduction of the congestion, or the minimization of energy consumption).
In the \emph{Path Encoding} category, we group all the papers that propose algorithms aiming to translate network paths into an SL. Specifically, we take as input a path in the form of a sequence of nodes and links that the generic path encoding algorithm provides as output a sequence of SIDs to be pushed in the packet header in order to steer the packet along the input path.
The sixth category is \emph{Network Programming}, where we inserted scientific works proposing solutions that exploit the programmability feature of SR. i.e. using service based SIDs to define the functions to be executed on packets crossing a specific segment list. A significant example of works falling into this category include works related to Service Function Chaining.  
Finally, we define a \emph{Miscellaneous} category, where we place all works not belonging to previous categories.


\begin{table}
\centering
\caption{\\Classification \#1 based on research categories}
\label{tab:res-taxonomy}
\begin{tabular}{|l|c|}
\hline
\multicolumn{1}{|c|}{\textbf{Category}}                                               & \textbf{References} \\ \hline
\textit{Monitoring (MON)}                                                                   & \cite{interoperable,scmon,li2018bandwidth,li2018ilp,polverini2018routing,cianfrani2018heuristic,polveriniNoF2018,xhonneux2018leveraging} \\ \hline
\textit{Traffic Engineering (TE)}                                                          & \cite{anefficient,novelsdn,ascalableand,energyefficient,ghuman2017per,trafficpmsr,ontraffic,optimizedte,roomi2018semi,defo1,gay2017expect,defo2,incrementaldeploy,moreno2017traffic,pereira2017optimizing,barakabitze2018novel,pang2017sdn,dugeon2017demonstration,hou2019optimization,settawatcharawanit2018segment,trimponias2019node,zhang2019bandwidth} \\ \hline
\textit{Failure Recovery (FR)}                                                             & \cite{timfa,trafficduplication,aubry2018robustly,optimizingrestoration,segmentfor,srdynamicrestoration,reliablesr,xhonneux2018flexible,foerster2018local} \\ \hline
\textit{\begin{tabular}[c]{@{}l@{}}Centrally Controlled\\ Architectures (CCA)\end{tabular}} & \cite{firstdemonstration,sdnandpce,paolucci2018network,paolucci2017service,castoldi2017segment,springopen,fressancourt2015sdn,li2017segment,lebrun2018software,duchene2018exploring,ventre2018sdn,demonstrationofsr,experimentalmulti,evolve,barakat2019busoni,eramo2019effectiveness} \\ \hline
\textit{Path Encoding (PEN)}                                                                & \cite{experimentaldemonstration,efficientlabel,pathencoding,labelencoding,pmsr,translating,liaoruo2018optimizing,guedrez2017new} \\ \hline
\textit{Network Programming (NP)}                                                          & \cite{srv62,17-vnf-chaining-sr,duchene2018srv6pipes,abdelsalam2018sera,srlb,desmouceaux2019content,desmouceauxzero,mayer2019efficient} \\ \hline
\textit{Performance Evaluation (PEV)}                                                                & \cite{ahmedperformance,teamsegment,leeperformance,abdelsalam2020srperf} \\ \hline
\textit{Miscellaneous (MISC)}                                                                & \cite{abdullah2018segment,cxp,pathlet,scalablesegment1,scalablesegment2,scalablesegment3,schuller2018practical,chi2018live,cao-industrial-iot,mayer-network-as-computer} \\ \hline
\end{tabular}
\end{table}

\begin{figure}
    \centering
    \includegraphics[trim=4cm 1.2cm 4cm 1cm,clip, width=1\columnwidth]{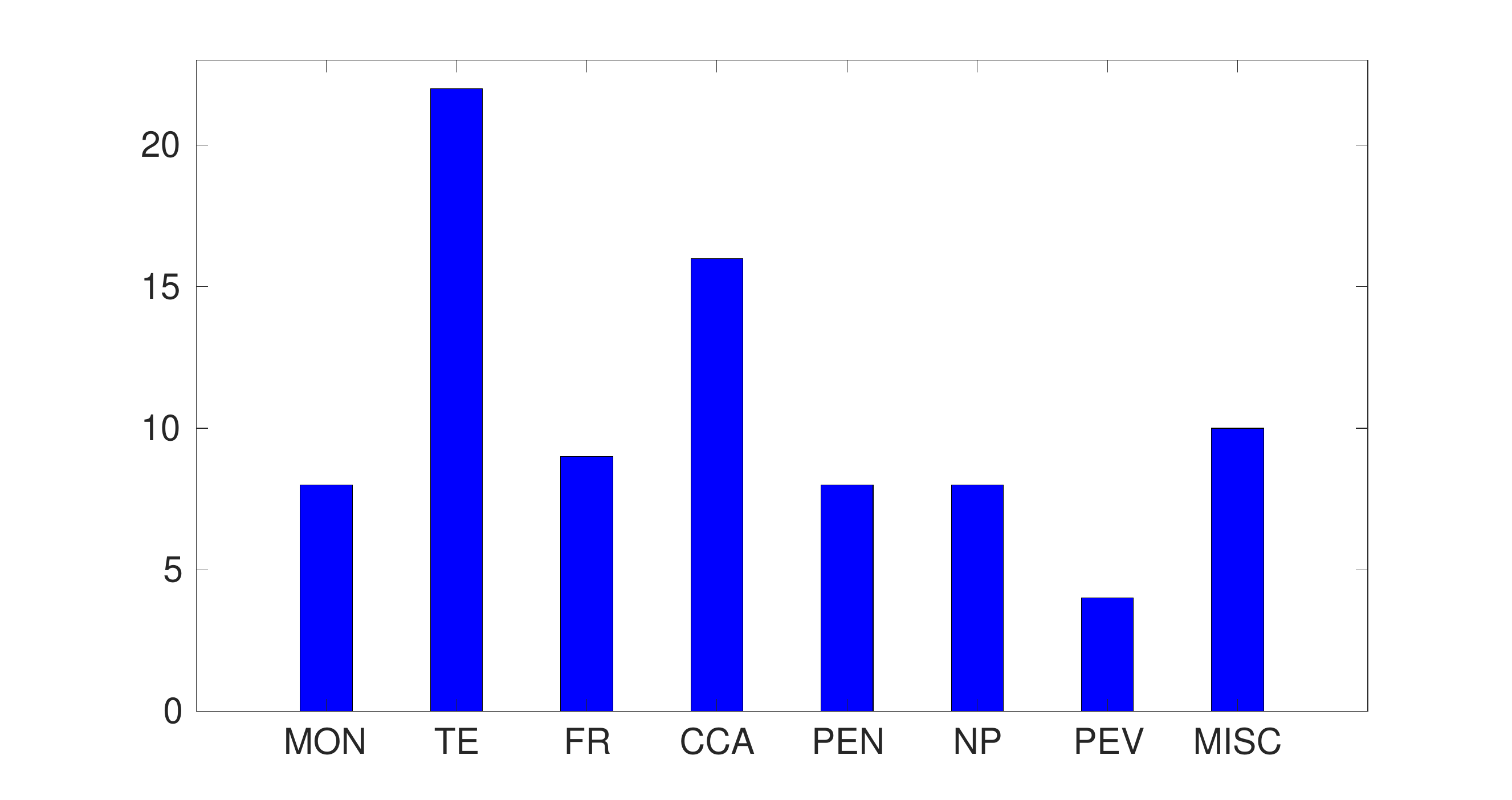}
    \caption{Number of references for each category of the defined taxonomy.}
    \label{fig:taxonomy1}
\end{figure}

In Fig.~\ref{fig:taxonomy1} we can see an histogram showing the number of references falling into each of the defined categories.
Analyzing Fig.~\ref{fig:taxonomy1}, it is evident that \emph{Traffic Engineering} and \emph{Centrally Controlled Architectures} are the most investigated subjects, while the other categories have been covered by almost the same number of works.
This behavior is quite expected since the main feature of SR is to define routing paths in a very flexible way and to make use of widely deployed data planes. This phenomenon is really appealing for the definition of new solution to TE problems and for the realization of overlay networks.
On the contrary, the number of papers related to \emph{Network Programming} could indicate a low interest in such a novel topic among the research community. We believe that \emph{Network Programming} represents a highly interesting research topic for near future, and that the lower number of available papers is only due to its early definition state.

In order to better investigate the research activity related to SR, we propose a second classification based on the specific SR topics considered in the research works. The new SR-related classification is reported in Tab.~\ref{tab:classification}. The different SR topics are also aggregated into three main groups:
\begin{itemize}
    \item SR feature exploitation;
    \item SR functions optimization;
    \item SR extensions.
\end{itemize}

The \emph{SR feature exploitation} group covers all works making use of SR features to solve classic networking problems, such as network resource optimization and performance improvement. The first SR topic is \emph{routing flexibility}, i.e., the possibility of steering a packet over a non trivial path (e.g., containing ECMP, loops, ect.).
The second feature is the \emph{source routing}, i.e. the capability of SR to instruct only the source node for the configuration of a specific network path.
The third SR feature is \emph{programmability}, i.e., the possibility to force a packet to go through a function by using specific SIDs.
All the remaining SR features are included in the \emph{other} topic.
Here, we merged the following topics together: i) the Adj-SIDs, used to force a packet to be forwarded on a specific output port; ii) the ECMP, i.e. the ability of SR to balance the traffic over multiple paths provided by the IGP routing protocols; iii) the Type Length Value (TLV) used to add optional data to the SR header; iv) the Binding SID (BSID) that allows a user to define SR tunnels in a transit node; v) the SR traffic counters (Base Pcounters and TM Pcounters), which collect traffic statistics based on the active segment carried in the packet headers; and vi) the spray policy, which allows the duplication of an incoming packet over multiple output ports and using different SLs. 

In the \emph{SR functions optimization} group, there is a prominence of research activities aimed at improving the inner functions of an SR network.
The first topic of the group is the \emph{SR Steering Policy}, i.e. the definition of policies to be installed on network devices in order to attach the proper SL to each incoming packet. The second topic is  \emph{SL length}, i.e. the number of SIDs of a segment list, that has an impact on SR header insertion and processing.

The last group, i.e. \emph{SR extensions}, is represented by a single topic (\emph{new functions}) and is related to new functionality implemented in SR to support advanced services.

Tab.~\ref{tab:classification} reports the classification of the research works according to the SR topics. Differently than Tab.~\ref{tab:res-taxonomy}, where the same reference can appear only once, in Tab.~\ref{tab:classification}, it can be classified under different categories. As a matter of example, \cite{translating} uses the \emph{Adj-SIDs} SR feature in order to perform \emph{SR functions optimization} from the point of view of \emph{SL length}.

\begin{table*}
\caption{\\Classification \#2 based on SR topics}
\label{tab:classification}
\begin{tabular}{|c|c|c|c|c|c|c|}
\hline
\multicolumn{4}{|c|}{\textbf{SR feature exploitation}}                                                                                                                                                                                                                                     & \multicolumn{2}{c|}{\textbf{SR functions optimization}}                                                                               & \textbf{Extend SR}                            \\ \hline
\multicolumn{1}{|c|}{\textit{Routing Flexibility}} & \textit{Source Routing}                                                        & \textit{Programmability}                                       & \textit{Others}                                                            & \multicolumn{1}{c|}{\textit{SR steering policy}}      & \textit{SL length}                                                     & \multicolumn{1}{c|}{\textit{New Functions}}   \\ \hline
\begin{tabular}[c]{@{}c@{}}\cite{scmon},\cite{polverini2018routing},\cite{cianfrani2018heuristic},\cite{novelsdn}\\ \cite{energyefficient},\cite{ghuman2017per},\cite{ontraffic},\cite{defo1}\\ \cite{gay2017expect},\cite{moreno2017traffic},\cite{hou2019optimization},\cite{settawatcharawanit2018segment}\\
\cite{trimponias2019node,eramo2019effectiveness,zhang2019bandwidth}\end{tabular}  & \multicolumn{1}{c|}{\begin{tabular}[c]{@{}c@{}}\cite{interoperable},\cite{scmon},\cite{li2018bandwidth}\\
\cite{li2018ilp},\cite{anefficient},\cite{novelsdn}\\
\cite{ascalableand},\cite{trafficpmsr},\cite{optimizedte}\\
\cite{roomi2018semi},\cite{defo1},\cite{gay2017expect}\\
\cite{incrementaldeploy},\cite{pereira2017optimizing},\cite{barakabitze2018novel}\\
\cite{pang2017sdn},\cite{dugeon2017demonstration},\cite{segmentfor}\\
\cite{firstdemonstration},\cite{sdnandpce},\cite{paolucci2018network}\\
\cite{paolucci2017service},\cite{castoldi2017segment},\cite{springopen}\\
\cite{fressancourt2015sdn},\cite{li2017segment},\cite{lebrun2018software}\\
\cite{duchene2018exploring},\cite{demonstrationofsr},\cite{experimentalmulti}\\
\cite{evolve}\end{tabular}} & \multicolumn{1}{c|}{\begin{tabular}[c]{@{}c@{}}\cite{mayer2019efficient},\cite{xhonneux2018leveraging},\cite{xhonneux2018flexible},\cite{srv62}\\\cite{17-vnf-chaining-sr},\cite{duchene2018srv6pipes},\cite{abdelsalam2018sera}\\\cite{srlb},\cite{desmouceaux2019content}\\
\cite{desmouceauxzero}\end{tabular}} & \multicolumn{1}{c|}{\begin{tabular}[c]{@{}c@{}}\textbf{BSID:}\cite{timfa},\cite{foerster2018local},\cite{lebrun2018software},\cite{duchene2018exploring},\cite{experimentalmulti}\\\cite{eramo2019effectiveness}\\ \textbf{ECMP:}\cite{optimizedte},\cite{moreno2017traffic},\cite{pereira2017optimizing}\\ \textbf{TLV:}\cite{xhonneux2018leveraging},\cite{timfa},\cite{xhonneux2018flexible},\cite{sdnandpce}\\ \textbf{Adj-SID:}\cite{pmsr},\cite{incrementaldeploy},\cite{efficientlabel},\cite{labelencoding},\cite{translating}\\ \textbf{Spray:}\cite{trafficduplication},\cite{aubry2018robustly}\\ \textbf{Traffic counters:}\cite{polveriniNoF2018}\end{tabular}} & \begin{tabular}[c]{@{}c@{}}\cite{ascalableand},\cite{gay2017expect},\cite{incrementaldeploy}\\ \cite{liaoruo2018optimizing},\cite{guedrez2017new}\end{tabular} & \multicolumn{1}{c|}{\begin{tabular}[c]{@{}c@{}}\cite{pmsr},\cite{anefficient}\\ \cite{trafficpmsr},\cite{incrementaldeploy}\\ \cite{moreno2017traffic},\cite{dugeon2017demonstration}\\ \cite{hou2019optimization},\cite{experimentaldemonstration}\\
\cite{efficientlabel},\cite{pathencoding}\\
\cite{labelencoding},\cite{translating}\\ \cite{liaoruo2018optimizing},\cite{guedrez2017new}\end{tabular}} & \begin{tabular}[c]{@{}c@{}}\cite{pmsr}, \cite{xhonneux2018leveraging}\\ \cite{segmentfor},\cite{srdynamicrestoration}\\ \cite{reliablesr},\cite{guedrez2017new}\\ \cite{abdelsalam2018sera},\cite{desmouceauxzero}\end{tabular} \\ \hline
\end{tabular}
\end{table*}

Tab.~\ref{tab:classification} shows that the most used SR feature is the \emph{source routing} one, while there is still a limited amount of works focusing on \emph{network programmability} and on the definition of \emph{new functions}.

\begin{figure}
    \centering
    \includegraphics[width=1\columnwidth]{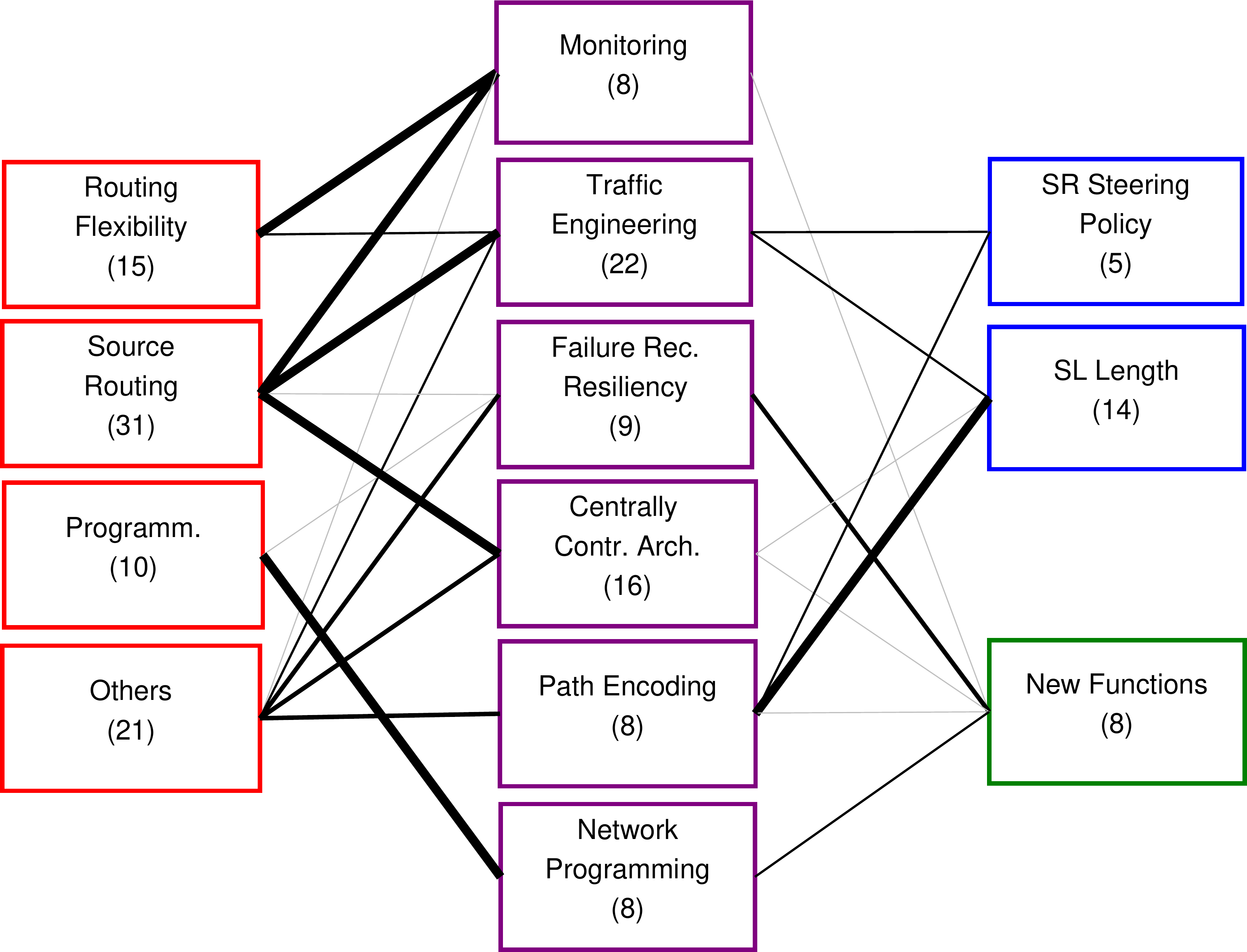}
    \caption{Graph showing the relation between the research categories and the SR topics.}
    \label{fig:taxonomy12}
\end{figure}

In order to obtain further insights into SR research activity, we defined a way to show the relationship between classifications reported in Tab.~\ref{tab:res-taxonomy} and Tab.~\ref{tab:classification}, respectively. In Fig.~\ref{fig:taxonomy12} we can see a graph which is defined in the following way: each node represents a research category (violet rectangles at the center of the figure) or an SR topic (divided in red, blue and green rectangles at the border of the figure), and an edge among a category and an SR topic is present only if both are present in the same work. Moreover, the thickness of an edge depends on the number of papers covering the specific category/topic pair. The graph in Fig.~\ref{fig:taxonomy12} shows several interesting outcomes: i) the works related to the \emph{Monitoring} category mainly exploit the routing flexibility and the source routing paradigm; ii) the same result is obtained for \emph{Traffic Engineering} works, but in this case SR steering policy and SL length topics are also covered; iii) in order to provide \emph{Failure Recovery} solutions, new functions generally need to be defined; iv) source routing is the most used SR feature for \emph{Centrally Controlled Architecture} solutions, since it allows the reduction of communication between the central controller and network devices; v) as expected, \emph{Path Encoding}  works are mainly focused on the optimization of the SL length (it is in any case interesting to note that the Adj-SID is the main SR feature used to achieve this scope); vi) finally, works falling into the \emph{Network Programming} category always exploit the programmability feature of SR, and, in some cases, new functions need to be defined.

\begin{figure}
    \centering
    \includegraphics[trim=4cm 1.2cm 4cm 1cm,clip, width=1\columnwidth]{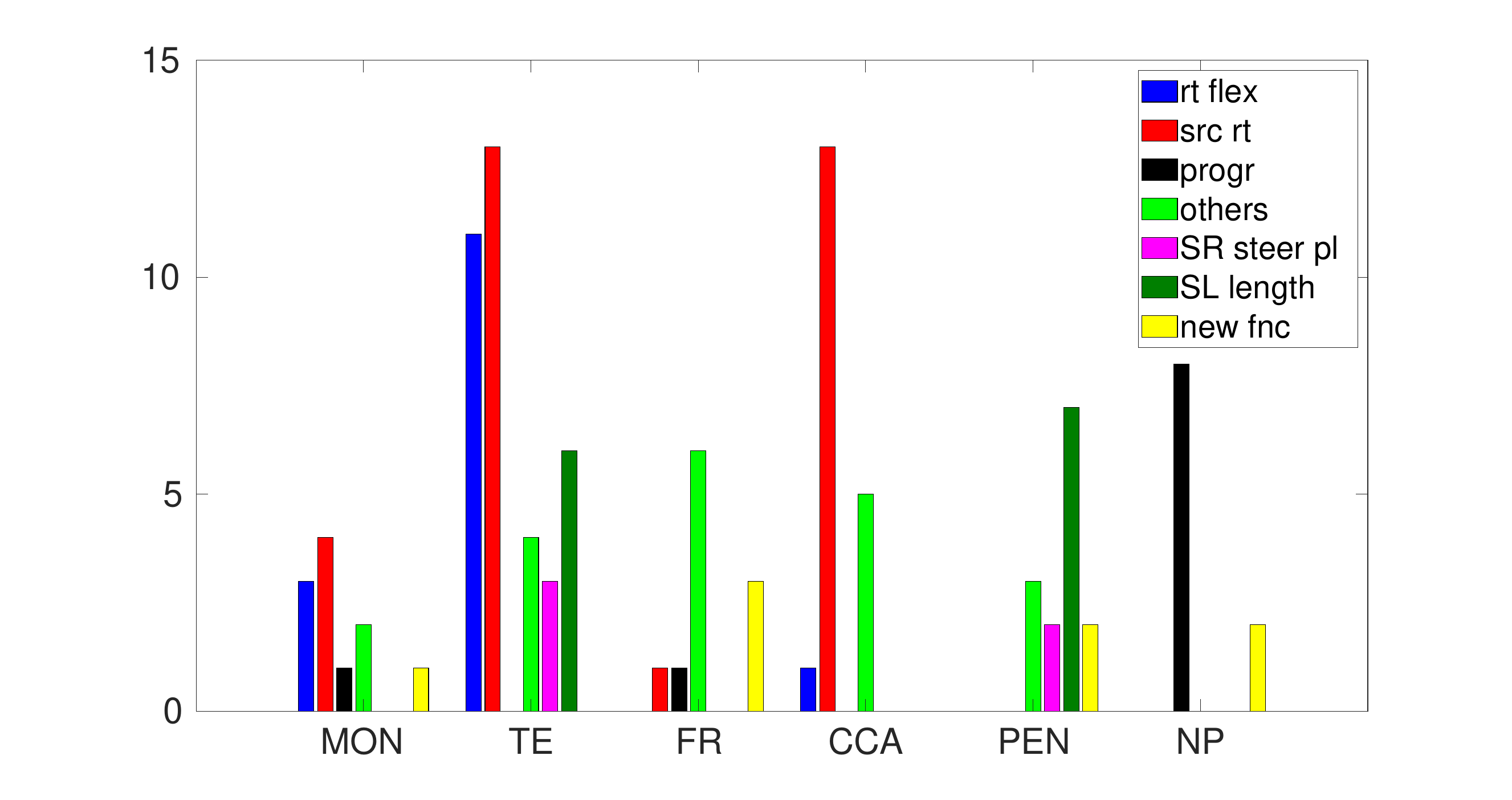}
    \caption{Histogram showing the relation between the Classifications defined in Tab.~\ref{tab:res-taxonomy} and Tab.~\ref{tab:classification}.}
    \label{fig:taxonomy12b}
\end{figure}

In Fig.~\ref{fig:taxonomy12}, we report for each category/SR topic the number of related references.

In the next subsections we briefly describe all research works, considering the classification proposed in Tab.~\ref{tab:res-taxonomy}.

\subsection{Monitoring}
\label{sec:mon}

\MonitoringPapers research works proposing monitoring solutions able to exploit SR have been defined, as shown in Table \ref{tab:monitoring}. These works have been classified on the basis of their main aim:
\begin{itemize}
    \item \emph{delay measurements}, aiming to obtain the delay for links and routers exploiting the possibility of modifying the SL at source nodes, i.e. source routing SR feature;
    \item \emph{health checking of network devices}, aiming to monitor the network state exploiting the capability of SR in defining ad-hoc routes, i.e. routing flexibility feature.
    \item \emph{traffic measurements}, aiming to assess the traffic matrix of a network exploiting routing flexibility and SR counters;
    \item \emph{traceroute}, aiming to implement the well-known traceroute utility in SR.
\end{itemize}

\begin{table}
\centering
\caption{\\Classification of the references related to Monitoring.}
\label{tab:monitoring}
\begin{tabular}{|c|c|}
\hline
\textbf{Objective}                                                                         & \textbf{References} \\ \hline
\textit{\begin{tabular}[c]{@{}c@{}}Delay measurements\end{tabular}} & \cite{interoperable},\cite{xhonneux2018leveraging} \\ \hline
\textit{\begin{tabular}[c]{@{}c@{}}Health checking \\ of network devices\end{tabular}}    & \cite{scmon},\cite{li2018bandwidth},\cite{li2018ilp} \\ \hline
\textit{Traffic measurements}                                                              & \cite{polverini2018routing},\cite{cianfrani2018heuristic},\cite{polveriniNoF2018} \\ \hline
\textit{Traceroute}                                                                        & \cite{xhonneux2018leveraging} \\ \hline
\end{tabular}
\end{table}

In the following paragraphs, we provide a brief overview of the references classified as monitoring related works.

\cite{interoperable} proposes a novel monitoring system powered by Segment Routing (SR), which is used for the provisioning of delay-aware network services spanning multiple-domains. Based on SR-MPLS principles, it enables delay measurements over multiple candidate routes without requiring related LSP signalling sessions. The authors consider two types of probes using SR-MPLS. The first type is originated and terminated by network stations, allowing the retrieval of round-trip measurements only. They also have less accuracy. Moreover, they are typically used for performance measurements over a single link. Instead, the second type relies on external monitoring components, which inject and receive timestamped probes routed according to the enforced SR segment list. The second type of measurement requires synchronization between the end-points, but also allows for the measurement of unidirectional delay, which is more useful when it is necessary to deploy services in the network. Along with most of related works, the project shares the objective of the reduction of control plane complexity through SR. However, from the paper is not clear which implementation the authors used for SR, or if they relied on any vendor solution.

The traffic steering capabilities of SR have been used in SCMon \cite{scmon}, a new solution for continuous monitoring of the data-plane. It allows the user to track the health of all routers and links: i) forcing ``monitoring probes to travel over cycles''; and (ii) testing ``parallel links and bundles at a per-link granularity''. The key insight is that network nodes compute a second network graph and calculate routes on this monitoring topology, which spans all network links. Nodes then carefully select ECMP paths and enforce packet forwarding through cycles leveraging SR in order to detect/localize failures and overloading of single/multiple links. 
A prototype implementation of SCMon has been evaluated on publicly available topologies and emulated networks. In the first experiment, the ratio of number of cycles over number of edges is evaluated to analyze the percentage of the networks covered. The authors then evaluate the time to detect black holes, showing that most of them are detected within less than 100 msec. The work's results are very interesting and can be applied to real use cases. However, an open-source implementation is not available at the time of writing.

In \cite{li2018bandwidth}, the authors focus on bandwidth-efficient monitoring schemes based on cycles. They propose four different algorithms to compute cycles which are designed to traverse/cover every link in the network. This optimization based on cycles allows network resources to be saved and the network to be monitored from a single point of advantage. Segment Routing is used as transport technology to forward the probes along the network. The paper builds upon the results of \cite{scmon}: the authors leverage the phase 1 of SCMon to build the monitoring topology of the network, and then apply their two-phase algorithms in order to minimize the cycle cover length. 
Performance evaluation shows the effectiveness of these algorithms in terms of cycle cover length, segment list depth and improvements with respect to the baseline (SCMon). 

Instead, \cite{li2018ilp} builds on \cite{li2018bandwidth}, proposing three ILP formulations for the construction of monitoring-cycles. A first ILP formulation solves optimally the problem of covering every link in the network using monitoring-cycles with minimum cycle cover length. To further conserve network bandwidth, the first formulation is extended to jointly minimize the total segment list size needed. Finally, since the time required to detect a network failure is affected by the longest cycle, the first formulation is also extended to minimize the length of the longest cycle jointly.

\cite{polverini2018routing} exploits the flexibility of SR to perform traffic measurements and obtain the Traffic Matrix.
A traffic measurement is performed by rerouting a flow and checking the load variations caused on the network links.
Even though the idea of measuring traffic through routing perturbations is not new, SR turns out to be an enabling technology for the applicability of such an approach.
In fact, while in the past traffic flows were rerouted by acting on the OSPF link weights, causing routing instability and performance degradation, SR allows the modification of a path by simply acting on the ingress node, then reducing the impact of a rerouting.
In \cite{polverini2018routing}, the problem of assessing the TM while minimizing the routing perturbations is formulated as an ILP and an heuristic algorithm called SERPENT is presented.
Due to the high computational complexity of SERPENT, a lighter greedy heuristic called PaCoB is proposed in \cite{cianfrani2018heuristic}.

An attractive feature of SR is the introduction of specific interface counters that allow statistics on network traffic flows to be obtained. If this feature is included in the hardware design of the router, the update of the traffic counters can be associated with the normal processing in the SR data plane, thus having a negligible impact on the router performance.
The simplest type of SR counters, named Base Pcounters, collect traffic statistics (byte/packets) passing through a router and having a specific active segment.
Enhanced SR counters, named TM (Traffic Matrix) Pcounters, allow the user to distinguish between traffic that is internal to the SR domain and flows that are injected into the SR domain.
Specifically, a TM Pcounter collects traffic statistics of traffic flows received by an interface marked as external (this is a configuration option for the network operator).
Since TM Pcounters can discriminate packets on the basis of the incoming interface, using TM Pcounters provides a thinner granularity with respect to using Base Pcounters and facilitates the estimation of the Traffic Matrix.
Starting from the availability of this new type of traffic related information, in \cite{polveriniNoF2018} the Traffic Matrix Assessment problem has been extended to include SR counter-measurements. The authors show that, depending on the structure of the Segment Lists used in the network, the Traffic Matrix can be assessed with no error.

\cite{xhonneux2018leveraging} extends the Linux kernel to run eBPF programs as in-network functions attached to the SRv6 SIDs; further details about the implementation are provided in the Section \ref{sec:tools}. The authors demonstrate the effectiveness of their approach by building three different applications. The first one offers  a passive monitoring solution of network delays (direct links or end-to-end paths) \cite{id-udp-pm}. The high level idea is that a small percentage of traffic is encapsulated with a special SRH carrying on additional information like timestamps. These timestamps are then used by the recipient nodes to calculate one-way or round-trip delays. 
The second application realizes a link aggregation group using SRv6. In particular, a weight-round-robin scheduling is realized to aggregate the bandwidth of two different links. Finally, an enhanced version of traceroute has been realized, implementing a new SRv6 behavior, the so-called \textit{End.OAMP}. This behavior, when triggered, performs a \textit{fib lookup} in the node and return to a destination address specified in the SRH all the ECMP next hops. If possible, this function is leveraged at each hop, otherwise the program reverts to the legacy ICMPv6 mechanism.

In the following paragraphs we provide a comparison of the works classified in the \emph{Monitoring} category.

Regarding the solutions related to delay measurements, both (\cite{interoperable} and \cite{xhonneux2018leveraging}) allow the one-way end to end delay between two points in the network to be obtained. While the former requires the use of an external monitoring tool to generate timestamped probes (the use of SR is limited to the creation of the end to end path), the latter does not. In any case, \cite{xhonneux2018leveraging} is suitable only for networks realized by means of Linux based SRv6 routers, since it exploits the definition of eBPF programs to perform the measurement.

There are two SR based monitoring tools for checking the health status of the network links. The first is (\cite{scmon}, and the second is a tool proposed in \cite{li2018bandwidth} and \cite{li2018ilp}). The approach they follow is similar, i.e., the creation of cyclic paths through SR where probe messages are sent. With respect to \cite{scmon}, \cite{li2018bandwidth} and \cite{li2018ilp} optimize the use of the monitoring resources needed to check the status of all the network elements. In any case, \cite{li2018bandwidth} and \cite{li2018ilp} do not support link bundles, while \cite{scmon} does.

Among the Traffic measurement tools based on SR, two different approaches can be identified in \cite{polverini2018routing,cianfrani2018heuristic, polveriniNoF2018}. The first approach aims at measuring traffic flows by causing link load variation through re-routing operations. This is exploited in \cite{polverini2018routing,cianfrani2018heuristic}. The second approach is based on the use of specific traffic counters, available in SR enabled nodes. This is exploited in \cite{polveriniNoF2018}. While both approaches offer comparable performances in terms of the quality of the assessed Traffic Matrix, the counter based solution does not affect the network performance. This is different from from the two re-routing based solutions.

\subsection{Traffic Engineering}
\label{sec:te}
Due to its appealing features in terms of routing flexibility, SR is widely used to face Traffic Engineering related problems.
During our investigation, we have found \tePapers papers exploiting SR to provide advanced TE solutions.
 TE research works  all employ the classic structure of an optimization problem: i) an objective function must be minimized/maximized, ii) taking into account a set of parameters, and iii) considering a specific network scenario.
Three different TE objectives have been covered by the literature, i.e., the minimization of network energy consumption, the optimization of congestion and the minimization of the number of rejected requests.
The high routing flexibility allowed by SR might cause the presence of complex and long network paths.
For this reason, while optimizing routing according to a specific objective, it is important to take into account the impact that excessively complex routing solutions might have on the network performance.
As well as end to end delay, some of the reviewed works also take into account SR related overhead, both in terms of bandwidth wasted due to the insertion in the packets of the SR header, and the number of SR steering policy to be configured in the edge routers.
Finally, the considered network can be a full SR one, i.e., all the nodes are SR capable, or a partially deployed SR, where only a subset of nodes can process the SR header.

Table \ref{tab:TE} show the classification of the TE related references.
It is interesting to note that most of the works consider the congestion minimization as the main objective, and that there are few solutions that can work also in a hybrid network scenario.

\begin{table}
\centering
\caption{\\Classification of the references related to Traffic Engineering.}
\label{tab:TE}
\begin{tabular}{|c|l|c|}
\hline
\multirow{3}{*}{\textbf{Objective}}                                                    & \textit{\begin{tabular}[c]{@{}l@{}}Energy\\ Consumption\end{tabular}}               & \cite{energyefficient},\cite{ghuman2017per} \\ \cline{2-3} 
                                                                                       & \textit{Link Bandwidth}                                                             & \begin{tabular}[c]{@{}c@{}}\cite{anefficient}, \cite{novelsdn},\cite{ascalableand},\cite{trafficpmsr},\cite{ontraffic} \\ \cite{optimizedte},\cite{roomi2018semi},\cite{defo1},\cite{gay2017expect},\cite{incrementaldeploy}\\ \cite{moreno2017traffic},\cite{pereira2017optimizing},\cite{barakabitze2018novel},\cite{pang2017sdn},\cite{dugeon2017demonstration}\\         \cite{hou2019optimization},\cite{settawatcharawanit2018segment}\cite{trimponias2019node},\cite{zhang2019bandwidth}\end{tabular} \\ \cline{2-3} 
                                                                                       & \textit{Rejected Requests}                                                          & \cite{anefficient},\cite{defo1}                                                   \\ \hline
\multirow{2}{*}{\textbf{\begin{tabular}[c]{@{}c@{}}Takes into\\ account\end{tabular}}} & \textit{Delay}                                                                      & \cite{novelsdn},\cite{defo1},\cite{dugeon2017demonstration},\cite{hou2019optimization} \\ \cline{2-3} 
                                                                                       & \textit{\begin{tabular}[c]{@{}l@{}}SR impact\\ (overhead, flow state)\end{tabular}} & \begin{tabular}[c]{@{}c@{}} \cite{anefficient},\cite{ascalableand},\cite{incrementaldeploy},\cite{moreno2017traffic},\cite{pereira2017optimizing} \\ \cite{barakabitze2018novel},\cite{pang2017sdn},\cite{dugeon2017demonstration},\cite{hou2019optimization}\end{tabular} \\ \hline
\multirow{2}{*}{\textbf{Scenario}}                                                     & \textit{Full SR}                                                                    & \begin{tabular}[c]{@{}c@{}}  \cite{anefficient},\cite{ascalableand},\cite{energyefficient},\cite{ghuman2017per},\cite{trafficpmsr}\\  \cite{ontraffic},\cite{optimizedte},\cite{roomi2018semi},\cite{defo1},\cite{defo2} \\  \cite{moreno2017traffic},\cite{barakabitze2018novel},\cite{pang2017sdn},\cite{dugeon2017demonstration},\cite{hou2019optimization}\\ \cite{trimponias2019node} \end{tabular} \\ \cline{2-3} 
                                                                                       & \textit{Partially deployed SR}                                                      & \cite{novelsdn},\cite{incrementaldeploy},\cite{zhang2019bandwidth}                                                   \\ \hline
\end{tabular}
\end{table}

In the following paragraphs we provide a brief overview of the references classified as TE related works.

\cite{anefficient} implements a TE algorithm with SR in a SDN infrastructure which builds a path with bandwidth guarantees and minimizes at the same time the possibility of rejecting traffic demands. With respect to other solutions, it takes into account the \dq{link criticality} and not only link residual bandwidth. Citing \cite{anefficient}: \dq{Link criticality is based on the concept of the minimal interference routing method}. It minimizes the possibility of rejecting requests when the network becomes overloaded. The proposed algorithm not only achieves the goal of balancing network traffic load, but it also promises to reduce network costs. Since it is based on SR principles, the proposed solution also considers the extra network overhead caused by the segment labels in the packet headers. The path length has been modeled as a constraint of the heuristic, adopting an extra hop limitation in order to save network resources - extra bandwidth used by the segment lists in the packet headers. According to the authors, the time complexity of the algorithm can meet the requirements of dynamic online routing. However, there are no open source implementations available, and only simulation results are provided by the authors.

Bahnasse et al. \cite{novelsdn} propose an SDN based architecture for managing MPLS Traffic Engineering Diffserv-aware networks, which bases its forwarding on SR-MPLS principles. This architecture is supposed also to support hybrid deployments whereby SDN equipment coexist with legacy devices, guaranteeing the same forwarding capabilities. Legacy devices are confined in the core of the network while SDN capabilities have to be supported by edge devices. In this way, once the controller has calculated the paths meeting the SLA parameters of the flows, it programs the network interacting with the edge and setting up the SR paths. Then over time the controller monitors the network and dynamically manages the SR-LSPs in order to ensure that the routing realized by segments does not violate end-to-end QoS constraints.

Segment Routing and Multicast are combined in \cite{ascalableand}. The authors propose a routing solution for Multicast based on SR and an heuristic for Multicast tree calculation with bandwidth guarantees, which takes into account the load of the links, the number of branching points and the state in the network. The objective is to minimize the number of requests being rejected. In particular, the SDN controller computes an explicit Multicast tree using the aforementioned heuristic, and then programs the source node of the tree and its branching points: each time a packet reaches a branching point it needs to be duplicated and forwarded on different paths, and a modification of the segment lists is performed. 
Simulation results show that the proposed method outperforms other routing algorithms, although an open source implementation is not available and deeper analysis is not feasible.

\cite{energyefficient} also deals with SR based TE. In particular, the authors design an online energy-efficient Traffic Engineering method. They use the SDN controller to selectively switch off and put in sleep mode a subset of links. They then dynamically adapt the number of the powered-on links to the traffic load.
In this work, SR is used to dynamically re-route traffic. First, a least-congested link technique is run to identify the eligible links that can be switched-off. At this point new routes are calculated, thus solving via heuristics the \dq{Multicommodity Flow Problem}. Finally, the SDN controller enforces new paths through SR or IGP forwarding is leveraged if the route corresponds to the shortest path. At this point it is not necessary to turn the links off. 
Similarly to other works, the problem is solved by identifying first the hop-by-hop paths and then mapping them into SR paths keeping the constraint of the fixed routing over the given hop-by-hop path. The authors of these works provide an interesting analysis implementing the solution in the OMNET++ simulator. However, only numerical results are given, and as far as we know an open source implementation is not available.

The flexibility in path selection achieved by Segment Routing is exploited in \cite{ghuman2017per} to propose a new energy efficiency routing strategy.
The main focus of the work is to switch off a subset of network links by properly selecting alternative paths for the traffic currently steered through the target links.
Clearly, the new paths must have enough bandwidth resources to handle the new traffic.
In order to better exploit the available capacity, differently from other energy aware routing strategies that work at the level of traffic flows, in \cite{ghuman2017per} the alternative path selection is performed at per-packet level.
This allows  a more accurate traffic splitting strategy to be defined which makes more efficient use of the available bandwidth, thus increasing the number of switched off links.

The authors in \cite{trafficpmsr} propose an architecture which combines SDN with SR-based TE. An open source implementation of SR-MPLS is provided together with the realization of an SDN control plane which deals with the calculation of the optimal SR paths in the network. The authors begin implementing a basic TE heuristic, which solves in approximate way the flow assignment problem. The latter allows  the overall network congestion to be minimized. This first procedure is also used as admission control for the next phase whereby the admitted paths are mapped onto SR paths using an heuristic of assignment (which has been described extensively in \cite{pmsr} - Section \ref{sec:path}). Performance evaluation analyzes the distribution of path lengths comparing TE paths with the shortest paths and the distribution of the segment list lengths, thus showing that most of the paths can be implemented using 1 or 2 SIDs. All developed code is open source and available at \cite{pmsrcode}

A theoretical analysis of the computational complexity of the Traffic Engineering problems in Segment Routing enabled networks is provided in \cite{ontraffic}.
Two different TE problems are considered: i) the throughput maximization, and the ii) maximum link load minimization.
As first the General Segment Routing paradigm is considered.
In such a scenario, segments are not constrained to follow shortest paths, but can represent any (possible) complex path.
The resolution of both the aforementioned TE problems results to be NP-hard.
This finding provides a theoretical foundation to the reason why, in Segment Routing, shortest paths are considered for each segment.
Following this, the complexity of the TE problems is studied for the case of Segment Routing with the shortest path.
An interesting outcome of this analysis is that, when the number of segments to be used for each segment list is fixed, the problem of minimizing the maximum link utilization can be solved in weakly polynomial time.
Despite this, when the length of the segment lists is only upper bounded (not fixed), then the investigated TE problems fall again in the NP-hard class.

In \cite{optimizedte}, the authors deal with SR-based TE designing solutions for the optimal allocation of traffic demands using an ECMP-aware approach. The authors propose two optimal solutions for online and offline optimization using a 2-segment allocation, i.e. limiting the length of Segment Lists to two SIDs. The latter consists of the computation for each flow the optimal segment list of two segments with the objective of minimizing overall network congestion. 
The key idea of this work is to minimize the worst-case link utilization by considering ECMP forwarding in the offline case. While in the online case, the traffic split values are properly computed also to minimize rejections of requests. Performance evaluation shows that the n-segment routing problem (``Multicommodity Flow Problem''), i.e. with no constraints on the Segment Lists length, is just slightly better than 2-segment routing problem but the computation complexity is higher due to more degrees of freedom. 

\cite{roomi2018semi} proposes an extension of the models presented in \cite{optimizedte}. In particular, the authors propose the 3-segments forwarding demonstrating that the one defined in \cite{optimizedte}, using two segments, is not sufficient to determine the optimal paths and leads to wasting bandwidth. 

DEFO (Declarative and Expressive Forwarding Optimizer) is a two layer architecture, described in \cite{defo1}, which is realized on top of a physical communication network, aiming at providing a flexible and highly programmable network infrastructure.
At the bottom of the architecture there is a connectivity layer, which is in charge of providing default paths between the network routers.
In DEFO, the connectivity layer is represented by an Interior Gateway Protocol (IGP).
By means of an optimization layer, the routing paths of a subset of traffic demands is deviated by the default behaviour, provided by the connectivity layer, and is steered through a set of optimized paths.
DEFO exploits the flexibility of Segment Routing to implement the optimization layer and configure optimized paths on top of underlying routing paths.
The Service Provider can program the network, thus leveraging a high level interface that allows  specific network goals to be defined through the use of Domain Specific Languages (DSL).
DEFO makes possible for a network operator to define multiple cost functions to be optimized. In the basic case, referred to as “Classic Traffic Engineering”, maximum link utilization is minimized;  in the “Tactical Traffic Engineering” case, the objective function is a combination of the  maximum link utilization and the number of modified paths.
Once the goal has been specified, DEFO starts the computation of optimized paths, by running an algorithm that exploits the concepts of Middle-point Routing (MR) and Constraint Programming (CP).

\cite{gay2017expect} faces the problem of reacting quickly to sudden traffic changes. In fact, these unexpected events, which occur at a low time scale, can create temporary congestion on links, thus degradating the network performance.
Classic solutions to this issue are based on MILP models or Constraint Programming, see \cite{optimizedte} and \cite{defo1}. Unfortunately, these approaches suffer with high computation time, since they work on a time scale of seconds or minutes, providing TE routing strategies that on average allow a reduction of the network congestion, but that might incur in-link overloading due to sudden traffic spikes.
For this reason, in \cite{gay2017expect} the authors present an algorithm which aims to mitigate link congestion under a hard time constraint.
The proposed solution exploits SR to re-route a subset of flows quickly and flexibly in order to decrease the maximum link utilization. The time constraint is taken into account under two different perspectives: i) it is directly considered as a hard constraint during the algorithm execution, i.e, it is a termination condition, and ii) the selected routing strategy has to be as close as possible to the current one in order to minimize the number of reconfigurations needed to make it work.
These two requirements are simultaneously satisfied by the proposition of a Local Search (LS) based heuristic, which takes as input an initial solution and iteratively goes from that solution to another one, by applying local changes called moves, until a stop criterion is met (e.g. the solution is good enough, or a time limit). 
Results show that the proposed algorithm overcomes MILP or Constraint Programming based heuristic, allowing for a significant reduction of network congestion with execution times lower than 1 second.

The Segment Routing Path variable is introduced in \cite{defo2} with the aim of reducing the memory space and the computation time to formulate and solve TE problems in SR networks.
In fact, while the memory space needed to instantiate classical TE problem formulations based on links, paths or nodes variables do not scale well with the size of the considered network and with the number of demands to be routed, SR path variables promise to increase efficiency in the problem resolution.
SR Path variables are based on the concept of a Forwarding Graph (FG).
An FG is a Direct Acyclic Graph (DAG) that originates from a node s and terminates on a sink node t.
The path followed by a demand in the network is encoded as a sequence of FGs, and this sequence is stored into an SR path variable.
Array-based sparse-sets are used to implement SR path variables.
A Large Neighborhood Search approach is used to compute optimized paths for the demands.
The idea is to iterative improve the best-so-far solution trying to reassign the value of a subset of SR path variables, related to demands that are critical (e.g. all the ones that are currently routed over the most loaded link).

\cite{incrementaldeploy} investigates the problem of migrating an IP network into a full SR-enabled one.
The idea is that the process of upgrading the system of IP routers to enable SR capabilities is carried out incrementally in order to reduce the chance of introducing possible misconfigurations or causing the unavailability of the service.
To do that, the Segment Routing Domain (SRD) is defined as the subset of SR capable nodes.
Depending on whether the SRD is a connected set or not, two different models are proposed: Single-SRD or Multiple-SRD.
Two main advantages of the S-SRD model are that it limits the number of flow states to be maintained at the edges of the SRD, and the average length of the segment list is restrained.
As a main drawback there is a potential decrease in the flexibility in the definition of network paths.
On the contrary, M-SRD allows  more complex paths to be defined at the cost of having a higher number of flow states and a higher average segment list length.
The Segment Routing Domain Design problem is formulated as an ILP, where the main goal is to maximize Traffic Engineering opportunities, i.e., the identification of a subset of nodes, of a given size, to be upgraded with SR capabilities, so that the highest possible flexibility is achieved in balancing the links load in the network.
The proposed formulation is able to capture both S-SRD and M-SRD models.
The work shares several design principles with other works reported in this survey, for example it considers incremental deployments of SR and deals with path-aware encoding of the segments list in order to guarantee that the SR path will follow exactly the hop-by-hop path decided by TE heuristics. With respect to other works, the authors also propose a loose forwarding solution whereby the packets belonging to the same flow can cross the network using different paths.

\cite{moreno2017traffic} proposes ILP models and heuristics for TE applications in SR-based networks. Three ILP models are proposed and they are only used as a benchmark for the heuristics due to their high computational complexity. The first is able to leverage ECMP forwarding, the second one computes single routes and implements a hop-by-hop forwarding. Finally, the third one is able to leverage the full capability of Segment Routing. An heuristic has been implemented as some instances of the ILP models require too much time to be solved. The heuristic computes an unique route for each flow and tries to keep the total and the maximum network utilization as low as possible. Moreover, it is able to guarantee that the maximum value of the segment list depth is not exceeded. 

\cite{pereira2017optimizing} proposes a TE solution for path computation that, leveraging at most three labels, is able to optimize link resources and avoid congestion in the network. Firstly, the SDN controller addresses the problem of properly computing the weights of the IGP Link State protocol using Evolutionary algorithms. Then, the traffic distribution is computed through Distributed Exponentially-weighted Flow SpliTting (DEFT) or Penalizing Exponential Flow-spliTting (PEFT), which assign the flows to a next-hop with a probability that decreases exponentially with the extra length of the path (with the respect to the shortest path). SR is used to achieve detours and implement traffic splitting. Performance evaluation shows that the TE solution delivers a lower congestion with respect to OSPF/ECMP with optimized configurations and is able to use all available links. 

The flexibility of SR in path selection, together with the higher throughput provided by the Multi Path TCP (MPTCP), are exploited in \cite{barakabitze2018novel} in order to optimize the throughput for large flows and cope with the explosive growth of multimedia traffic in 5G networks.
The proposed architecture uses a centralized control plane, with a central controller in charge of managing the Quality of Experience of each MCTCP connection.
In particular, the central controller finds multiple paths for each connection, checks the bandwidth requirements and installs specific flow rules at the ingress nodes of the considered 5G network.
When a new MPTCP connection is established, the controller must find a path for each of the subflows belonging to this connection.
For each subflow, it first checks the flow path and resource database, in order to check whether there is already a pre-computed path that has enough bandwidth to support the new subflow.
If this check fails, the controller calculates a new path, encodes it into a segment list and installs a new flow entry in the ingress switch.
The algorithm used to compute the new paths, named QoE-Centric Multipath Routing Algorithm (QoMRA), attempts to find multiple disjoint paths while considering QoE requirements.

\cite{pang2017sdn} proposes the use of Multi Path TCP (MPTCP)  in tandem with SR-MPLS to maximize the throughput of traffic flows in a Data Center network. MPTCP allows  a single connection to be split over several paths, thereby increasing  the total throughput. SDN-based MPTCP solutions are considered by the authors to achieve fine-grain control over TCP connections. However, this drastically increases the overall number of  traffic flows that need to be stored in the devices, by causing scalability issue to arise. SR is used to reduce the number of flow rules needed to steer the single TCP connections over disjoint paths and to save precious space in the Ternary Content Addressable Memory (TCAM) of the devices. The architecture envisages a reactive approach for flow allocation: each time a new subflow is \dq{generated} by MPTCP, a Packet-In is sent to the SDN controller which allocates a new disjoint path whenever is possible, and then installs the necessary flowrules to support this subflow in the edge devices. SR allows  the state in the core device to be reduced. Unfortunately from the paper is not clear how the authors can avoid the explosion of the state at the edge of the network due to the matching conditions at the transport layer.

In \cite{dugeon2017demonstration}, ELEANOR, a northbound application for the OpenDayLight (ODL) Software Defined Network (SDN) controller is presented.
ELEANOR considers an MPLS-SR data plane, where a Maximum Stack Depth (MSD) constraint, i.e., an upper bound on the number of sids that can be stacked in a segment list, has to be considered.
The main components of ELEANOR are: i) a path computation module, and ii) a label stack optimization module.
When a new request arrives at the controller, it first finds a suitable path in order to meet specific QoS requirements (bandwidth, delay, etc.), then the appropriate SL is produced.

In \cite{hou2019optimization}, the authors propose two routing algorithms based on SR for realizing TE applications in SDN networks. These algorithms  search for an appropriate selection of the link weights for optimizing path costs and balancing load across the links. This is obtained through the multiple objective particle swarm optimization (MOPSO) algorithm. Two objective functions are used to measure path cost and load balancing respectively.
According to the authors, their algorithms not only reduce the cost of the paths, better balance the network load, and decrease the maximum link utilization rate, but also can improve the satisfaction ratio with respect of shortest path first (SPF), shortest widest path (SWP), widest shortest path (WSP), minimum interference routing algorithm (MIRA), and the Lee algorithm.


\cite{settawatcharawanit2018segment} proposes the Bounded Stretch constraint to boost the resolution of the SR-TE problem. The Bounded Stretch is used to shrink the set of intermediate node candidates, which are selected during the building of the SL. This allows the space of the solutions to be reduced. The high level idea of the Bounded Stretch is that when an intermediate node is too far away from a source node $i$ to a destination node $j$ then this node should not be considered as a candidate. The selection is achieved comparing the length of the intermediate shortest paths with the length of the end-to-end paths scaled by a given constant. The authors demonstrate through experiments that the constraint helps in reducing the computation time at cost of having a slightly higher utilization over the links.

In \cite{trimponias2019node}, the node constrained TE problem is defined and analyzed, and SR is claimed to be an enabling technology for such routing strategies.
This problem consists in the optimization of one of these two objectives: maximizing network throughput, or minimizing maximum congestion.
Different routing strategies are considered. 
In the most general case, end to end paths are constrained to go through a set of middle-points, leaving the freedom to choose whatever path between two middle-points.
This problem is formulated and proven to be NP-hard.
Next, the feasibility region is limited by forcing the path between two middle-points to be the shortest one.
The derived problem formulation turns out to be solved in weakly polynomial time.
In any case, since the solutions of the previous problem can contain routing loops (the same link is crossed two times), the authors of \cite{trimponias2019node} also consider a variant whereby the solution is constrained to select only acyclic end to end paths.
This last variant of the node constrained TE problem is shown to be NP-hard.
A further contribution of \cite{trimponias2019node} is the proposition of the concept of \emph{flow centrality} as a design parameter to select the most suitable middle-points.
The flow centrality is expressed as the maximum percentage of traffic that can go through a node.
This concept is further enhanced by defining the \emph{group flow centrality}, which is a generalization of the flow centrality over a set of N middle-points.

The work in \cite{zhang2019bandwidth} proposes a traffic engineering solution (path computation and bandwidth allocation) for a hybrid IP/SR network able to maximize a utility function reflecting user satisfaction. User satisfaction is computed as a logarithmic function of the bandwidth assigned to flows. Routing paths are constrained to be the shortest ones in the IP domain, while SR routers can choose among a set of allowable paths in the SR domain. After defining the optimization problem, a two-step iterative algorithm is proposed: at each step of the iteration, link weights inside the SR domain are updated. Two main assumptions are made: i) each flow can be forwarded on a single path, and i) a packet cannot cross the SR domain more than one time.

Independently from the aim of the optimization and of the considered constraints, the main goal of a TE strategy is to find a routing configuration for a given set of input demands. Here we outline some of the differences between the SR related research works classified in the category \emph{Traffic Engineering}. Specifically, we point out five different criteria to compare them: i) the type of approach (optimization based, heuristic based, both), ii) the demand granularity level (Origin-Destination or Ingress-Egress), iii) the path computation strategy, iv) how the TE routing is translated into a set of SLs, and v) the traffic split policy (single SL or multiple SLs).

Most of the works on \emph{Traffic Engineering} with SR propose a heuristic algorithm to determine a set of paths to satisfy a given objective. Only in \cite{ontraffic}, \cite{optimizedte}, \cite{incrementaldeploy}, \cite{moreno2017traffic}, \cite{settawatcharawanit2018segment} and \cite{zhang2019bandwidth} a problem formulation is also presented.
Another interesting difference between the research works on \emph{Traffic Engineering} is the model considered to describe the traffic demands. In fact, while \cite{trafficpmsr}, \cite{trimponias2019node}, \cite{defo1}, \cite{gay2017expect}, \cite{barakabitze2018novel}, \cite{pang2017sdn}, \cite{dugeon2017demonstration}, and \cite{settawatcharawanit2018segment} consider an Origin-Destination (OD) model, where multiple demands can enter and leave the network from the same pair of Ingress-Egress (IE) nodes, all the others assume an IE model (there is a single demand between each pair of IE node that is representative of the aggregation of many OD flows).
This affects both the flexibility of the solution, allowing for a thinner optimization, and the complexity of the algorithms, since they have to deal with a larger number of variables.

Different strategies are used to create the end to end paths. Some of them are based on the definition of parameters able to catch the current network status. The paths are then selected according to a Least Cost rule.
As an example, in \cite{anefficient} are defined the link \dq{criticality} and the link congestion index, \cite{trafficpmsr} defines the concept of network crossing time, \cite{trimponias2019node} reduces the feasible path space by imposing that only nodes with high centrality can be used as middle points. In \cite{zhang2019bandwidth} single path routing is considered, and the objective function of the optimization problem is the user satisfaction calculated as a logarithmic function of the bandwidth allocated to the user flows. Furthermore, also in \cite{barakabitze2018novel} the path search strategy is based on the link criticality, while \cite{pang2017sdn} exploits the concept of delay index.
The remaining works use more sophisticated techniques, such as constraint programming (\cite{defo1,defo2}), ILP based search (\cite{energyefficient,ghuman2017per,optimizedte,roomi2018semi,incrementaldeploy}, local search heuristic (\cite{gay2017expect}), and constrained Shortest Path First algorithms (\cite{dugeon2017demonstration}).

Another interesting difference of the path selection strategies adopted by papers falling into the \emph{Traffic Engineering} category is related to the way they generate the SLs associated to the determined path.
Specifically, two different approaches are generally used: i) the path is found and then encoded into a SL (\cite{energyefficient,incrementaldeploy,moreno2017traffic,barakabitze2018novel,pang2017sdn,dugeon2017demonstration}, or ii) the path is directly constructed as a SL (\cite{optimizedte,roomi2018semi,defo1,defo2,gay2017expect,pereira2017optimizing,settawatcharawanit2018segment,trimponias2019node}).

The final comparison that we propose is related to the possibility to split the traffic demands over multiple SLs.
This option, which is allowed by properly configuring the SR policies, is explicitly used in the algorithms described in \cite{optimizedte,roomi2018semi,incrementaldeploy,pereira2018segment,barakabitze2018novel,pang2017sdn}.
Clearly, having the possibility to split the same flow over multiple SLs increases the flexibility of the routing strategy at the cost of increasing the information to be stored in the head end nodes, where SR policies are installed.

\subsection{Failure Recovery}
\label{sec:fail}

\FailurePapers research works dealing with SR for Failure Recovery and Network Resiliency have been published in recent years. 
The proposed solution can be classified considering the type of failure they are able to recover from:, i.e.link or node failures.
Table \ref{tab:failure-recovery} shows the classification of the covered papers.
In the following paragraphs we provide a brief overview of the references classified as Failure Recovery related works.

\begin{table}
\centering
\caption{\\Classification of the references related to Failure Recovery.}
\label{tab:failure-recovery}
\begin{tabular}{|c|c|}
\hline
\textbf{Node failure} & \cite{trafficduplication},\cite{aubry2018robustly},\cite{optimizingrestoration},\cite{segmentfor},\cite{srdynamicrestoration},\cite{reliablesr}                                             \\ \hline
\textbf{Link failure} & \begin{tabular}[c]{@{}c@{}}\cite{timfa},\cite{trafficduplication},\cite{aubry2018robustly},\cite{segmentfor},\cite{srdynamicrestoration},\cite{reliablesr} \\ \cite{xhonneux2018flexible},\cite{foerster2018local}\end{tabular} \\ \hline
\end{tabular}
\end{table}

\cite{timfa} deals with resilient SR forwarding. In particular, the authors focus on static fast failover solutions for Segment Routing, not requiring any interaction with the control plane. The algorithm proposed, referred to as Topology Independent Multi-Failure Alternate (TI-MFA), is an improvement of the Topology Independent Loop Free Alternate (TI-LFA), described in \cite{id-segment-routing-ti-lfa} and elaborated upon in Section \ref{sec:key_standard}. TI-MFA has interesting performance guarantees and it is also resilient to multiple failures, while traditional SR fast failover based on TI-LFA can work only with a limited number of failures. Firstly, the authors demonstrate that TI-LFA loops indefinitely also for two link failures. Then, a robust but inefficient solutions is shortly presented which pre-compute routing rules considering destinations, incident failures and incoming ports of the packets. Even if this solution can provide better resiliency guarantees, it introduces some inefficiencies (in terms of path lengths) even if only one link failure occurs. Finally, the authors present their solution which basically proposes to store the already hit failures in the packet header and, each time a new failure is faced in the network, the segment list is re-computed using the pre-computed local table entries and the state stored in the packets. 

Traffic duplication through disjoint paths is explored in \cite{trafficduplication}. In particular, the authors leverage SRv6 to realize a traffic duplication service which can guarantee an 1+1 protection scheme through the use of disjoint paths. The main difference from other protection mechanisms is that with 1+1 protection both channels are active and data is sent over both paths. The authors use mirroring behavior in the Linux kernel to realize the traffic duplication. The work builds upon the results of \cite{srv61}, in particular it leverages the Linux implementation of SRv6. Then, the authors propose an algorithm that is able to compute disjoint paths with the least latency and that can be implemented with a number $k$ of segments (they set an upper bound limit on their number and use only node segments). 

\cite{aubry2018robustly} builds on \cite{trafficduplication} with the introduction of robustly disjoint paths. The authors built, extending routing theory and leveraging configuration synthesis, an automated compiler which is proactive, fast and self-healing (no external intervention are required): it computes pairs of disjoint paths for given sources and destination routers, which are robust in the way that they remain disjoint even upon a set of failures. This is achieved without requiring any intervention thanks to SR technology. Indeed, SR allows sequences of segments to be written, which map to different disjoint paths even when there are topological changes. Finally, leveraging the results of \cite{trafficduplication}, Aubry et al. added to the compiler the capability of limiting the number of segments (i.e. path encoding problem) and computing paths that do not degrade data-plane performance (finding low latency SR paths - addressing also TE aspects). 

In \cite{optimizingrestoration}, Hao et al. propose a linear programming model to optimize the restoration in SR based networks. The key idea of the optimized restoration is to share the remaining bandwidth properly when several failures happen. This is addressed through an optimal configuration of the initial segments, knowing in advance the traffic matrix and the network topology. In particular, the authors develop an efficient primal-dual algorithm, which can handle single link failures and multiple logical link failures at the same time (including node failures). Moreover, with a simple randomized rounding scheme they can also take into account ECMP forwarding in the network. 

A logically centralized implementation of the SR control plane (SDN based) is leveraged in \cite{segmentfor}, \cite{srdynamicrestoration} and \cite{reliablesr}. They describe a method to  recover the network from link or node failures dynamically. 
The failover mechanism envisages a failover table for each interface of the node, and when a port goes down automatically the related secondary table is used to implement a loop-free backup path from the point of failure to the destination node. Firstly, the node pops all the labels in the segment list except the last label (which represents the final destination), and then the packet processing is passed to the correct failover table. The authors provide an implementation based on the  OpenFlow (OF) protocol leveraging the OF Group tables feature for monitoring and backup actions. (\cite{grouptutorial} explains how to use OF Fast-Failover Group). In \cite{reliablesr} and in \cite{srdynamicrestoration}, the SR path encoding algorithm can lead to longer segment lists than the one generated by the algorithm proposed in \cite{segmentfor}. In general, a low number of labels is necessary to implement most of the backup paths. \cite{reliablesr} and \cite{srdynamicrestoration} implement a simple detour from the node detecting the failure towards the next-hop or the next-next-hop.

The same authors of \cite{xhonneux2018leveraging} propose an open source implementation of SRv6 TI-LFA in \cite{xhonneux2018flexible} using the extensible Berkeley Packet Filter framework (a thorough article explaining eBPF concepts in the Linux kernel is reported here \cite{lwn-ebpf}). The fast rerouting solution is implemented as a custom BPF program compiled on the fly and attached to a route. In particular a program is loaded into the kernel for each link to be protected. The repair list associated to the route is computed by the control plane and then hard-coded in the eBPF program, which is subsequently compiled and installed in the kernel. The solution has been complemented with a robust failure detection architecture, which implements the Bidirectional Forwarding Detection (BFD) \cite{rfc5880} echo mode using SRv6. In particular, the architecture envisages for each link a master node, periodically sending probes which can activate the fast rerouting mechanism described so far. 
The probes are sent with a special segment list which allows the redirecting of traffic to a special BPF program (BPF slave) on the remote peer and the looping back of probes to the master. The BPF slave, running in the remote peer, handles the prob packets and can activate the SRv6 TI-LFA mechanism for its side of the link. The master uses SRv6 type-length-value to insert sequence numbers and timestamps, which also allow the remote peer to detect failures. The evaluation in the paper considers the number of false positives due to an overloaded CPU, the BPF implementation of the peer nodes reduces the false positives almost to zero, even when the failure detection is less than $10ms$. Instead, the master node is still largely affected by the overloading of the CPU since it uses an user space process for sending the probes.

\cite{foerster2018local} study the problem of defining a backup path scheme that is robust to the presence of multiple link failures. 
Specifically, the main contributions are: i) the introduction of a polynomial-time fast rerouting algorithm which allows a backup path scheme for resilience to be defined under $k$ simultaneous link failures, in particular the case of a hyper cube topology, and ii) the formalization, by means of an ILP formulation, of the problem of defining a backup path scheme that maximizes the number of allowed simultaneous link failures in general graphs.

Let us provide a comparison among the works classified in the \emph{Failure Recovery} category.
\cite{xhonneux2018flexible} proposes an implementation of the TI-LFA mechanism \cite{id-segment-routing-ti-lfa} in a Linux based SRv6 node. \cite{timfa} extends TI-LFA mechanism adding the capability of handling multiple failures.
Among the solutions based on SR to recover from network failures, there are only two works (\cite{timfa,foerster2018local}) which are aimed at dealing with multiple failures. While the first is more focused on implementation aspects, the latter is more theoretical and focused on the definition of algorithms to find loop-free re-routing strategies. Not all other recovery mechanisms are explicitly declared as being able to deal with such a failure scenario.
Another interesting difference between failure recovery solutions is the method they are based on. Specifically, most of them (\cite{timfa,segmentfor,srdynamicrestoration,reliablesr,xhonneux2018flexible,foerster2018local}) are based on a re-routing strategy, i.e., the packets are detoured over a pre-computed alternative path once a link is declared as failed. On the other hand, \cite{trafficduplication,aubry2018robustly} are based on a traffic duplication scheme.
Without going into the details, the main difference between \cite{timfa,foerster2018local} and \cite{segmentfor,srdynamicrestoration,reliablesr, xhonneux2018flexible} is on the number of simultaneous failures they allow recovery from.

\subsection{Centrally Controlled Architectures}
\label{sec:central_control}

The definition of Centrally Controlled Architectures exploiting the SR architecture has been widely investigated in literature, resulting in \ccaPapers different works.
The solution proposed has been classified on the basis of three different aspects, as reported in Table \ref{tab:central-control}. Considering the SR data plane, a high number of works makes use of the SR-MPLS data plane while only three works are based on SRv6. Moreover, some works do not explicitly consider a specific SR implementation, but simply exploit the SR possibility of inserting the flow state in the packet header.
A further aspect used to classify the research papers is the protocol used for the southbound communication between the network devices and the central controller.
The considered protocols are Openflow, PCEP or others.
The last component that differentiates the proposed architectures is the underlay network.
Specifically, the network devices can be IP/MPLS routers or Openflow switches.
In the following we provide a brief overview of the references classified as Centrally Controlled Architectures related works.

\begin{table}
\centering
\caption{\\Classification of the references related to Centrally Controlled Architectures.}
\label{tab:central-control}
\begin{tabular}{|c|l|c|}
\hline
\multirow{3}{*}{\textbf{SR data plane}}                                                   & \textit{SR concept} & \cite{fressancourt2015sdn},\cite{li2017segment},\cite{experimentalmulti},\cite{eramo2019effectiveness} \\ \cline{2-3} 
                                                                                         & \textit{SR-MPLS}    & \begin{tabular}[c]{@{}c@{}}  \cite{firstdemonstration},\cite{sdnandpce},\cite{paolucci2018network},\cite{paolucci2017service}\\ \cite{castoldi2017segment},\cite{springopen},\cite{demonstrationofsr},\cite{evolve} \end{tabular} \\ \cline{2-3} 
                                                                                         & \textit{SRv6}       & \cite{ventre2018sdn},\cite{lebrun2018software},\cite{duchene2018exploring},\cite{barakat2019busoni}                                              \\ \hline
\multirow{3}{*}{\textbf{\begin{tabular}[c]{@{}c@{}}Southbound\\ Interface\end{tabular}}} & \textit{Openflow}   & \cite{firstdemonstration},\cite{castoldi2017segment},\cite{springopen},\cite{li2017segment} \\ \cline{2-3} 
                                                                                         & \textit{PCEP}       & \cite{sdnandpce},\cite{paolucci2018network},\cite{paolucci2017service},\cite{demonstrationofsr} \\ \cline{2-3} 
                                                                                         & \textit{Other}      & \cite{ventre2018sdn},\cite{demonstrationofsr},\cite{evolve}                                              \\ \hline
\multirow{2}{*}{\textbf{\begin{tabular}[c]{@{}c@{}}Underlay\\ data plane\end{tabular}}}   & \textit{IP/MPLS}    & \begin{tabular}[c]{@{}c@{}} \cite{sdnandpce},\cite{paolucci2018network},\cite{castoldi2017segment} \\ \cite{springopen},\cite{demonstrationofsr}\end{tabular} \\ \cline{2-3} 
                                                                                         & \textit{IP/SDN}     & \begin{tabular}[c]{@{}c@{}} \cite{ventre2018sdn},\cite{firstdemonstration},\cite{fressancourt2015sdn},\cite{li2017segment},\cite{lebrun2018software}\\ \cite{duchene2018exploring},\cite{barakat2019busoni}\end{tabular}\\ \hline
\end{tabular}
\end{table}

\cite{firstdemonstration} implements an SDN based SR-MPLS architecture in a multi-layer packet-optical network. In particular, the authors demonstrate optical bypass upon traffic load variations without requiring GMPLS operations. The RYU SDN controller \cite{ryu} has been extended to control the labels stack configuration at the edge nodes (Open vSwitch based). The SDN controller utilizes OF 1.3 to program the edge devices. Open vSwitch has been modified to increase the maximum MPLS stack depth from 3 to 15 labels. The OF protocol has been modified as well to push all the required labels as a single flow entry and with a single action. Commercial Reconfigurable Optical Add-Drop Multiplexer (ROADM) devices have been used to provide an optical bypass between nodes and provide alternative paths during path computation. 
In the performance evaluation, the authors asses through emulation the influence of the 15-label deep stack, evaluating the flows' setup time and packet forwarding in the devices. The latter is not influenced at all, while the setup time is almost triple.

Later in \cite{sdnandpce}, SR has been implemented in a multi-layer network using a PCE architecture instead of SDN/OF. In this scenario nodes consist of commercially available IP/MPLS routers and the SR Controller is an extended version of a PCE stateful control plane. Extensions to the PCE protocol allow a centralized PCE to control the label stacking configuration. PCC (PC client - devices) uses the SR-PCE-CAPABILITY type-length-value for specifying the capability of handling SR-enabled Label Switched Paths and the capability to perform SR computation. The Explicit Routing Object (ERO) carried out in the Path Computation Reply message contains the list of computed SIDs and/or the Node or Adjacency Identifier depending on the SID type. On the device the agent collects the information derived from IGP protocol and configures the related shortest path entries and the SID labels. When a new PC replay message is received, the label stack is properly configured. Also this SR implementation using PCE is able to perform dynamic packet rerouting (with optical bypass capabilities), by enforcing different segment-lists at source node, without any signalling protocol and with no packet loss.

In \cite{paolucci2018network} Segment Routing is proposed as a solution to realize Network Service Chaining (NSC) in a metro-core network scenario, where service chain requests are represented by the so called micro flows, i.e., a huge number of low or medium bit-rate flows.
In this scenario, classic solutions based on MPLS or pure SDN fail due to scalability issues.
On the contrary, SR moves the flow state into the packet header, reducing the configuration costs (and time) to the installation of the encapsulation rule at the ingress point of the network (the one used to add the segment list to each packet of the considered flow).
Based on this consideration, \cite{paolucci2018network} describes an SR Path Computation Element (SR-PCE) which is in charge of orchestrating connection setup/release/modification, and is made up of two main modules: i) the flow computation element, having the goal of find a path with available resources to serve a micro flow and ii) flow steering API that is responsible of installing the SR encapsulation rule in the ingress node.
An experiment evaluation of the proposed architecture is proposed in \cite{paolucci2017service}.

\cite{castoldi2017segment} proposes an SR-based Software Defined Network (SDN) architecture which is able to perform load balancing among ECMP and non-ECMP routes in multi-layer networks including an IP/MPLS layer over an optical network layer. In particular, two solutions are described: i) Centralized-SR; ii) Preconfigured-SR. The former leverages a SDN controller to steer traffic over alternative paths upon network failures. Instead, the second solution uses OpenFlow load-balance groups to actively forward the traffic on several routes. With the second solution the data plane layer can autonomously react to a network impairment removing the failed output port, while the first solution always requires the intervention of the SDN controller, but results are more generic with regards of the second one. Both solutions push the SID of the destination node and leverage the available ECMP paths. The recovery property of the architecture has been validated simulating network failures. According to the authors, some packet losses have been recorded and the recovery time was around $170ms$.

The SPRING-OPEN project \cite{springopen} is an ONOS \cite{onos} use case, which provides an open source SDN-based implementation of SR. Its architecture is based on a logically centralized control plane, built on top of ONOS, and it drastically eliminates the IP/MPLS Control Plane from the network. Part of this work converged later in the Trellis project \cite{trellis}, an SDN based leaf-spine fabric, built using bare-metal hardware, open-source software from the OCP \cite{ocp} and ONOS projects, and OpenFlow-Data Plane Abstraction (OF-DPA)  \cite{ofdpa}, an open-API from Broadcom \cite{brcm} to program merchant-silicon ASICs. The leaf-spine fabric is based on SR-MPLS principles. However, it does not implement fully-fledged SR architecture, as it just uses global significant Node-SIDs. These MPLS labels are statically configured in the SDN control plane and are used to globally identify the ToR switches of the fabric, routing the traffic towards them using a single MPLS label.

In order to provide resiliency against link and node failures for Cloud Service Provides (CSPs), an overlay infrastructure realized by means of Segment Routing and Software Defined Networking control plane is proposed in \cite{fressancourt2015sdn}.
The main idea of the proposed architecture is to substitute the dedicated physical infrastructure that interconnects Data Centers of a CSP, with an overlay network realized on top of an underlay infrastructure, represented by the interconnection of many Internet Service Provider (ISP) networks.
The prerequisite is the availability of multi-homed connections for each Data Center of the CSP.
Thus, the logical components of the proposed overlay infrastructure include: i) a central controller that monitors the underlay network status and determines new paths in case of failure, ii) the egress points that are responsible for routing the traffic flows leaving a Data Center toward the most appropriate ISP, and iii) the routing inflection points that are special nodes that manage the routing between two distant Data Centers, by using SR encapsulation.

\cite{li2017segment} proposes the use of SR in a hybrid IP/SDN network as a technique to mitigate the problem of limited storage space in the flow tables of SDN switches.
The main goal is to use SR to optimize the use of the flow tables and link capacity.
In the considered scenario, every node supports both the OpenFlow operations and the normal IP forwarding operations.
When a packet enters a node, it is first classified in order to decide through which pipeline it has to be steered, then it is processed accordingly.
Operations attributed to the normal IP pipeline are decided using a routing protocol (eg. OSPF).
Alternatively, the path to be followed by traffic flows steered through the OpenFlow Switching Layer (OFSL) are decided by the central SDN controller.
In order to limit the number of flow entries to install in order to configure a path, SR is exploited.
In this way, a portion of the flow state information is moved from the flow table of the switches to the packet header.
In order to insert the SR related information, i.e., the segment list, in the packet header, \cite{li2017segment} proposes the use of unused fields (eg. VLAN tag or optional fields).

Software Resolved Networks (SRNs) is a new SDN architecture recently proposed for IPv6 enterprise networks \cite{lebrun2018software}\cite{duchene2018exploring}; further details about the implementation are provided in the Section \ref{sec:tools}. The network is managed by a logically centralized controller, which interacts with the end-hosts through an extended DNS protocol: applications are allowed to embed traffic and/or path requirements in their requests, and the controller is able to return the appropriate path to the applications satisfying their needs. SRv6 is used as data plane technology to steer traffic on a specific path according to network policies. Each component that can be reached through the network is always referenced through a DNS name. The default DNS resolver in the hosts is modified to interact with the controller of the architecture. SRN also provides a mechanism for the dynamic registration of the end-points. In this way, the DNS database can be properly updated and the name resolution can be performed by the clients. The connections are always unidirectional, thus it is necessary to establish two paths in order to enable the communication between the endpoints. A binding segment is used to implement a path id, and it is automatically translated in a SRv6 policy in the access node. Path segmentation is performed using the algorithm illustrated in \cite{scmon}, which allows a given policy to be matched and the minimal list of segments to be guaranteed. In order to optimize the interaction with the controller, upon a request the controller computes the two paths to support the communication and then instruct the access device of a node to add also the reverse binding segment in the SRH. In this way, the reply can be simply echoed back. Software Resolved Network has been implemented on Linux end-hosts, routers and controllers. 

In \cite{ventre2018sdn} a novel SDN architecture is proposed for SRv6 technology; Section \ref{sec:tools} provides further information about the implementation and where it is possible to download the code. The data plane is constituted by Linux based SRv6 nodes built from open source components which expose an open API towards the SDN controller. As a result, the nodes become hybrid as they envisage the coexistence of a legacy IP control plane with an SDN control plane. The authors present the design and implementation of the Southbound API between the SDN controller and the SRv6 devices, which is used to instantiate SRv6 policies in the network. Specifically, they propose a data-model and provide four different implementations of the API, respectively based on gRPC, REST, NETCONF and remote Command Line Interface (CLI). Topology discovery is also addressed by actively extracting the topology database from the IP routing daemons running in the network nodes. 

A hierarchical multi-domain control plane for SDN networks based on SR has been demonstrated in \cite{demonstrationofsr}. The control plane is composed by an orchestrator application, which runs on top of multiple SDN controllers and leverages their NB APIs to create multi-domain SR based services. BGP-LS and PCEP are used as southbound in the SDN controllers. They provide network topology respectively and the creation of MPLS SR tunnels. IS-IS is used inside the domain to exchange reachability information and SIDs between nodes. SDN controllers do not exchange any reachability information nor SIDs. The orchestrator interacts with the SDN controllers and builds a global network view that will be used to perform the path computation and to instantiate SR services. A practical demonstration has been realized using software routers. 

In \cite{experimentalmulti} two solutions for multi-domain SR are proposed: end-to-end Segment Routing and per-domain Segment Routing. Both methods leverage a non-standard east/west interface between peer controllers, thus relying on a flat control-plane architecture and not using signaling sessions in the data plane. In the first approach, the segment list already contains the end-to-end path crossing several domains. The originator domain sends a request to the destination domain. The destination controller computes the segment list to reach the destination from its ingress router and sends it back to the previous domain. Each intermediate domain applies the same procedure stitching the segment list computed by the downstream until the reply is received back by the originator. Conversely, in the second approach the end-to-end path is obtained stitching several SR paths: in each domain the segment list contains a virtual label as the bottom of the stack, which triggers a modification of the segment list in the ingress border node of the next domain. In this case, there is no global view of the network, and controllers do not know the domain sequence to reach the destination. Scalability of the proposed schemes is evaluated in terms of segment list depth. Results show that per-domain SR is able to encode 60\% of the paths using at most 3 labels while end-to-end SR just the 12\%. In general, the average SL depth is 5.34 and 3.36 respectively for end-to-end SR and per-domain SR.

\cite{evolve} proposes an advance of Carrier Ethernet architecture and envisages an approach mixing SR and SDN technologies. It results in a trade-off between fully-distributed control planes and centralized approaches: an inventory database is maintained in the SDN controller with the configuration for each device. The SDN controller provides the IP configuration and SR configuration such as the loopback address, node label, label range, gateway label information via a southbound API such as NETCONF/Yang. For any communication inside the domain, the network will use the IGP based forwarding by design without the need of the SDN controller and will impose on the traffic a single MPLS label (loopback SID). 
Multiple labels can be used to realize TE applications.
Instead, in a multi-domain scenario several SDN controllers are deployed and exchange reachability information in order to properly program the edge nodes. In this way, an inter-domain path can be established simply with a label stack that includes local and remote border router labels, plus the end node label. 
Similarly to other works, further analysis is not possible as the code of the control plane is not open source. As for the data plane, the solution relies on Carrier Ethernet hardware.

Busoni \cite{barakat2019busoni} is an orchestration framework for Segment Routing based networks, which automates and simplifies many aspects of the network management. From an architectural standpoint, Busoni sits on top of a SDN controller and benefits from the information exported by the controller to feed its data-store. The latter is a graph database, which is used to persist data. In particular, Busoni leverages it to keep track of the SIDs advertised in the network, the installed policies, and to respond to any dynamic event. The framework provides users with programming tools to compose and manage SR policies . It operates efficiently even under multi-tenancy environments. Finally, Busoni automatically updates the nodes and the edges of the graph database whenever there is an update in the network and reflects these changes on the installed policies. This allows their resilience to dynamic events to be maintained.

\cite{eramo2019effectiveness} proposes a scalable centralized controlled architecture for the management of SFC Routing and Cloud Bandwidth resource Allocation (SRCBA) based on SR.
The proposed solution is thought to be applied in a multi domain scenario, where a transport network interconnects a set of private cloud infrastructures, possibly owned by different providers.
In particular, instead of using classic approaches to solve the SRCBA problem, which require a detailed knowledge of either the transport network or the cloud infrastructure, the proposed architecture exploits the BSID concept to abstract the services provided by a single cloud infrastructure to the external network.
The resulting Orchestrator is then divided into two logical components: i) a centralized Network Service Orchestrator (NSO), which is in charge of collecting SFC requests and managing bandwidth in the transport network, as well as deciding to which datacenter to assign the processing of the incoming requests, ii) and a set of local Resource Orchestrators that are in charge of managing the network and cloud resources in the context of a single infrastructure.
In this way, the centralized NSO can rely on summarized information to take decisions while solving the SRCBA problem.
This allows for a great reduction of computation time, while assuring comparable performance in terms of efficiency in the use of the resources.

In the following paragraphs we provide a comparison among the works classified in the \emph{Centrally Controlled Architecture} category. An interesting aspect to compare the research works falling into this category is related to the way they use SR. For instance, some solutions exploit SR to overcome some limitations existing when other technologies are adopted. Other centrally controlled architectures make use of SR to realize specific functions, which are more complex if realized by means of other paradigms. Finally, some works falling into this category have the goal of proposing an implementation, and provide a demonstration of the feasibility and the performance achieved.

Among the works that exploit SR to overcome limitations of other existing approaches, there are \cite{paolucci2018network,li2017segment,barakat2019busoni,eramo2019effectiveness}. Specifically, in \cite{paolucci2018network} SR allows for the reduction in the complexity of configuring and updating the path for an incoming SFC request, thanks to the adopted source routing paradigm. In \cite{li2017segment}, they exploit the same principle to show that, by adopting SR, the stringent TCAM size constraint can be overcome. Finally, \cite{barakat2019busoni} proposes an orchestration framework to simplify the policy management. Similarly, \cite{eramo2019effectiveness} proposes a scalable centralized controlled architecture for the management of the SFC Routing and Cloud interconnect bandwidth allocation.

Different functions at the network level are realized in centrally controlled architectures proposed in \cite{experimentalmulti,castoldi2017segment,fressancourt2015sdn}. As an example, in \cite{experimentalmulti} SR is used to realize an optical bypass, while it is exploited to overcome node and link failures in \cite{castoldi2017segment}. Finally, an overlay network to interconnect geographical distributed DCs is realized through SR in \cite{fressancourt2015sdn}.

As previously stressed, the aim of some of the works falling into the \emph{Centrally Controlled Architecture} category is to propose possible implementations of SR, and to provide a demonstration. This is the case in \cite{ventre2018sdn,springopen,demonstrationofsr}. In \cite{ventre2018sdn}, SRv6 nodes based on the Linux implementation are considered in the data plane, while the other works are focused SR-MPLS. Furthermore, \cite{demonstrationofsr} demonstrates how to build an end to end paths in a multi domain SR network, while in \cite{springopen} a single domain scenario is considered.

\subsection{Path Encoding}
\label{sec:path}

The translation of a network path, resulting from a specific TE objective, into a sequence of SIDs, i.e. a Segment List, is a key operation for the deployment of SR in a real network. This operation is usually referred to as Path Encoding.
\PathencPapers different papers have been defined for the definition of a proper a solution to the path encoding problem.
The different algorithms differ according to the nature of the network path, i.e. the path to be encoded.
Specifically, the input path might be computed on top of a network with uniform link weight or arbitrary ones.
Moreover, it may or may not include ECMP.
A further aspect for the path encoding algorithms is the possibility of requiring additional device configurations, as the insertion of a new policy.
Finally, path encoding algorithms can differ depending on the possibility of encoding the input path with one or more SLs.
Table \ref{tab:path-encoding} shows the paper classification according to the aforementioned aspects.
In the following we provide a brief overview of the references classified as Path Encoding related works.

\begin{table}
\centering
\caption{\\Classification of the references related to Path Encoding.}
\label{tab:path-encoding}
\begin{tabular}{|c|l|l|}
\hline
\multirow{4}{*}{\textbf{\begin{tabular}[c]{@{}c@{}}Path to be\\ encoded\end{tabular}}}                        & \textit{\begin{tabular}[c]{@{}l@{}}uniform\\ IGP weights\end{tabular}}   & \cite{experimentaldemonstration},\cite{efficientlabel},\cite{pathencoding} \\ \cline{2-3} 
                                                                                                              & \textit{\begin{tabular}[c]{@{}l@{}}arbitrary\\ IGP weights\end{tabular}} & \cite{labelencoding},\cite{pmsr},\cite{translating},\cite{liaoruo2018optimizing},\cite{guedrez2017new} \\ \cline{2-3} 
                                                                                                              & \textit{\begin{tabular}[c]{@{}l@{}}ECMP\\ not allowed\end{tabular}}      &  \cite{pmsr},\cite{experimentaldemonstration},\cite{pathencoding},\cite{labelencoding} \\ \cline{2-3} 
                                                                                                              & \textit{\begin{tabular}[c]{@{}l@{}}ECMP\\ allowed\end{tabular}}          & \cite{efficientlabel},\cite{translating},\cite{liaoruo2018optimizing},\cite{guedrez2017new} \\ \hline
\multirow{2}{*}{\textbf{\begin{tabular}[c]{@{}c@{}}Requires \\ further configuration\end{tabular}}} & \textit{no}                                                              & \begin{tabular}[c]{@{}l@{}}  \cite{experimentaldemonstration},\cite{efficientlabel},\cite{pathencoding} \\ \cite{labelencoding},\cite{pmsr},\cite{translating}\end{tabular} \\ \cline{2-3} 
                                                                                                              & \textit{yes}                                                             &  \cite{liaoruo2018optimizing},\cite{guedrez2017new} \\ \hline
\multirow{2}{*}{\textbf{Encoded path}}                                                                        & \textit{single SL}                                                       & \begin{tabular}[c]{@{}l@{}} \cite{experimentaldemonstration},\cite{efficientlabel},\cite{pathencoding},\cite{labelencoding} \\  \cite{pmsr},\cite{liaoruo2018optimizing},\cite{guedrez2017new} \end{tabular} \\ \cline{2-3} 
                                                                                                              & \textit{multiple SLs}                                                    & \cite{translating}                                              \\ \hline
\end{tabular}
\end{table}

In \cite{experimentaldemonstration}, SDN and PCE based implementations of the SR controller share a common path engine, which performs the hop-by-hop path computation and SR path assignment. As regards the path computation engine, the controller selects the least congested path on a set of candidate paths. Then, the proposed SR path assignment algorithm provides the shortest Segment list considering a unique path towards the destination and avoiding load balancing through ECMP. The algorithm uses two pointers, $i$ and $j$, to navigate the target path from source to destination. Firstly, $j$ is incremented until the considered sub-path is no more the unique shortest path. At this point, the SID of the node $j-1$ is inserted into the Segments list, the two pointers are both set to $j-1$, and then the procedure restarts. This cycle is repeated until the node $j$ is equal to the destination. The algorithms provide the segment list of minimum depth, but the solution only considers global Node-SID, and therefore it cannot be applied to topologies with arbitrary IGP link costs.

The authors of \cite{efficientlabel} propose a Segment list encoding algorithm to express a given path, which minimizes (enforcing a given threshold) the Segment list depth in SR-based networks. It considers ECMP forwarding by default, but can also introduce constraints to support a deterministic hop-by-hop path. With respect to other efforts (such as \cite{pmsr}), the solution is not able to support arbitrary hop-by-hop paths when arbitrary IGP link costs are used. 
The core of the algorithm consists of the creation of a graph, whose arcs model SR related instructions (node and adjacency SIDs). Specifically, an arc connecting two nodes represents the shortest path in the original network between the same pair of nodes. 
The so-called auxiliary graph is built firstly using the physical links of the paths, which are computed by an external TE heuristic. Subsequently, virtual links are added representing the ECMP paths between two nodes. Each virtual link is annotated with the metrics and the number of the ECMPs between the two nodes. Using this auxiliary graph, a new path computation is performed adding a new constraint related to the number of hops: each path having a hop number greater than the maximum Segment list depth is rejected. The candidate paths are then sorted firstly according to their original metrics, secondly according to their length, and then according to the number of ECMPs. Finally, the paths are translated into SIDs using this approach: physical links are changed with their respective Adjacency SID, virtual links are mapped with the Node SID of the destination node. If the Adjacency SID are not local the algorithm first inserts the Node SID of the source and then the Adjacency SID. 

\cite{pathencoding} proposes two algorithms for the computation of the segment list; the algorithms are optimal in terms of stack depth when a unique hop-by-hop path has been computed. The key idea is to consider at each iteration bigger sub-paths and verify whether a unique shortest path exists. It it exists, substitute the sub-path with the tail node SID before moving to the remaining part of the path. The main difference between the algorithms is how to vary the breadth of the sub-paths until considering the original hop-by-hop path. The first algorithm navigates the target path starting from the source node toward the destination, while the second one leverages the opposite direction. The analysis of the overhead in the packet headers shows that reverse algorithm introduces less overhead compared to the direct algorithm, as the computed segment list typically includes nodes that are near the source. 
Other works (for example \cite{trafficpmsr, pmsr}) share the same objective of reducing the overhead of the packet headers, but with respect to the solutions described above, they start from the original hop-by-hop path and then evaluate the sub-paths breadth reduction at each iteration.

In \cite{labelencoding}, two algorithms for an efficient paths encoding are proposed. The algorithms, referred to SR-LEAs, take as input an explicit shortest path and then compute the relative segment list having as constraint a given maximum segment list depth. They are composed of two main steps: i) computation of successive shortest paths; ii) label replacing. Specifically, they compute the subpaths of the original shortest path and take into account the limitation of the hardware to reduce the number of subpaths. In the second step, the subpaths composed of three or more nodes are replaced by the Node-SID of their tailnode. SR-LEA replaces two node subpaths using the Adj-SID. The variant SR-LEA-A is very similar to the above algorithm but takes advantage of the global Adj-SID to further reduce the depth of the label stack. Of course, it requires the advertisement of these SIDs in order to work properly. Simulation results show that the algorithms are able to compute segment lists with an average length lower than 3. SR-LEA-A delivers best results and can improve the segment list computation compared to SR-LEA, but the gain is around 5\% in terms of average length.

The authors of \cite{pmsr} propose an optimal SR path assignment algorithm and prove that it is optimal in terms of the number of used segments. The algorithm takes as an input a hop-by-hop path resulting from a TE algorithm and define an SR path for it composed by the minimum number of segments. 
The SR assignment algorithm evaluates at each iteration if a shortest path between the current source and the current destination exists, considers different subpaths and updating the current source or the current destination. 
Each time a new segment is found, the current source is updated with the previous destination, and the current destination is replaced by the destination of the original hop-by-hop path. In the worst case, the link between the current source and its next hop is evaluated, and if the link is not part of the shortest path, an adjacency segment is pushed in the segment list. 

The authors of \cite{translating} propose a two-step method for translating a given Traffic Engineering (TE) path into an SL. In the first step, an auxiliary graph is created. The aim of the auxiliary graph is to represent all the Interior Gateway Protocol (IGP) paths available, i.e. forwarding paths, for the specific TE path to be encoded. In the second step, a MILP problem over the auxiliary graph is defined: the MILP problem allows for the minimization of the overall Segment List length given the target TE path. One of the main features of the proposed solution is multi-path support, i.e. the ability to split a source-destination flow among a weighted set of segment lists. 

In \cite{liaoruo2018optimizing} the problem of optimizing the performance of an SR network under the maximum Segment List Depth (max-SLD) constraint is studied.
In fact, especially when SR is realized on top of the MPLS data plane, the constraint on the max-SLD is particularly limiting.
A possible solution is to create new LSPs, thus increasing the availability of alternative paths in the underlay network, and consequently enabling shorter segment lists to be written.
Despite the fact that the creation of new LSPs reduces the length of the segment lists, its main drawback is the increase in the number of required forwarding rules to be installed in the routers' forwarding tables.
In order to mitigate this negative effect, \cite{liaoruo2018optimizing} defines the concept of panel based forwarding: a panel refers to a set of node-disjointed LSPs that can be represented by the same label.
Furthermore, an ILP formulation is proposed to solve the path encoding problem, minimizing both the number of new defined LSPs and the installed rules, while respecting the max-SLD constraint.

To overcome the Maximum Stack Depth (MSD) constraint in SR-MPLS, a new type of SID, named Targeted SID (TSID), is defined in \cite{guedrez2017new}.
A TSID is a local segment that is associated with a sequence of SIDs. 
The instruction related to a TSID consists of replacing it in the SL with the associated sequence of SIDs.
By using a TSID, it is possible to reduce the length of an SL at the cost of introducing a new flow state in the node that implements the instruction related to the TSID.
For this reason, \cite{guedrez2017new} proposes two different optimization problems that allow a trade-off between the benefits and the costs of the TSID tool.
The first optimization problem takes as input the set of paths with an SL overcoming the MSD bound and aims to minimize the number of defined TSIDs. 
The focus is then the reduction of the number of extra flow states to be maintained by the network nodes.
A second optimization problem is presented with the goal of minimizing the PCEP sessions that have to be maintained between the central controller (responsible for the TSID installation) and the nodes where the TSIDs have to be installed.
In this case, the main idea is to install as many TSIDs as possible in the same node.

A possible key to comparing the existing SL encoding techniques is to focus on the method they use to obtain the final list of SIDs.
In particular, two different methods are exploited. The first one is based on the Bellman-Ford principle, which states that each sub path contained into a shortest path is in itself a shortest path. Based on this consideration, these algorithms explore the path to be encoded, starting from the source node up to an intermediate one.
Until the path between the source and the intermediate node is found to be a shortest path, they move on by considering the next node in the path.
When this condition does not hold, the first SID is found.
The process goes on until the full path is explored.
SL encoding tools presented in \cite{pmsr,experimentaldemonstration,pathencoding,guedrez2017new} use this approach. 
Furthermore, it is interesting to emphasize that all the works using this approach do not allow the use of ECMP in the underlay.

The second type of approach consists of the creation of an auxiliary graph to represent the underlay paths.
Specifically, an arc of the auxiliary graph is an entire path in the underlay.
This type of approach is used in \cite{efficientlabel,translating}.
The main difference between the solutions presented in these two works is that, in \cite{efficientlabel} the end to end path is encoded by applying the Dijkstra algorithm over the auxiliary graph, while in \cite{translating} the problem is modeled as a Multi Commodity Flow over the auxiliary graph. As a consequence, \cite{translating} also allows the use of multiple SLs to steer a single flow.

\subsection{Network Programming}
\label{sec:netprog}

Despite the fact that Network Programming is the most attractive and innovative feature of SR, we have found only \networkProgPapers papers related to this topic.
As shown in table \ref{tab:network-programming}, these works can be classified in Service Function Chaining related ones, or operational function.
The latter are aimed at implementing operational functions, such as a firewall, a load balancer and a zero-loss VM migration tool, by exploiting the network programming feature of SR.
In the following paragraphs we provide a brief overview of the references classified as Network Programming related works.

\begin{table}
\centering
\caption{\\Classification of the references related to Network Programming.}
\label{tab:network-programming}
\begin{tabular}{|c|c|}
\hline
\textbf{\begin{tabular}[c]{@{}c@{}}Service Function\\ Chaining\end{tabular}} & \cite{mayer2019efficient},\cite{srv62},\cite{17-vnf-chaining-sr},\cite{duchene2018srv6pipes}, \\ \hline
\textbf{Operational Function}                                                & \cite{abdelsalam2018sera},\cite{srlb},\cite{desmouceaux2019content},\cite{desmouceauxzero} \\ \hline
\end{tabular}
\end{table}

In \cite{srv62} the Linux SRv6 implementation described in \cite{srv61} is enhanced introducing the support for Service Function Chaining. In more detail, the Linux kernel provides an API to map a service segment, i.e. the identifier of a network service running on a Virtual Machine. The experimental evaluation shows that the impact of SR operations and service segment processing on the packet forwarding capabilities is limited, i.e. with a reduction lower than $10\%$.

The architecture of a network domain supporting Service Function Chaining (SFC) through SRv6 is investigated in \cite{17-vnf-chaining-sr}.
The authors mainly focus on the implementation of a VNF node able to host multiple VNF instances. The main components of the VNF nodes are the SR/VNF connector, in charge of logically connecting the SR routing with local VNFs, and the VNFs, supporting specific network functions. The VNFs can be SR-aware or SR-unaware: i) SR-aware VNFs can process the
information contained in the SR header (SRH) of incoming packets; ii) SR-unaware VNFs are not able to handle the SRH. Thus, in order to correctly apply the VNF to the original packet, the SR/VNF connector must pre-process the packet by removing the SR encapsulation, and then re-apply it when the packet is returned by the VNF.
The authors propose a Linux-based implementation of a VNF node supporting both SR-aware and SR-unaware VNFs. In more detail, using the netfilter framework, a new kernel module called srext (Segment Routing EXTensions) is implemented to act as a SR/VNF connector and to support SR-unaware VNFs.
A virtualized testbed based on Virtualbox ang Vagrant is realized to evaluate the processing overhead introduced by the proposed implementation with respect to a classic IPv6 forwarding solution. 

SRV6Pipes \cite{duchene2018srv6pipes} is an extension of the IPv6 implementation of Segment Routing in the Linux kernel, which enables chaining and operation of in-network functions operating on streams. SRv6 is used to enforce an end-to-end path between the client and the server passing through the equipment hosting the networking functions. 
The rationale behind SRv6Pipes lies in the fact that some network functions need to include a TCP implementation to work on the streams. SRv6Pipes leverages ``the TCP stack that is already present in the Linux kernel'' and implements a transparent TCP proxy to offload the TCP functionalities to it and terminate the TCP connections where a network function is deployed. In this way, it is possible to expose to the network functions the bytestreams they need to process. Special addressing is used to specify network functions and their parameters. In particular each proxy exposes an 80 prefix. The first 80 bits are used to traverse the proxy, the following 16 bits are used to identify the network function and the remaining ones are used to specify the parameters of the function. 

\cite{abdelsalam2018sera} defines the concept of SR Aware VNF, as an application that is able to process the SR information in the packet.
Moreover, \cite{abdelsalam2018sera} also proposes the implementation an SR Aware (SERA) Firewall application.
The SERA Firewall is able to work both as a legacy firewall (basic mode), or define filtering rules that also include condition on the SR fields (advanced mode).
In the basic mode, it can apply the normal firewall processing to the original packets even if they have an SR based SFC encapsulation.
It means that the filtering rules are applied to the original values of the packet header (the SR related fields are transparent).
In the advanced mode, SERA offers new matching capabilities and new SR specific actions that allow  fields in the SR Header to be modified. 

Segment Routing Load Balancing (SRLB) \cite{srlb} is an Application-aware load-balancer that avoids cost due to monitoring tasks.
It is thought to work in the context of a Data Center network, where several instances of the same application are instantiated in different host machines.
Each host machine is equipped with a VPP based virtual router which dispatches packets between physical NICs and application-bound virtual interfaces.
The Load Balancer (LB) is located at the edge of the Data Center network.
A request for an application is represented by the first packet sent from the client to the application server (generally a syn message). 
When a new request arrives at the LB, the Service Hunting function, which consists of finding a server that can serve the current request, is executed.
LB exploits SR to query a subset of servers that host the requested application.
Specifically, LB encodes the set of randomly selected potential servers in the segment list before encapsulating the first packet of the request.
When the first server receives the packet, it can decide whether to deliver it to the application, and consequently assign to it the processing of this request, or it can forward the packet to the next server in the segment list.
The decision about whether to accept/refuse a request is taken according to a connection acceptance policy which takes into account the internal state of the application (CPU usage, memory usage, etc.).

\cite{desmouceaux2019content} proposes a new architecture for the Content Delivery Networks (CDNs) building upon the results of \cite{srlb}. In this new paradigm, CDN decisions (cache vs origin servers) are offloaded to the data plane. This architecture is based on two fundamental pieces: i) chunk-level content addressing and ii) in-network server selection. The first is realized by assigning a unique and globally routable IPv6 address to each chunk. Instead, the in-network server selection leverages these identifiers exposed as IP addresses to make in-band forwarding decisions, which are later bound to a SRv6 steering policy. In particular, upon arrival of a request at a CDN proxy, the IPv6 identifier is used by the prediction engine to perform cache admission by estimating the popularity of requests with a Least-Recently-Used (LRU) filter. If this is not available (cache miss), 6LB \cite{srlb} is used to forward requests directly to the origin servers instead of proxying them at the cache. According to the authors, this mechanism allows the load on the edge cache to be reduced and negative effects on the Quality of Experience to be avoided.

\cite{desmouceauxzero} proposes an SR based live migration technique achieving zero packet loss.
The starting point is an SRv6 enabled Data center network.
Two new SR functions are defined and used in the process of VM live migration.
The first one is \emph{forward to local if present}, which is a conditional version of the END behavior described in section~\ref{sec:sr_net_prog}. Specifically, in case the last SID in the SL is locally available, then the packet is directly processed, thus ignoring possible intermediate SIDs.
The second one is \emph{buffer and forward to local}.  This forces the node to inspect the last SID, and if it is not locally available, the packet is buffered until the SID becomes available.
Assuming that a VM is migrated from host A to host B, all requests received by the Data center gateway are then tunneled during the migration process using an SL whereby: i) the first SID points at the \emph{forward to local if present} function implemented at A, ii) the second SID points at the \emph{buffer and forward to local} function implemented at B, and iii) the last SID is referred to the VM.
In this way, until the VM is available at A, the packets are directly delivered to the VM thanks to the \emph{forward to local if present} function. Then, during the downtime, the packets directed to the VM are buffered at B thanks to the \emph{buffer and forward to local}. Finally, when the VM becomes available at B, the buffered packets, as well as new arrivals, are directly delivered to the VM at B.
In this way it is possible to implement VM live migration having no packet loss.

SRNK \cite{mayer2019efficient} is an SR-proxy for legacy VNFs which are unaware of SRv6 technology and expect to process traditional IP packets. SRNK extends the implementation of SRv6 in the Linux kernel \cite{lebrun2017implementing} adding the support for the \textit{End.AS} and \textit{End.AD} behaviors. The performance of the proposed solution has been evaluated by the authors, who identified a poor scalability with respect to the number of VNFs to be supported within an NFV node. With an enhancement of the Linux Policy Routing framework, they provided a second design SRNKv2, which does not depend on the number of supported VNFs in a node. They compared the performance with a reference scenario not performing the encapsulation and decapsulation operation and demonstrated that the overhead of SRNKv2 is very small, on the order of 3.5\%.

\cite{srv62, 17-vnf-chaining-sr, duchene2018srv6pipes, mayer2019efficient} have in common the Service Function chaining topic. At the same time, there are some important differences that have to be highlighted. \cite{srv62} adds an API in the SRv6 implementation of the Linux kernel to map the service segments. It mainly deals with SR-aware VNFs. Instead, \cite{mayer2019efficient} proposes a SR-proxy for SR-unaware VNFs. \cite{17-vnf-chaining-sr} implements a solution able to deal with both SR-aware and SR-unaware VNFs, but with respect to the previous works it extends the netfiler framework, while the previous solutions are extensions of the SRv6 implementation in the Linux kernel. 
Finally, \cite{duchene2018srv6pipes} is still an extension of the SRv6 implementation in the Linux kernel, but it the enables chaining of in-network functions operating on streams, thus working at a higher level (Transport layer) with respect to the previous works.

\subsection{Performance evaluation}
\label{sec:pe_eval}

In this subsection, we report \perfevalPapers papers dealing with the performance evaluations of SRv6 implementations.

SRPerf \cite{ahmedperformance} \cite{abdelsalam2020srperf} is a performance evaluation framework for software and hardware implementations of SRv6. SRPerf is able to perform different benchmarking tests such as throughput and latency. At the time of writing, the framework supports Linux kernel and VPP implementations of SRv6. The architecture of SRPerf can be easily extended to support new benchmarking methodologies as well as different SRv6 implementations. The framework supports two different metrics to characterize the throughput of a SRv6 enabled node: No-Drop Rate (NDR) and Partial Drop Rate (PDR). PDR is defined as the highest throughput achieved without dropping packets more than a predefined threshold. NDR, which corresponds to the Throughput defined by RFC 1242~\cite{rfc1242}, can be described as PDR with a threshold of 0\%. The framework orchestrates all aspects of an experiment starting from the setup of the testbed to the enforcement of SRv6 configurations, thus relieving the experimenter from a significant configuration effort. SRPerf has been used to evaluate the performance of SRv6 implementations in the Linux kernel and in VPP. Moreover, in \cite{abdelsalam2020srperf}, the authors propose the evaluation of an enhanced Linux kernel, which has been obtained by adding the implementation of missing behaviors and fixing the implementation of existing ones.

\cite{teamsegment} presents a solution whereby low-level network functions, such as SRv6 encapsulation, are offloaded to Intel FPGA programmable cards. In particular, the authors partially offload the SRv6 processing from a VPP software router to the NICs of the servers, thus increasing data-path performance and at the same time saving resources. These precious CPU cycles are made available for VNFs or for other workloads in execution on the servers. Tests results of the \textit{End.AD} use case show in the worst scenario a CPU saving of 67.5\%. Moreover, the maximum throughput achievable by a pure VPP solution with 12 cores is obtained by the accelerated solution using only 6 cores.

\cite{leeperformance} studies SRv6 as alternative user plane protocol to GTP-U \cite{gtpu}. Firstly, the authors propose an implementation of the GTP-U encap/decap functions and of the SRv6 stateless translation behaviors defined in \cite{id-srv6-mobile-uplane}. These behaviors guarantee the coexistence of the two protocols, which is crucial for a gradual roll-out. The authors used programmable data center switches to implement these data plane functionalities. Since it is hard to get telemetry from a commercial traffic generator when a translation takes place, the authors injected timestamp with a resolution of nanoseconds to measure the latency of SRv6 behaviors. Finally, they measured throughput and packet loss under light and heavy traffic conditions in a local environment. Results show no huge performance drop due to the SRv6 translation. Moreover, the latency of the SRv6 behaviors is similar to the GTP-U encap/decap functions.

\subsection{Miscellaneous}

In this section, we have included all works not belonging to previous categories. In general, these works are very different among them and have only Segment Routing in common.

\cite{abdullah2018segment} provides a tutorial on Segment Routing and a survey on research activity covering more than 50 scientific papers. The tutorial part focuses on the SR-MPLS data plane part and does not consider the recent SRv6 data plane. SR standardization efforts, implementation activities and deployments are not specifically analysed in \cite{abdullah2018segment}.

The Control Exchange Point (CXP) defined in \cite{cxp}, even if not explicitly based on SR, proposes the concept of \textit{pathlets} \cite{pathlet}, which closely resembles the idea of the list of instructions present in the SR architecture. The main goal of Control Exchange Point (CXP) is to provide services with QoS constraints across domains. This is achieved stitching the \textit{pathlets} which are partial paths advertised by domains. An ISP abstracts its network as a set of \textit{pathlets} connecting the network edges, and then advertises these on the northbound. More specifically, this abstraction is realized with tunnels instantiated with OF, MPLS, optical paths and so on. The \textit{pathlet} abstraction is bundled with properties that the ISP provides such as latency, costs, available bandwidth and so on. A CXP is an external entity acting as brokering layer and providing inter-domain routing coordination based on SDN APIs. The general idea seems to be in line with SR architecture presented so far, and it can be implemented using SR data-plane technologies.

\cite{scalablesegment1}, \cite{scalablesegment2} and \cite{scalablesegment3} propose an alternative implementation of SR architecture through Omnipresent Ethernet (OE), which is a modification of Carrier Ethernet architecture. It is based on source-routed, binary-routed labels embedded in an Ethernet frame. \cite{scalablesegment3} provides details on the implemented SR header. 
The authors of the aforementioned works address scalability issues of SR in the context of multi-domain scenarios from two different points of view: packet header size and the number of table entries required at the edge nodes. They realize also a testbed to validate the implementation of the proposed schemes. However, further analysis is not possible since the solution builds upon Carrier Ethernet hardware.

\cite{schuller2018practical} analyzes the TE problem for SR from a different perspective: the authors evaluate the influence of the metrics used to define the cost of the links. Key findings of the study indicate that the routing metric can influence shortest path based TE. Very simple and reasonable metrics such as inverse capacity work as well as a complex optimized metric. They allow this to be close to the optimum with respect to the most common traffic engineering objective of minimizing maximum utilization. Finally, if other objectives are introduced besides optimizing link utilization, the choice of the metric is more important and there are significant differences between the metrics. Although \cite{schuller2018practical} deals with Traffic Engineering, it does not provide a new heuristic nor a different way to address the problem, rather it is simply a study on the influence of the routing metrics. Therefore we have not classified it in the TE category.

In \cite{chi2018live}, the authors take advantage of Segment Routing to build an SR-based Multicast delivery mechanism to efficiently 
provide live video streaming services for 5G users. SR is used to alleviate the rule update overhead and to avoid the explosion of routing tables in the devices. The work builds upon an interesting problem, although it does not provide a lot of details about SR implementation. The work mainly focuses on the problem of building an efficient Multicast tree to further reduce the rules update due to the user movements (Handover-aware Multicast tree).

\cite{cao-industrial-iot} considers an Industrial IoT (Internet of Things) scenario with an extremely large number of objects to be connected, also taking into account their mobility. The proposed solution relies on Segment Routing for enabling scalability and flexibility in packet forwarding. This is in particular to bypass the overloaded links and to achieve load balance.

\cite{mayer-network-as-computer} proposes a novel distributed processing model for IoT, based on the extension of the SRv6 Network Programming Model. The idea is that each IoT node offers an abstract machine that can be programmed using an Instruction Set Architecture. The program can be embedded in an SRv6 segment list. An SRv6 packet carries both the program and the execution state. It can travel across IoT nodes, reading and writing the I/O ports of the device and executing computations as dictated by the program in the packet itself.

\section{SR implementations and deployments}
\label{sec:tools}

In this section, we describe the implementation results related to SR. We will mostly focus on the SRv6 version, which is attracting a lot of interest and development efforts. The SR-MPLS version is already in a mature development phase, well supported by the main core router vendors (e.g. Cisco, Huawei, Juniper). SR-MPLS can be incrementally deployed in current IP-MPLS backbones, as it only requires software updates to networking devices. Operators can migrate to SR-MPLS to simplify the control plane operations and improve the scalability. As for the SRv6 data plane, there are two main Open Source data plane implementations for software routers: the Linux kernel implementation (described in Subsection \ref{sec:linux}) and the implementation done by the FD.io VPP project (described in Subsection \ref{sec:vpp}).  Section \ref{sec:rest} presents other open source implementations, mostly related to research activities. Finally, in Subsection \ref{sec:hw_interop} we briefly analyze the hardware implementations of SRv6, the inter-operability efforts done by several vendors and the current deployments of SRv6 in large production networks.

\subsection{Linux kernel}
\label{sec:linux}

The SRv6 capabilities were first added in Linux kernel 4.10 \cite{lebrun2017implementing}. Kernel 4.10 includes the support for some \textit{SRv6 Policy Headend} behaviors (formerly known as \textit{transit} behaviors) (e.g., \textit{H.Insert} and \textit{H.Encaps}). The \textit{SRv6 Policy Headend} behaviors are implemented as Linux Lightweight Tunnel (LWT). The implementation of the iproute2~\cite{iproute2} user space utility is extended to support adding a \textit{localsid} associated with an \textit{SRv6 Policy Headend} behavior~\cite{srv6-impl-basic}. SRv6 \textit{localsids} with \textit{SRv6 Policy Headend} behavior are added as IPv6 FIB entries into the kernel main routing table. Kernel 4.14 is another important milestone for the SRv6 support in Linux: a set of SRv6 \textit{endpoint} behaviors have been implemented by adding a new type of LWT~\cite{lebrun2017reaping}. The supported SRv6 \textit{endpoint} behaviors are \textit{End.X}, \textit{End.T}, \textit{End.DX2}, \textit{End.DX4}. \textit{End.DX6}, \textit{End.DT6}, \textit{End.B6}, and \textit{End.B6.Encaps}. Some new \textit{SRv6 Policy Headend} behaviors have been added (e.g., \textit{H.Encaps.L2}). The iproute2 implementation was extended as well \cite{iproute2} \cite{srv6-impl-adv}. The SRv6 capabilities in Linux kernel were extended in kernel 4.16 \cite{kernel4-16} to include the netfilter framework \cite{netfilter}. A new iptables match extension, named \texttt{srh}, was added to the kernel to support matching of SRH fields. The \texttt{srh} match extension is a part of the SERA firewall~\cite{abdelsalam2018sera} and supports matching all the fields of the SRH. The implementation of iptables user space utility \cite{wiki-iptables} is extended with a new shared library (\texttt{libip6t\_srh}) that allows to define iptables rules with \texttt{srh} options.

Linux Kernel 4.18 \cite{kernel4-18} has added few more features both in the core SRv6 stack and in the netfilter framework. In the netfilter framework, the \texttt{srh} match is extended to provide the matching of SRH's \textit{Previous SID}, \textit{Next SID}, and \textit{Last SID}. The iptables user space utility is updated as well to support the new matching options. Instead, a new feature is added in the Linux SRv6 stack to support custom SRv6 network functions implemented as small eBPF \cite{lwn-ebpf} programs. \cite{xhonneux2018leveraging} extends \cite{id-srv6-network-prog} introducing a new \textit{End} behavior the so called \textit{End.BPF}. From an implementation point of view, a new hook for BPF is added to the SRv6 infrastructure that can be used by network operators to attach small programs written in \textit{C} to SRv6 SIDs which have direct access to the Ethernet frames. Moreover, specific SRv6-BPF helpers have been provided in order to allow \textit{End.BPF} functions to execute basic SRv6 actions (\textit{End.X}, \textit{End.T} and many others) or adding TLVs. This allows custom \textit{SRv6 Policy Headend} behaviors to be implemented (mainly to extend SRv6 encapsulation policies implemented by the kernel). The tutorial about eBPF extensions to SRv6 is available at \cite{srv6-ebpf-tutorial}. The source code of the sample applications described in \cite{xhonneux2018leveraging} is freely available at \cite{srv6-ebpf-code1}. Instead, the eBPF-based fast-reroute and failure detection schemes described in \cite{xhonneux2018flexible} is available at \cite{srv6-ebpf-code2}.

SR-MPLS has not received the same attention of the SRv6 implementation in the Linux kernel. All the features which are available are mostly related to the well-established MPLS forwarding. They have been made available from the version 4.1 of the kernel. In particular, kernel v4.1 has seen the introduction of the MPLS Label Switching Router (LSR) behavior. MPLS capabilities have been extended later in the kernel v4.3. LWT framework and MPLS tunnel were added allowing the implementation of the MPLS Label Edge Router (LER) behavior. Finally, MPLS multipath functionality has been added only in the version 4.5 of the kernel.

In general, the Linux kernel lacks of the support of the SR policy framework which is instead available for FD.io VPP implementation (Subsection \ref{sec:vpp}). This means that at the time of writing is not possible to create  an SR policy (both MPLS and IPv6) and associate a \textit{BindingSID} to it nor instantiate SR-MPLS/SRv6 steering rules pointing to SR-MPLS/SRv6 policies.

\subsection{FD.io VPP}
\label{sec:vpp}

FD.io Vector Packet Processing (VPP) \cite{fd-io-vpp} platform is an extensible framework that provides out-of-the-box production quality switch/router functionality that can run on commodity CPUs. VPP  17.04 included the support for the \textit{SRv6 Policy Headend} (formerly known as \textit{transit}) behaviors and most of the \textit{endpoint} behaviors defined in \cite{id-srv6-network-prog}. These behaviors are implemented in dedicated VPP graph nodes. The SRv6 graph nodes perform the required SRv6 behaviors as well the IPv6 processing (e.g. decrements Hop Limit). Whenever an SRv6 segment is instantiated, a new IPv6 FIB entry is created for the segment address pointing to the corresponding VPP graph node. Release 17.04 also brought SR headend capabilities to VPP by introducing the concept of SR policy in the SRv6 implementation. In VPP, an SR policy is uniquely identified by its \textit{BindingSID} address, which serves as a key to a particular SR policy. This is not compliant with the SR policy definition \cite{id-segment-routing-policy}, but a reasonable shortcut considering the absence of control-plane capabilities in VPP. 

The SR policies in VPP support several SID lists with weighted load-balancing of the traffic among them. When a new segment list is specified for an SR policy, VPP pre-computes the rewrite string that will be used upon steering traffic into that SID list, either via  an \textit{SR Policy Headend} behavior or a \textit{BindingSID}. VPP then initializes one FIB entry for the SR policy \textit{BindingSID} in the FIB and an entry in a hidden FIB table for the IPv6 traffic steered into the SR policy via  an \textit{SR Policy Headend} behavior. Each one of these FIB entries points to the SR policy object, which in turn recurses on the weighted segment lists.

Traffic can be steered into an SR policy either by sending it to the corresponding \textit{BindingSID} or by configuring a rule, called steering policy, that directs all traffic transiting towards a particular IP prefix or L2 interface into  an SRv6 policy. The latter mechanism is implemented as FIB entry for the steered traffic in the main FIB to be resolved via the FIB entry of the SR policy in the hidden FIB table. In this way, a hierarchical FIB structure is realized: the traffic is not directly steered over an SR policy, but instead directed to a hidden FIB entry associated with the policy. This allows the SR policy to be modified without requiring any change to the steering rules that point towards it.

Release 17.04 has also seen the introduction of the SRv6 LocalSID development framework and the SR-MPLS implementation. The former is an API which allows developers to create new SRv6 \textit{endpoint} behaviors using the VPP plugin framework. The principle is that the developer only codes the actual behavior, i.e. the VPP graph node. Instead, the segment instantiation, listing and removal are performed by the existing SRv6 code. The SR-MPLS framework is focused on the SR policies, as well on its steering. Likewise in SRv6, an SR policy is defined by a MPLS label representing the \textit{BindingSID} and a weighted set of MPLS stacks of labels. Spray policies are a specific type of SR-MPLS policies where the packet is replicated on all the SID lists, rather than load-balanced among them. To  steer packets in transit into an SR-MPLS policy, the user has to  create an SR-MPLS steering policy. Instead, others SR-MPLS features, such as for example adjacency SIDs, can be achieved using the regular VPP MPLS implementation. In release 18.04, service programming proxy behaviors \textit{End.AS}, \textit{End.AD} and \textit{End.AM} were introduced as VPP plugins leveraging the framework described before.

\subsection{Other open source implementations}
\label{sec:rest}

Several research efforts analyzed in section~\ref{sec:research} have released the components and the extensions realized for SR as open source. Some of the them build upon the implementations described in the previous Subsections, while other ones propose alternative solutions. The SREXT module (\cite{17-vnf-chaining-sr}) provides a complementary implementation of SRv6 in Linux based nodes. When it was designed, the Linux kernel only offered the basic SRv6 processing (\textit{End} behavior). SREXT complemented the SRv6 Linux kernel implementation providing a set of behaviors that were not supported yet. Currently most of the behaviors implemented in SREXT are supported by the mainline of Linux kernel (with the exception of the SR proxy behaviors). SREXT provides an additional local SID table which coexists with the one maintained by the Linux kernel. The SREXT module registers itself as a callback function in the \textit{pre-routing} hook of the netfilter \cite{netfilter} framework. Since its position is at beginning of the netfilter processing, it is invoked for each received IPv6 packet. If the destination IPv6 address matches an entry in the local SID table, the associated behavior is applied otherwise the packet will follow the normal processing of the routing subsystem. The source code of SREXT together with the Vagrant box are available at \cite{srext-home}. Using the Vagrant box, it is possible to bootstrap a small testbed in few minutes and start the experiments on SREXT features.

FRRouting (FRR) \cite{frr} is an open source routing protocol stack for Linux forked from Quagga \cite{quagga}. In FRR, there is an experimental support \cite{frr-sr} of the draft \cite{ietf-ospf-segment-routing-extensions} which defines the OSPFv2 extensions for Segment Routing (SR-MPLS). At the time of writing, there is no support for SRv6.

The SPRING-OPEN project \cite{springopen} provides an SDN-based implementation of SR-MPLS. The architecture is based on a classic SDN control plane (logically centralized), built on top of ONOS. Part of this work converged later in Trellis \cite{trellis}, an open-source multi-purpose leaf-spine fabric supporting distributed access networks, NFV and edge cloud applications. Trellis includes also the support of P4/P4Runtime \cite{p4} \cite{p4runtime} devices as well as Stratum \cite{stratum} enabled devices. Trellis has been used as underlay/overlay fabric in the CORD project \cite{cord} which aims at redesigning central-office architectures. Recently, it has been integrated in the SEBA project \cite{SEBA} which targets residential-access networks. All the software stack and the documentation is freely available on \cite{trellis-doc}. Moreover, a tutorial together with a ready-to-go VM can be downloaded from \cite{trellis-tutorial}.

PMSR (\cite{pmsr} and \cite{trafficpmsr}) provides an open source implementation of SR-MPLS together with the realization of  an SDN control plane. The data plane leverages the OSHI architecture (\cite{oshi1}, \cite{oshi2}) which combines  an SDN data plane, implemented with Open vSwitch \cite{ovs}, and OSPFv2 control logic, realized with Quagga. This architecture is extended in PMSR with the introduction of a Routes Extraction entity which connects to Quagga and receives routes update using the FPM interface provided by Quagga \cite{quagga}. These routes are then translated in SIDs and installed in the SDN data plane as OpenFlow MPLS forwarding rules. Authors provide a set of management tools \cite{mantoo} which assist experimenters and relieve them from a huge configuration effort. A tutorial to start working with PMSR is available on \cite{pmsr-tutorial}; instead a ready-to-go VM with all the dependencies installed can be downloaded from \cite{oshi-home}.

Software Resolved Network (SRN) (described in \cite{lebrun2018software} and \cite{duchene2018exploring}) is a variant of the SDN architecture. The network controller is logically centralized and co-located with a DNS resolver and uses extensions of the DNS protocol to interact with end-hosts. The Open vSwitch Database Management Protocol (OVSDB) \cite{rfc7047} is used to enable the communication between SDN controller and the network nodes: i) the latter populates the distributed database with the topology information and TE metadata; ii) the former once computed the path, upon a request, populates the OVSDB instance with the SRv6 Segment list matching the desired requirements. Finally, this is pulled by the access device which enables the communication of the end-hosts. The source code is freely available at \cite{srn-code}. An overview of the architecture can be found in \cite{srn-overview}. A ready-to-go VM with packaged experiments can be created using the instructions in \cite{srn-vm}.

\cite{ventre2018sdn} proposes a classical SDN architecture for SRv6 technology: a centralized logic takes decisions on the Segment Lists that need to be applied to implement the services, then the SDN controller, using a southbound API, interacts with the SR enabled devices to enforce the application of such Segment Lists. The code related to the SDN architecture, i.e. the four different implementations of the Southbound API and the topology discovery, can be downloaded from the page of the project \cite{srv6-sdn}. In addition the authors, to support both the development and testing aspects, have realized an Intent based emulation system to build realistic and reproducible experiments relieving the experimenters from a huge configuration effort. The emulation tools are available at \cite{rose}.

SRV6Pipes \cite{duchene2018srv6pipes} is an extension of the SRv6 implementation in the Linux kernel \cite{lebrun2017implementing} which enables chaining and operation of in-network functions operating on streams. SRv6 policies are installed using the SRN architecture \cite{lebrun2018software} described earlier in this section. However, SRN components are not mandatory since SRv6 policies can be installed in the edge nodes using the \textit{iproute} utility. SRv6Pipes is composed by multiple components that necessarily need to be installed in the target machines: TCP proxy, patched Kernel and user space utilities. The minimum components can be downloaded from the repository of the project \cite{srv6pipes-code}. The complete code of the experiments together with a walkthrough can be found in \cite{srv6pipes}.

SRNK \cite{mayer2019efficient} extends the implementation of SRv6 in the Linux kernel \cite{lebrun2017implementing} adding the support for the \textit{End.AS} and \textit{End.AD} behaviors. The source code is freely available at \cite{srnk-home}, where it is possible to download the patched Linux kernel (starting from 4.14.0 branch) and the patched iproute2 (starting from iproute2-ss171112 tag). Instead, the detailed configurations steps of the SR-proxy are reported in the Appendices A and B of \cite{mayer2019efficient}.

\cite{aubry2018robustly} describes the implementation of a path computation element able to compute robust disjoints SR paths which remain  disjoint even after an input set of failures without the need of configuration changes. The java implementation of the algorithms, the public topologies used for the experiments, the experimental results and a detailed walkthrough to replicate the experiments of the paper are available at \cite{robustly-home}.

\subsection{Hardware implementations, inter-operability efforts and deployments for SRv6}
\label{sec:hw_interop}

\cite{matsushima-spring-srv6-deployment} provides an overview of IPv6 Segment Routing implementations, describes some interoperability scenarios that have been demonstrated in public events and reports a list of recent SRv6 deployments in production networks.


With regards to the hardware implementations  \cite{matsushima-spring-srv6-deployment} mentions 8 vendors that declare production support of SRv6 in their hardware. The platforms ASR 1000, ASR 9000, NCS 5500, NCS 540 and NCS 560 are reported as the Cisco Routing platforms supporting SRH processing; ASR 9000 and NCS 5500 being deployed in production networks. As for Huawei, the reported platforms are: ATN, CX600, NE40E, ME60, NE5000E, NE9000 and NG-OLT MA5800, all with VRPV8 shipping code. The programmable devices based on Tofino chipset \cite{barefoot} can be programmed to support SRH processing. This is also true for the reference software implementation of the P4 devices \cite{bmv2}, the Stratum based devices \cite{stratum} and all the programmable chipsets (Cavium Xpliant \cite{cavium} to give an example). Other SRv6 hardware implementations are reported for the Prestera family of Ethernet switches by Marvell, for the SkyFlux UAR500 router by UTStarcom. Spirent and Ixia also support SRv6, respectively in their TestCenter and IxNetwork testing platforms. Finally, as reported in \cite{rfc8754}, Juniper's Trio and vTrio NPUs have an experimental support of SRH (SRH insertion mode and \textit{End} processing of interfaces addresses). 

Three interoperability events are listed in \cite{matsushima-spring-srv6-deployment}. The first (in chronological order) interoperability testing scenarios showcased at the 2017 SIGCOMM conference \cite{interop-demo}. The set of experiments included a L3 VPN scenario augmented with TE functionality and services function chaining processing. SREXT, VPP, Linux kernel, Barefoot Tofino, Cisco NCS5500 and Cisco ASR1000 routers were the network devices implementing SRv6 behaviors. Iptables (firewall) and Snort (Intrusion Detection System) have been used as service functions. Finally, Wireshark and tcpdump have been leveraged to verify the proper operations of the network. The second set of interoperability test scenarios were run in March 2018 by the European Advanced Networking Test Center (EANTC) and their results \cite{eantc-2018} were presented at the MPLS + SDN + NFV World Congress conference in April 2018. In these tests, the implementations of CISCO and UTStarcom and the testing platforms of Ixia and Spirent were involved. The tests concerned Layer 3 IPv4 VPNs based on SRv6 (also including Traffic Engineering features in the SRv6 underlay) and SRH based Topology Independent (TI-LFA) Fast Reroute mechanisms. The third set of interoperability test scenarios were run in March 2019 and their results \cite{eantc-2018} were presented at the MPLS + SDN + NFV World Congress conference in April 2019. In these tests, a routing platform from Cisco (NCS 5500) and two from Huawei (NE9000-8 and NE40E-F1A) were involved. The tests concerned Layer 3 IPv4 and IPv6 VPNs based on SRV6, the validation of some SRv6 behaviors, SRv6 based fast reroute and OAM procedures (Ping and traceroute) based on \cite{id-srv6-oam}.

Eight large scale deployments of SRv6 are listed in \cite{matsushima-spring-srv6-deployment}, involving the following nationwide operator networks: Softbank (Japan), China Telecom (China), Iliad (Italy), LINE Corporation (Japan), China Unicom (China), CERNET2 (China), MTN Uganda Ltd (Uganda), NOIA Network. We refer the reader to \cite{matsushima-spring-srv6-deployment} for the details about these deployments.


The draft \cite{matsushima-spring-srv6-deployment} elaborates also on the open source applications supporting the processing of the IPv6 Segment Routing header, among which we mention the well known Wireshark \cite{wireshark}, tcpdump \cite{tcpdump}, iptables \cite{iptables}, nftables \cite{nftables} and snort \cite{snort}.

\cite{teamsegment} reports the implementation of a solution that offloads the SRH encapsulation, decapsulation operations and the SRv6 cache handling from the servers to the Intel FPGA programmable cards. Open Programmable Accelerator Engine \cite{opae} developed by Intel and the VPP software router are used as basic frameworks. The authors defined a new functional splitting of the SRv6 behaviors between the hardware and the software and the inter-working is realized introducing new graph nodes in VPP. These extensions to the VPP graph are able to encode/decode/process metadata exchanged between VPP and the Intel cards. For example, in the ingress direction, the card performs a SID lookup, and based on lookup result can strip the outer IPv6/SRH headers and add its own metadata header. Since VPP is augmented with new nodes is able to process this meta-header and the inner packets. As explained  above, the solution partially offloads the functions to the hardware for example Route lookup and in general control plane functions are still done in software by VPP leveraging the CPU of the servers.


\section{Lessons learned}
\label{sec:lesson}

In this section we describe the main outcomes learned during the survey activity. We first focus on some general considerations, then we elaborate on research related outcomes.

Despite the fact that the concept of Source Routing had already been proposed in the past (late 70’), its deployment in IPv4 networks was not supported, mainly due to security issues. Novel SR architecture provides a secure solution for the deployment of the concept of source routing in operator's networks. SR also provide a scalable solution: i) the number of flow states to be maintained in network nodes is highly reduced, ii) only border nodes are involved by classification procedures and transit nodes do not maintain any flow state information.

SR-MPLS was the first deployment of SR concepts, only in the last few years SRv6 architecture has been defined.
SR-MPLS architecture has, in our view, the main merit of not requiring any change to the MPLS forwarding plane. In this way, SR-MPLS represents a simple solution for the service provider with an IPv4 infrastructure. The same considerations apply to SRv6, the transition strategy from a classical IPv6 network to an SRv6 one can be realized in successive steps with no service disruption.

We have also noted that SR-MPLS implementation has not received the same attention as SRv6 is receiving from the research community. We believe that having available different open source implementations (Linux kernel and VPP) from the beginning has contributed decisively to this. Another reason is to be found in the required control plane extensions. Even though SR-MPLS has been the first implementation, the researchers had to struggle to have something to work with. MPLS data plane implementation has been unstable for a long time in the Linux kernel, and a completely revised implementation has been merged quite recently. Finally, SR-MPLS requires extensions to the routing protocols and there were no such extensions at the time that SR was initially being studied.

One of the main features of SR is the native support of NFV/SFC scenarios. Furthermore, the Network Programming model of SRv6 offers the possibility of virtualizing any service by combining the basic functions in a network program that is embedded in the packet header (implementing a network programming model on MPLS dataplane would not be scalable despite its feasibility). In this context, we believe that SRv6 can simplify network architectures with respect to SR-MPLS, since the use of different tunneling solutions can be avoided. 
As such, SRv6 is an attractive choice for operators that are deploying new networks or planning the evolution of their networking architectures.

During the revision of the Internet Drafts, RFCs and patents we noticed that there is a clear trend in the standardization activities to \dq{follow} patents. In facts, most patents have also been standardized later or have been included in several standardization activities led by the same vendors. The most recent patents also highlight that SR can have a significant role in the deployment of networking core technologies, such as network slicing, 5G and cloud-based networking.  

Regarding the outcomes of the research activity and following the classification provided in Section IV, we will briefly describe the most significant lessons coming from different research categories.


Research activity related to the \emph{Monitoring} category (section \ref{sec:mon}) highlights a significant interest towards solutions to improve existing monitoring tools or to define new tools exploiting SR features. Contrarily, there are lack of works focused on SR failure monitoring, such as misconfigurations, undefined SIDs, SR related black holes, packet loss, etc.

Moving to the \emph{Traffic Engineering} category (section \ref{sec:te}), the main outcome of the reviewing activity is that one of the interesting features of SR routing, i.e. the capability of splitting the traffic over several SLs, is not exploited in practice. This unexpected outcome can be motivated by the capability of obtaining an high routing flexibility also with a single SL for each flow; in any case, the use of multiple SL per flow can be considered in the future to further improve TE solutions.



An additional consideration comes from the \emph{Centrally Controlled Architecture} category (section \ref{sec:central_control}): the scientific community is oriented towards a centralized control plane for SR-MPLS networks, even if a distributed control plane is also supported. However, this approach requires extensions in the routing protocols in the case of MPLS, which represented a major barrier, thus causing many works to decide to leverage the openness of the SDN approach to overcome such limitations. It is also interesting to note that same trend cannot be found with works related to SRv6, where such extensions in routing protocols are not strictly required.

The \emph{Path Encoding} (section \ref{sec:path}) related works highlight two interesting outcomes. As first, the large effort taken by researchers in reducing the overhead in the packet headers demonstrates that is seen as a limiting factor for SR deployments. Two schools of thought can be found: i) optimize the translation of the paths into SLs; ii) reduce its impact using a smarter encoding of the paths into segment lists.

The second lesson learned is related to the adj-sids; they are largely used by path encoding works to increase routing flexibility, while they are not fully exploited in TE works: the reason could be that including adj-sids leads to an increase in complexity for optimization models used by TE algorithms. It will be interesting to investigate the impact of adj-sids in novel SR-based TE solutions both in terms of performance and complexity.

The works related to the \emph{Network Programming} category (section \ref{sec:netprog}) suggest that this SRv6 feature opens up a wide range of research opportunities: it can be applied in different use cases, from Service Function Chaining to the implementation of complex operations (such as the management of VM migration in a DC network). We believe that network programming will attract significant attention in the coming years.

Finally, an interesting consideration of the \emph{Performance Evaluation} category (section \ref{sec:pe_eval}) is related to the topics of the analyzed works. All papers concentrate on the data plane, while the only work studying control plane performance is \cite{ventre2018sdn} (actually only a performance evaluation of different southbound implementations has been conducted).

\section{Future research directions}
\label{sec:future}

Most of the SR works we have reviewed have focused on the definition of novel solutions for classical network problems (such as Monitoring, Traffic Engineering and Failure Recovery) or on optimization of specific SR procedures (such as Path Encoding). In general, these works showed that SR can provide significant enhancements with respect to other solutions and we believe that there is still room and interest for extending the achieved results in these areas. In addition, we try to identify and discuss a set of research directions for Segment Routing that are definitely worth exploring in the near future: i) Service Function Chaining support, ii) SRv6 end-host implementation aspects, iii) Cloud Orchestration, iv) Integration with Applications, v) 5G, vi.) IoT. All these research areas are based on SRv6, i.e. on Segment Routing over the IPv6 data plane, as we believe that the future evolution of Segment Routing will be based on SRv6.
\subsection{Service Function Chaining support}
The Programmability feature of SRv6 represents an enabling factor for the implementation of Network Function Virtualization (NFV) and Service Function Chaining (SFC) in provider networks. In this regard, new abstraction models for the management of Network Functions by means of dedicated SRv6 control procedures could be studied.
\subsection{SRv6 end-host implementation aspects}
Another interesting topic is related to the implementation of SRv6 in end-hosts. One aspect is related to moving SRv6 functions in end-hosts from the software closer to the hardware with SmartNICs. Programmable NICs allow to implement network traffic processing on the NIC instead of using the CPU of the end-nodes/devices. Another aspect is related to exploiting the recent advances in Linux kernel networking for fast packet processing, namely eBPF \cite{lwn-ebpf} and XDP \cite{xdp-wikipedia} for implementing SRv6 functions. 
\subsection{Cloud Orchestration}
The third research opportunity regards integration of the SRv6 technology into Cloud orchestrators like OpenStack \cite{openstack} and Kubernetes \cite{kubernetes}. Considering Data Center networking scenarios, it will be possible to replace actual data plane mechanism based on legacy tunneling mechanisms like VXLAN with SRv6, with a drastic simplification of the network stack: the needed information will be integrated into SRv6 SIDs and/or in the TLV field, with no need of dedicated headers for tunneling.
\subsection{Integration with Applications}
Allowing direct interaction of applications with SRv6 features could enable innovative services and improve the efficiency of existing ones. Applications could use SRv6 SIDs to express their service requirements and to interact with network features, dynamically participating in the definition/composition of network services. To achieve this interaction, first the APIs of the operating system (e.g. Linux) need  to be extended, then SRv6 aware applications need  to be developed. 
\subsection{5G}
SRv6 is being considered for the data plane of future releases of 5G networks thanks to its stateless traffic steering and programmability features. On one hand, SRv6 could support 5G features like network slicing, on the other hand, it will be important to evaluate the performance of SRv6 based data plane, to verify that strict 5G constraints on latency are met.
\subsection{Internet of Things}
Considering the problem space of Internet of Things, which includes scalability aspects, routing aspects, interactions between networking and application layers, the application of the SRv6 architecture seems very promising. 

\section{Conclusion}
\label{sec:conclusion}

Segment Routing technology is based on source routing and tunneling paradigms. Segment Routing supports services such as Traffic Engineering, Virtual Private Networks, Fast Restoration in IP backbones and datacenters, and has proved to be flexible in supporting new use cases. Moreover, SR architecture reduces the amount of state information that needs to be configured in the core nodes. 

SR-MPLS and SRv6 are the two data plane instantiations of SR architecture. This is the first tutorial and survey work covering in detail the novel SRv6 solution (i.e. SR over IPv6 data plane), which represents the most promising implementation for future research activity. SRv6 provides a consistent solution for solving long-term problems in IP networks, simplifying protocol stacks and improving scalability with respect to current solutions.

In the survey we covered standardization work, patents and research activities related to Segment Routing. We also considered the recent deployments of SR in real networks and existing SR implementations, with a focus on the open source tools that can support SR research and development activities.

As for research activities, we covered about 90 scientific papers related to SR and proposed a taxonomy for their classification. One of the main outcomes of the classification was to identify relationships between SR features and research topics. For instance, the source routing paradigm has turned out to be the key enabling feature for the implementation of Traffic Engineering solutions in an SR network, while the routing flexibility feature is mainly used to realize network monitoring tools. We also identified the most interesting SR standardization documents and provided a taxonomy for their classification. A number of patents related to Segment Routing has also been discussed.

The review of SR implementations has highlighted the maturity of open source solutions based on Linux kernel and VPP. Both Linux and VPP, widely used by the research and developer community, allow the easy deployment of a virtual SR playground. As part of our survey activity we have also reported our vision and our experience in terms of lessons learned and future research topics.

We hope that this tutorial and survey work will draw further attention from the research community to Segment Routing technology and motivate new researchers to join the development of new use cases and standardization efforts. We have anticipated future research directions which can be taken as starting points.

New versions of this survey will be available at \cite{ventre2019survey}. Moreover, we strongly encourage the community to provide feedback and updates as new research works and SR deployments come out and technology evolves. For this reason we have created a public repository\footnote{https://github.com/netgroup/sr-survey}, where interested researchers can contribute and update this documentation.

\section*{Acknowledgments}
This work has received funding from the Cisco University Research Program Fund and from the EU H2020 5G-EVE project.

\bibliographystyle{IEEEtran}
\bibliography{references}

\extended{
\vspace{-4em}
\begin{IEEEbiography}[{\includegraphics[width=1in,height=1.25in,clip,keepaspectratio]{fig/bio/ventre.jpg}}]{Pier Luigi Ventre}
is a Member of Technical Staff at the Open Networking Foundation (ONF), where he works on Trellis - the leading open-source leaf-spine fabric. Before joining ONF, Pier Luigi Ventre worked at CNIT as a researcher and post-doctoral researcher on several projects funded by the EU. He received his PhD in Electronics Engineering in 2018 from University of Rome "Tor Vergata". From 2013 to 2015, he was one of the beneficiaries of the scholarship "Orio Carlini" granted by the Italian NREN GARR. His main interests focus on Software Defined Networking, Network Function Virtualization, Virtualization and Segment Routing.
\end{IEEEbiography}
\begin{IEEEbiography}[{\includegraphics[width=0.9in,height=1.10in,clip,keepaspectratio]{fig/bio/salsano.png}}]{Stefano Salsano}
(M'98-SM'13) received his PhD from the University of Rome \dq{La Sapienza} in 1998. He is Associate Professor at the University of Rome Tor Vergata. Since July 2018 he has been the Coordinator of the Bachelor's Degree \dq{Ingegneria di Internet} and of the Master's Degree \dq{ICT and Internet Engineering} He has participated in 16 research projects funded by the EU, being project coordinator for one of them and technical coordinator for two. He has been PI in several research and technology transfer contracts funded by industries. 
His current research interests include SDN, Network Virtualization, Cybersecurity, Information Centric Networking. He is co-author of an IETF RFC and of more than 160 peer-reviewed papers and book chapters.
\end{IEEEbiography}
\begin{IEEEbiography}[{\includegraphics[width=0.9in,height=1.10in,clip,keepaspectratio]{fig/bio/marco.jpeg}}]{Marco Polverini} received a Master's Degree in Telecommunications Engineering and a Ph.D. Degree in Information and Communication Engineering from the University of Rome La Sapienza in 2010 and 2014 respectively, where he is currently a Research Fellow at the Department of Information, Electronic and Telecommunications Engineering. His main research interests are routing protocols for energy saving in IP networks, network traffic monitoring and measurement in next generation routing technologies.
\end{IEEEbiography}
\begin{IEEEbiography}[{\includegraphics[width=0.9in,height=1.10in,clip,keepaspectratio]{fig/bio/cianfrani.pdf}}]{Antonio Cianfrani} received a master's degree in telecommunications engineering and a Ph.D. degree in information and communication engineering from the University of Rome La Sapienza in 2004 and 2008 respectively. He is currently an Associate Professor in the DIET Department, University of Rome La Sapienza. His fields of interest include routing algorithms, network protocols, performance evaluation of software routers and green networks. His current research interests are focused on segment routing and traffic matrix assessment. He is co-author of more than 70 peer-reviewed papers and book chapters. He serves on the Editorial Boards of the IEEE Transactions on Green Communications and Networking.
\end{IEEEbiography}
\begin{IEEEbiography}[{\includegraphics[width=1in,height=1.25in,clip,keepaspectratio]{fig/bio/ahmed.jpeg}}]{Ahmed Abdelsalam} is a software engineer in the Segment Routing v6 at Cisco. He is also a PhD student in computer science at the Gran Sasso Science Institute (GSSI). His main research interests focus on IPv6 Segment Routing (SRv6), Network Function Virtualization (NFV), Software Defined Networks (SDN), Service Function Chaining (SFC), and container networking. He has contributed to many Open sources including the Linux kernel, Tcpdump, IPTables, and Snort.
\end{IEEEbiography}
\begin{IEEEbiography}[{\includegraphics[width=1in,height=1.25in,clip,keepaspectratio]{fig/bio/filsfils.jpg}}]{Clarence Filsfils}
is a Cisco Systems Fellow, has a 20-year expertise leading innovation, productization, marketing and deployment for Cisco Systems. He invented the Segment Routing Technology and is leading its productization, marketing and deployment. Previously, he invented and led the Fast Routing Convergence Technology and was the lead designer for Cisco System's QoS deployments. Clarence is a regular speaker at leading industry conferences. He holds over 130 patents and is a prolific writer, both in academic circles and in the standardization or books.
\end{IEEEbiography}
\begin{IEEEbiography}[{\includegraphics[width=1in,height=1.25in,clip,keepaspectratio]{fig/bio/fclad.jpeg}}]{Francois Clad} received an M.Sc and Ph.D. degree in computer science from the University of Strasbourg, France in 2011 and 2014 respectively. He spent one year as a Post-Doctoral researcher with the Institute IMDEA Networks, Madrid, Spain, before joining Cisco in 2015. His research activities are focused on IP routing and in particular evolving the Segment Routing technology.
\end{IEEEbiography}
\begin{IEEEbiography}[{\includegraphics[width=1in,height=1.25in,clip,keepaspectratio]{fig/bio/pcamaril.JPG}}]{Pablo Camarillo} is one of the engineers behind Segment Routing v6 at Cisco. He is coauthor of various IETF drafts, holds several patents and has developed SR implementation in FD.io VPP. Prior to joining Cisco, he was a research engineer at the IMDEA Networks Institute, where he prototyped a BGP route server in ExaBGP and researched the algorithmic of TI-LFA (SR Topology Independent Loop Free Alternates).
\end{IEEEbiography}}
\end{document}